%%%%%%%%%%%%%%%%%%%%%%%%%%%%%%%%%%%%%%%%%%%%%%%%%%%%%%%%%%%%%%%%%%%%%%%%%
%                                                                        %
%                                                                        % 
%                  by J. Bouttier and  E. Guitter                        %
%                TEX file, using lanlmac.tex macros                      %
%                                                                        %
%                                                                        %
%%%%%%%%%%%%%%%%%%%%%%%%%%%%%%%%%%%%%%%%%%%%%%%%%%%%%%%%%%%%%%%%%%%%%%%%%%
\input lanlmac
\def\href#1#2{{#2}}
\def\hhref#1{{#1}}
\input epsf.tex

\overfullrule=0mm

\newcount\figno
\figno=0
\def\fig#1#2#3{
\par\begingroup\parindent=0pt\leftskip=1cm\rightskip=1cm\parindent=0pt
\baselineskip=11pt
\global\advance\figno by 1
\midinsert
\epsfxsize=#3
\centerline{\epsfbox{#2}}
\vskip 12pt
{\bf Fig.\ \the\figno:} #1\par
\endinsert\endgroup\par
}
\def\figlabel#1{\xdef#1{\the\figno}}
\def\encadremath#1{\vbox{\hrule\hbox{\vrule\kern8pt\vbox{\kern8pt
\hbox{$\displaystyle #1$}\kern8pt}
\kern8pt\vrule}\hrule}}

%Macros 
%%%%%%%%%%%%%%%%%%%%%%%%%%%%%%%%%%%%%%%%%%%%%%%%%%%%%%%%%%%%%%%%%

\def\IR{\relax{\rm I\kern-.18em R}}
\font\cmss=cmss10 \font\cmsss=cmss10 at 7pt

\def\q#1{\left[#1\right]_x}

\font\cmss=cmss10 \font\cmsss=cmss10 at 7pt
\def\IZ{\relax\ifmmode\mathchoice
{\hbox{\cmss Z\kern-.4em Z}}{\hbox{\cmss Z\kern-.4em Z}}
{\lower.9pt\hbox{\cmsss Z\kern-.4em Z}}
{\lower1.2pt\hbox{\cmsss Z\kern-.4em Z}}\else{\cmss Z\kern-.4em Z}\fi}
\def\IN{\relax{\rm I\kern-.18em N}}
\def\b{\circ}
\def\n{\bullet}

\def\gbbbb{\Gamma_4^{\hbox{$\scriptstyle \b \b$}\kern -8.2pt
\raise -4pt \hbox{$\scriptstyle \b \b$}}}
\def\gnnnn{\Gamma_4^{\hbox{$\scriptstyle \n \n$}\kern -8.2pt  
\raise -4pt \hbox{$\scriptstyle \n \n$}}}
\def\gnnnnnn{\Gamma_6^{\hbox{$\scriptstyle \n \n \n$}\kern -12.3pt
\raise -4pt \hbox{$\scriptstyle \n \n \n$}}}
\def\gbbbbbb{\Gamma_6^{\hbox{$\scriptstyle \b \b \b$}\kern -12.3pt
\raise -4pt \hbox{$\scriptstyle \b \b \b$}}}
\def\gbbbbc{\Gamma_{4\, c}^{\hbox{$\scriptstyle \b \b$}\kern -8.2pt
\raise -4pt \hbox{$\scriptstyle \b \b$}}}
\def\gnnnnc{\Gamma_{4\, c}^{\hbox{$\scriptstyle \n \n$}\kern -8.2pt
\raise -4pt \hbox{$\scriptstyle \n \n$}}}
\def\Rbud#1{{\cal R}_{#1}^{-\kern-1.5pt\blacktriangleright}}
\def\Rleaf#1{{\cal R}_{#1}^{-\kern-1.5pt\vartriangleright}}
\def\Rbudb#1{{\cal R}_{#1}^{\circ\kern-1.5pt-\kern-1.5pt\blacktriangleright}}
\def\Rleafb#1{{\cal R}_{#1}^{\circ\kern-1.5pt-\kern-1.5pt\vartriangleright}}
\def\Rbudn#1{{\cal R}_{#1}^{\bullet\kern-1.5pt-\kern-1.5pt\blacktriangleright}}
\def\Rleafn#1{{\cal R}_{#1}^{\bullet\kern-1.5pt-\kern-1.5pt\vartriangleright}}
\def\Wleaf#1{{\cal W}_{#1}^{-\kern-1.5pt\vartriangleright}}
\def\Cleaf{{\cal C}^{-\kern-1.5pt\vartriangleright}}
\def\Cbud{{\cal C}^{-\kern-1.5pt\blacktriangleright}}
\def\Crleaf{{\cal C}^{-\kern-1.5pt\circledR}}

%%%%%%%%%%%%%%%%%%%%%%%%%%%%%%%%%%%%%%%%%%%%%%%%%%%%%%%%%%%%%%%%%
%%%%%%%%%%%%%%%%%%%%%%%%%%%%%%%%%%%%%%%%%%%%%%%%%%%%%%%%%%%%%%%%%%%%%

\magnification=\magstep1
\baselineskip=12pt
\hsize=6.3truein
\vsize=8.7truein
 at 8truept
 at 8truept
 at 10truept
%\footline={\footsc the electronic journal of combinatorics
%   {\footbf 11} (2004), \#R00\hfil\footrm\folio}

%%%%%%%%%%%%%%%%%%%%%%%%%%%%%%%%%%%%%%%%%%%%%%%%%%%%%%%%%%%%%%%%%%%%%%%%
\font\bigrm=cmr12 at 14pt
\centerline{\bigrm Confluence of geodesic paths and separating loops}
\medskip
\centerline{\bigrm in large planar quadrangulations}

\bigskip\bigskip

\centerline{J. Bouttier and E. Guitter}
  \smallskip
  \centerline{Institut de Physique Th\'eorique}
  \centerline{CEA, IPhT, F-91191 Gif-sur-Yvette, France}
  \centerline{CNRS, URA 2306}
\centerline{\tt jeremie.bouttier@cea.fr}
\centerline{\tt emmanuel.guitter@cea.fr}

  \bigskip

     \bigskip\bigskip

     \centerline{\bf Abstract}
     \smallskip
     {\narrower\noindent
We consider planar quadrangulations with three marked vertices and
discuss the geometry of triangles made of three geodesic paths joining 
them. We also study the geometry of minimal separating loops, 
i.e.\ paths of minimal length among all closed paths passing by one of the 
three vertices and separating the two others in the quadrangulation. 
We concentrate on the universal scaling limit of large quadrangulations, 
also known as the Brownian map, where pairs of geodesic paths or
minimal separating loops have common parts of non-zero macroscopic
length. This is the phenomenon of confluence, which distinguishes the
geometry of random quadrangulations from that of smooth surfaces. We 
characterize the universal probability distribution for the lengths of 
these common parts.
\par}

     \bigskip

%references
\nref\QGRA{V. Kazakov, {\it Bilocal regularization of models of random
surfaces}, Phys. Lett. {\bf B150} (1985) 282-284; F. David, {\it Planar
diagrams, two-dimensional lattice gravity and surface models},
Nucl. Phys. {\bf B257} (1985) 45-58; J. Ambj\o rn, B. Durhuus and J. Fr\"ohlich,
{\it Diseases of triangulated random surface models and possible cures},
Nucl. Phys. {\bf B257} (1985) 433-449; V. Kazakov, I. Kostov and A. Migdal
{\it Critical properties of randomly triangulated planar random surfaces},
Phys. Lett. {\bf B157} (1985) 295-300.}
\nref\DGZ{for a review, see: P. Di Francesco, P. Ginsparg and 
J. Zinn--Justin, {\it 2D Gravity and Random Matrices},
Physics Reports {\bf 254} (1995) 1-131.}
\nref\ADJ{J. Ambj\o rn, B. Durhuus and T. Jonsson, {\it Quantum Geometry:
A statistical field theory approach}, Cambridge University Press, 1997.}
\nref\MARMO{J. F. Marckert and A. Mokkadem, {\it Limit of normalized
quadrangulations: the Brownian map}, Annals of Probability {\bf 34(6)}
(2006) 2144-2202, arXiv:math.PR/0403398.}
\nref\LEGALL{J. F. Le Gall, {\it The topological structure of scaling limits 
of large planar maps}, invent. math. {\bf 169} (2007) 621-670,
arXiv:math.PR/0607567.}
\nref\LGP{J. F. Le Gall and F. Paulin,
{\it Scaling limits of bipartite planar maps are homeomorphic to the 2-sphere},
arXiv:math.PR/0612315.}
\nref\MierS{G. Miermont, {\it On the sphericity of scaling limits of 
random planar quadrangulations}, Elect. Comm. Probab. {\bf 13} (2008) 248-257, 
arXiv:0712.3687 [math.PR].}
\nref\AW{J. Ambj\o rn and Y. Watabiki, {\it Scaling in quantum gravity},
Nucl.Phys. {\bf B445} (1995) 129-144.}
\nref\GEOD{J. Bouttier, P. Di Francesco and E. Guitter, {\it Geodesic
distance in planar graphs}, Nucl. Phys. {\bf B663}[FS] (2003) 535-567, 
arXiv:cond-mat/0303272.}
\nref\CS{P. Chassaing and G. Schaeffer, {\it Random Planar Lattices and 
Integrated SuperBrownian Excursion}, 
Probability Theory and Related Fields {\bf 128(2)} (2004) 161-212, 
arXiv:math.CO/0205226.}
\nref\MW{G. Miermont and M. Weill, {\it Radius and profile of random planar
maps with faces of arbitrary degrees}, Electron. J. Probab. 
{\bf 13} (2008) 79-106, arXiv:0706.3334 [math.PR].}
\nref\STATGEOD{J. Bouttier and E. Guitter, {\it Statistics of
geodesics in large quadrangulations}, J. Phys. A: Math. Theor. {\bf 41} 
(2008) 145001 (30pp), arXiv:0712.2160 [math-ph].}
\nref\Mier{G. Miermont, {\it Tessellations of random maps of arbitrary
genus}, arXiv:0712.3688 [math.PR]}
\nref\LEGALLGEOD{J.-F. Le Gall, {\it Geodesics in large planar maps and 
in the Brownian map}, arXiv:0804.3012 [math.PR].}
\nref\THREEPOINT{J. Bouttier and E. Guitter, {\it The three-point function 
of planar quadrangulations}, J. Stat. Mech. (2008) P07020, 
arXiv:0805.2355 [math-ph].}
\nref\Aoki{H. Aoki, H. Kawai, J. Nishimura and A. Tsuchiya, {\it Operator
product expansion in two-dimensional quantum gravity}, Nucl. Phys. 
{\bf B474} (1996) 512-528, arXiv:hep-th/9511117.}
\nref\MS{M. Marcus and G. Schaeffer, {\it Une bijection simple pour les
cartes orientables} (2001), available at 
\hhref{http://www.lix.polytechnique.fr/Labo/Gilles.Schaeffer/Biblio/};
see also G. Schaeffer, {\it Conjugaison d'arbres
et cartes combinatoires al\'eatoires}, PhD Thesis, Universit\'e 
Bordeaux I (1998) and G. Chapuy, M. Marcus and G. Schaeffer, 
{\it A bijection for rooted maps on orientable surfaces}, 
arXiv:0712.3649 [math.CO].}
\nref\CORV{R. Cori and B. Vauquelin, {\it Planar maps are well labeled trees},
Canad. J. Math. {\bf 33(5)} (1981) 1023-1042.}
\nref\DELMAS{J.-F. Delmas, {\it Computation of moments for the length of 
the one dimensional ISE support}, Elect. Journ. of Probab. {\bf 8(17)} 
(2003) 1-15.}
\nref\BMISE{M. Bousquet-M\'elou, {\it Limit laws for embedded trees. 
Applications to the integrated superBrownian excursion},
Random Structures and Algorithms {\bf 29(4)} (2006) 475-523, 
arXiv:math.CO/0501266.}
\nref\JBPHD{J. Bouttier, {\it Physique statistique des surfaces al\'eatoires
et combinatoire bijective des cartes planaires}, PhD Thesis (2005).}
\nref\JainMa{S. Jain and S. Mathur, {\it World-sheet geometry and
baby universes in 2D quantum gravity}, Phys. Lett. {\bf B 286} (1992) 239-246, 
 arXiv:hep-th:9204017.}
\nref\BaFlScSo{C. Banderier, P. Flajolet, G. Schaeffer and M. Soria, {\it 
Random Maps, Coalescing Saddles, Singularity Analysis, and Airy Phenomena},
Random Structures and Algorithms {\bf 19} (2001) 194-246.}
\nref\DSKPZ{B. Duplantier and S. Sheffield, {\it Liouville Quantum Gravity 
and KPZ}, arXiv:0808.1560 [math.PR].}
\nref\DBKPZ{F. David and M. Bauer, {\it Another derivation of the 
geometrical KPZ relations}, arXiv:0810.2858 [math-ph].}
\nref\BOUKA{D. Boulatov and V. Kazakov, {\it The Ising model
on a random planar lattice: the structure of the phase 
transition and the exact critical exponents}, Phys. Lett. {\bf B186} (1987)
379-384.}
\nref\BMS{M. Bousquet-M\'elou and G. Schaeffer,{\it The degree distribution
in bipartite planar maps: application to the Ising model},
arXiv:math.CO/0211070.}
\nref\HPCOM{J. Bouttier, P. Di Francesco and E. Guitter. 
{\it Combinatorics of Hard Particles
on Planar Graphs}, Nucl.Phys. {\bf B655} (2003) 313-341, 
arXiv:cond-mat/0211168.}
\nref\HObipar{J. Bouttier, P. Di Francesco and E. Guitter. 
{\it Combinatorics of bicubic maps 
with hard particles}, J.Phys. A: Math.Gen. {\bf 38} 
(2005) 4529-4560, arXiv:math.CO/0501344.}
\nref\MOB{J. Bouttier, P. Di Francesco and E. Guitter. {\it 
Planar maps as labeled mobiles},
Elec. Jour. of Combinatorics {\bf 11} (2004) R69, arXiv:math.CO/0405099.}
\nref\FOMAP{J. Bouttier, P. Di Francesco and E. Guitter. {\it Blocked edges 
on Eulerian maps and mobiles: Application to spanning trees, hard particles 
and the Ising model}, 	J. Phys. A: Math. Theor. {\bf 40} (2007) 7411-7440, 
arXiv:math.CO/0702097.}

%text
\newsec{Introduction}
Understanding the geometry of large random quadrangulations is a 
fundamental issue relating combinatorics, probability theory and
statistical physics. Indeed random quadrangulations, or more generally 
random maps, provide natural discrete models for random surfaces,
for instance in the context of two-dimensional quantum gravity 
[\xref\QGRA-\xref\ADJ], and may mathematically be viewed as metric 
spaces endowed with the graph distance. 
In the same way that discrete random walks converge to 
the Brownian motion in a suitable scaling limit, it is expected that 
random planar quadrangulations converge to the so-called Brownian map 
[\xref\MARMO,\xref\LEGALL] in the 
scaling limit where the size of the quadrangulation becomes large 
jointly with the fourth power of the scale at which distances are measured.
This Brownian map is moreover expected to be the universal scaling
limit for many models of planar maps such as random planar triangulations
or more generally maps with arbitrary bounded face degrees or even maps
coupled to non-critical statistical models. It can be constructed as
a random metric space and has been shown to be homeomorphic to the
two-dimensional sphere [\xref\LGP,\xref\MierS].

A number of local properties of the Brownian map can be derived from a
detailed analysis of discrete maps. 
In this spirit, the simplest observable in the Brownian map is
the distance between two points. The statistics of this distance 
is characterized by the so-called two-point function and was obtained
in Ref.~\AW\ via scaling arguments for large triangulations, and 
in Ref.~\GEOD\ via an exact computation of the discrete two-point
function for planar quadrangulations. A related quantity is the
radius, whose law was studied in Refs.~[\xref\CS,\xref\MW]. 
The question of estimating the number
of geodesics (i.e.\ paths of shortest length) between two points was 
addressed later \STATGEOD\ and it was found that for typical points, all
geodesics coalesce into a unique macroscopic geodesic path in
the scaling limit [\xref\Mier,\xref\LEGALLGEOD]. 
\fig{A schematic picture of the phenomenon
of confluence for the geometry of triangles (a) and separating
loops (b) in the scaling limit of large maps. In (a), the three geodesics 
(represented as thick blue lines) linking the three points 
$v_1$, $v_2$ and $v_3$ have common parts of macroscopic length. 
The triangle is therefore characterized by six lengths 
(as indicated by double arrows) and by the area of the two domains
delimited by its open part. In (b), a minimal separating loop
passing by $v_3$ and separating $v_1$ from $v_2$ also has a
common part of macroscopic length and is therefore characterized 
by two lengths (as indicated by double arrows) and by the area of the two 
domains delimited by its open part.}{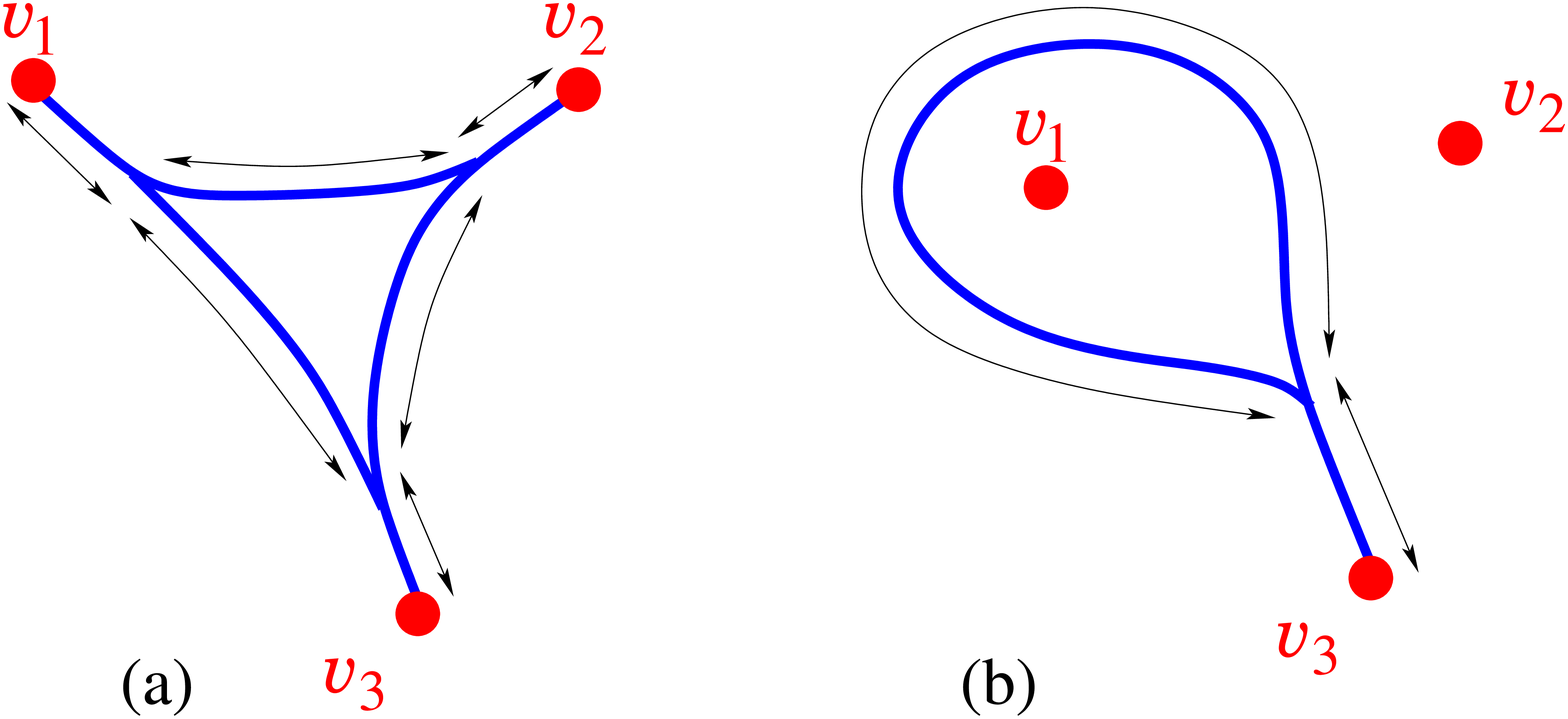}{11.cm}
\figlabel\confluence
Properties involving three points on the map give a much richer 
geometric information. For instance, we may consider the ``triangle"
made by the three geodesics between these points.  
In a previous paper \THREEPOINT, the authors have 
computed the joint probability 
distribution for the pairwise distances between three uniformly 
chosen random vertices in a random quadrangulation. In the scaling limit, 
this yields the so-called three-point function of the Brownian map, 
which can interpreted as the joint law for the three side lengths
of the triangle. The three-point function was considered previously 
in Ref.~\Aoki\ where an expression involving two distances only was obtained 
and used as a basis for an Operator Product Expansion analysis in the limit 
where two of the points approach each other. The full dependence 
on the three distances was found in Ref.~\THREEPOINT\ as a corollary 
of the exact discrete expression for quadrangulations. 
On the other hand, it was recognized by Le Gall that 
geodesics exhibit a phenomenon of {\it confluence} \LEGALLGEOD. In our
setting, this means that any two sides of the triangle merge
before reaching their common endpoint, and hence have a common part 
of non-zero macroscopic length. This is quite unlike smooth
surfaces where two sides of a triangle only meet at their
endpoint. Thus a full characterization of the geometry of
triangles involves six lengths, which are those of the three
segments proper to each side and of the three segments common 
to two sides (see Fig.~\confluence-(a)), 
as well two areas for the two domains in the map delimited 
by the triangle.

Beyond triangles, another interesting geometric construction
involving three points is what we call a minimal separating
loop, defined as follows: given three distinguished points, say $v_1$, $v_2$ 
and $v_3$, we define a separating loop as a closed path passing
through $v_3$ and {\it separating $v_1$ from $v_2$}, in the 
sense that any path from $v_1$ to $v_2$ necessarily intersects
it. A minimal separating loop is such a separating loop 
with minimal length. We expect the minimal separating loop 
to be unique at a macroscopic level, and to have a finite macroscopic
length (note that, if we relax the condition that the loop passes through
$v_3$ or that it separates $v_1$ from $v_2$, then clearly we can find 
loops of arbitrarily small length). Moreover, its two halves 
are geodesic paths and we again expect a 
phenomenon of confluence, namely the two halves share a macroscopic 
common segment (see Fig.~\confluence-(b)).
The characterization of the geometry of minimal separating
loops involves therefore the lengths of its common and
``open" parts, as well as the areas of the two 
domains delimited by the loop.

In this paper, we derive the probability distributions for the
above parameters characterizing triangles and loops when
the three points are chosen uniformly at random. 
This is done by explicit computations of the discrete 
counterparts of these distributions in the framework
of planar quadrangulations, using the methodology 
developed in Ref.~\THREEPOINT\ and based on the Schaeffer \MS\ and
Miermont \Mier\ bijections between quadrangulations and 
well-labeled maps.

The paper is organized as follows: in Section 2, we give a precise
definition of minimal separating loops in triply-pointed planar
quadrangulations and compute the generating function for such 
quadrangulations with a {\it prescribed value for the loop length}. 
To this end, we provide in Section 2.1 two alternative bijections
based on the Schaeffer and Miermont constructions relating
the desired class of triply-pointed quadrangulations with suitable 
classes of well-labeled trees or maps. In Section 2.2, we calculate 
their generating functions by expressing them in terms of basic
building blocks already computed in Ref.~\THREEPOINT. Section 2.3 is devoted
to the analysis of the scaling limit, with a particular emphasis on
the universal probability law for the length of the minimal separating
loop, as well as its correlation with the distances between the marked
vertices. In Section 3, we turn to the phenomenon of confluence,
which we analyze by a refinement of the above enumeration. In Section
3.1, we give the probability law for the length of the part common 
to two geodesics leading to the same vertex. We then investigate
the phenomenon of confluence for minimal separating loops in Section 
3.2 where we derive the probability distribution for the parameters
characterizing the geometry of these loops. Section 4 is devoted to the
geometry of triangles. There we revisit the bijection of 
Ref.~\THREEPOINT\ and solve a refined enumeration problem in order to keep 
track of the six lengths characterizing the triangle. We deduce their
joint law in the scaling limit, and provide explicit expressions for
a number of marginal laws. We discuss our results and conclude
in Section 5.

\fig{A quadrangulation with three marked vertices $v_1$, $v_2$ and $v_3$,
represented as a traffic network, i.e.\ a ribbon graph with roundabouts. 
In (a), we show (red thick lines) a
particular separating loop of length $6$. In (b), the indicated 
separating loop is minimal, i.e.\ has a minimal length 
(here 4) among the loops passing through $v_3$ and separating $v_1$
from $v_2$.}{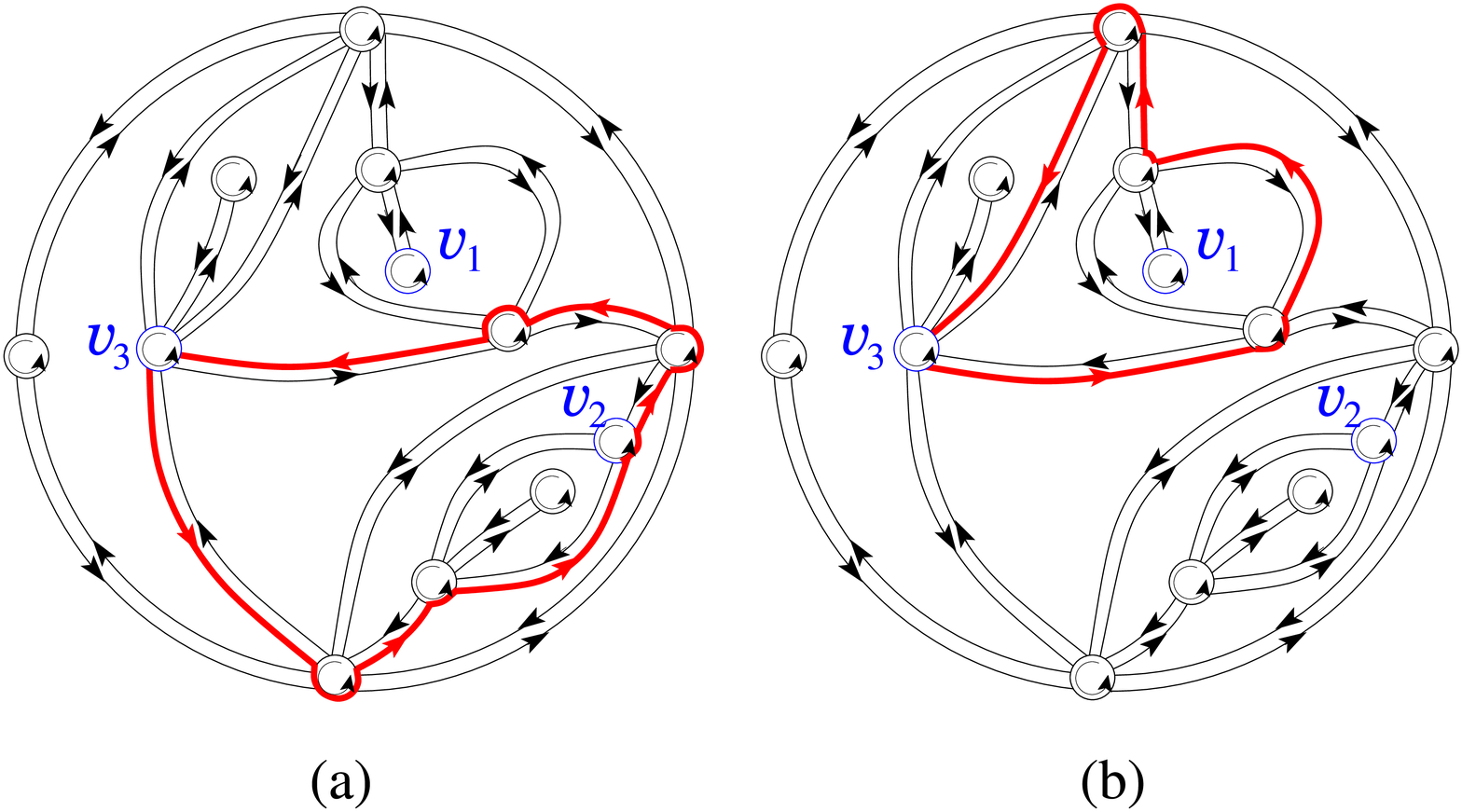}{11.cm}
\figlabel\separloop
\newsec{Minimal separating loops}
Consider a quadrangulation of the sphere, i.e.\ a planar map whose
faces all have degree four, equipped with three marked distinct vertices
$v_1$, $v_2$ and $v_3$. As customary for orientable maps, we may represent
the map as a ribbon graph by splitting each edge of the quadrangulation 
into two oriented half-edges (with opposite orientations) so that half-edges 
are oriented clockwise around each face (see Fig.~\separloop). 
It is also convenient to place a small counterclockwise oriented roundabout 
around each vertex so that the map looks like a traffic network. 
We can then consider (oriented) paths on this traffic 
network, and in particular loops made of a closed non-intersecting 
circuit starting from and returning back to the marked vertex $v_3$. 
Any such loop separates the sphere into two simply connected domains. 
Note that any vertex along the loop naturally belongs to exactly one of these
domains by following the roundabout convention. The circuit is called a 
{\it separating loop} if the marked vertices $v_1$ 
and $v_2$ do not lie in the same domain (see Fig.~\separloop\ for an 
illustration). The length of a circuit is the 
number of half-edges it passes through. A {\it minimal separating loop} 
is a separating loop of minimal length. 

Clearly the length $l_{123}$ of a minimal separating loop is strictly 
positive and, from the bipartite nature of planar quadrangulations, 
it is even. Also,
if we call $d_{13}$ (respectively $d_{23}$) the graph distance from
$v_1$ (respectively $v_2$) to $v_3$, following a geodesic path back and 
forth from $v_3$ to the closest vertex $v_1$ or $v_2$ forms a separating 
loop of length $2 \min(d_{13},d_{23})$, therefore:
\eqn\condu{l_{123} \leq 2\min(d_{13},d_{23})\ .}
The purpose of the next sections is to enumerate triply-pointed 
quadrangulations whose three marked vertices have prescribed values
of $d_{13}$, $d_{23}$ and $l_{123}$.

An alternative definition of separating loops, mentioned in the introduction,
consists in taking arbitrary (possibly self-intersecting) closed paths
passing through $v_3$ and such that any path from $v_1$ to $v_2$
necessarily intersects them. This gives rise to a broader set 
of minimal separating loops but does not affect the minimal length
since any such minimal separating loop can be transformed into
a non self-intersecting circuit of the same length by ``undoing"
the crossings.

\subsec{Combinatorics}
\noindent{\it Approach via the Schaeffer bijection}

\fig{The quadrangulation of Fig.~\separloop\ with a marked origin 
(corresponding to $v_3$ in Fig.~\separloop) and its coding (a) by
a well-labeled tree (blue thick lines). The quadrangulation is recovered
from the well-labeled tree by connecting each corner to
its successor (dashed red arrows in (b)).}{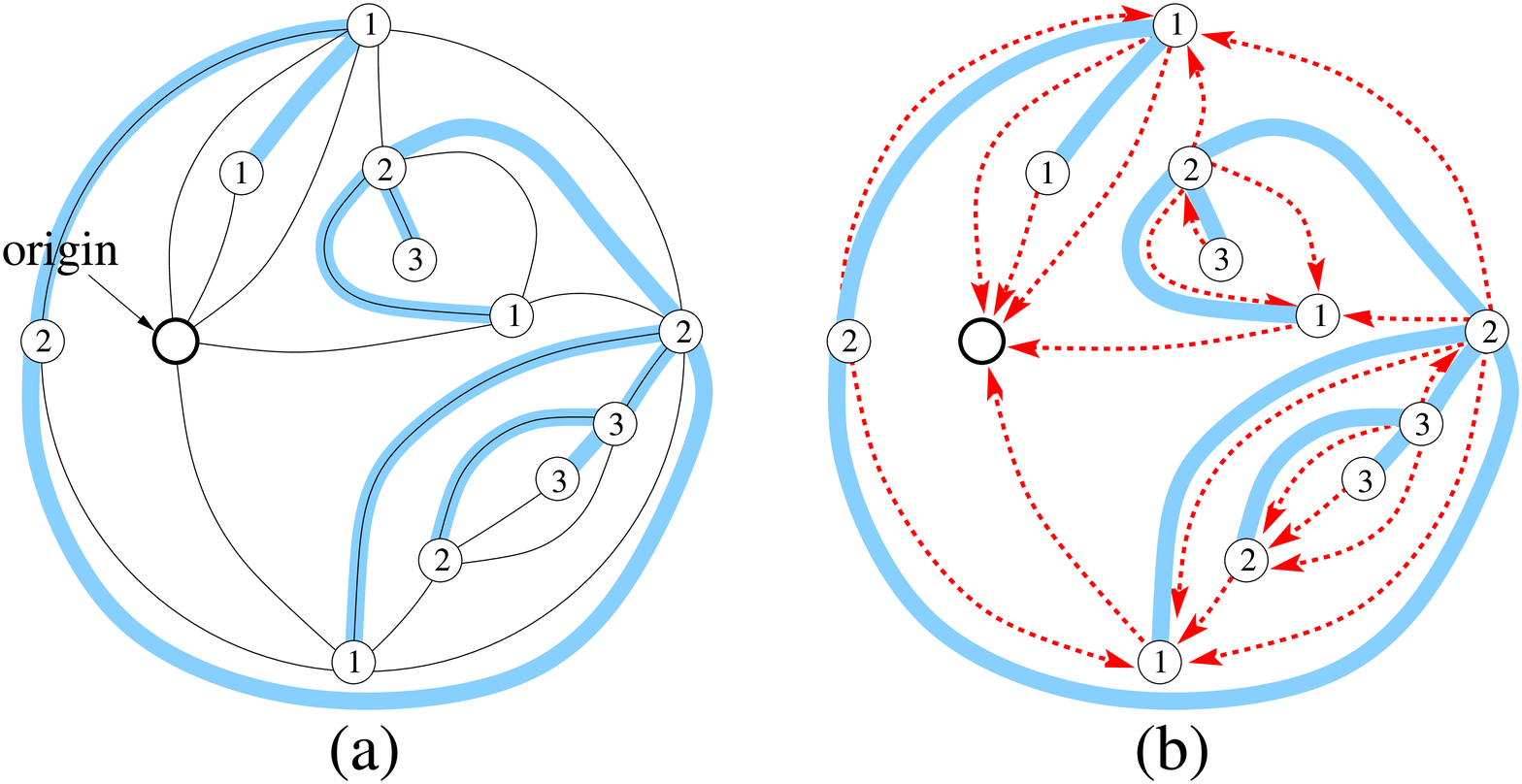}{11.cm}
\figlabel\schaefferbij
It is well-known \CORV\ that any planar quadrangulation with $n$ faces and a 
marked origin vertex is in one-to-one correspondence 
with a {\it well-labeled tree} with $n$ edges and with minimal label $1$. 
Here we define a well-labeled tree as a plane tree with vertices 
carrying integer labels $\ell$ satisfying
\eqn\welllab{\vert \ell(v)-\ell(v') \vert \leq 1 \ \ \hbox{if
$v$ and $v'$ are adjacent in the tree\ .}}
As shown by Schaeffer \MS, this tree can be drawn directly on the 
quadrangulation by applying 
local rules which associate with each face of the quadrangulation 
an edge of the tree (see Fig.~\schaefferbij). The tree spans all vertices 
of the quadrangulation
except the origin, and the label of each vertex is nothing but its
graph distance to the origin in the quadrangulation.
Conversely, to recover the quadrangulation from the well-labeled tree,
we draw non-crossing arches connecting every {\it corner} of the tree to
its {\it successor}. Recall that a corner is the sector between two
consecutive edges around a vertex, and the successor of a corner with 
label $\ell>1$ is the first corner with label $\ell-1$ encountered after
it clockwise along the contour of the tree, while all corners with 
label $1$ have the same successor which is an extra vertex added 
in the external face (see Fig.~\schaefferbij-(b)).
The arches form the edges of the quadrangulation and the added vertex
is the origin. Note that the {\it chain of successors} of a given corner 
(i.e.\ its successor, the successor of its successor, and so on until
the origin) provides a geodesic path from the associated vertex
to the origin.

\fig{A well-labeled tree with two marked vertices $v_1$ and $v_2$.
The edges of the branch from $v_1$ to $v_2$ are represented as
magenta thick lines and the other edges as light-blue thin solid
lines. The vertex $v_3$ is the origin added in the external face.
We consider a vertex of minimal label (here $2$) on the branch from 
$v_1$ to $v_2$ and represent the chains of successors (dashed red arrows) 
starting from two of its corners, one on each side of the branch. These 
form a minimal loop separating $v_1$ from $v_2$ and passing through 
$v_3$.}{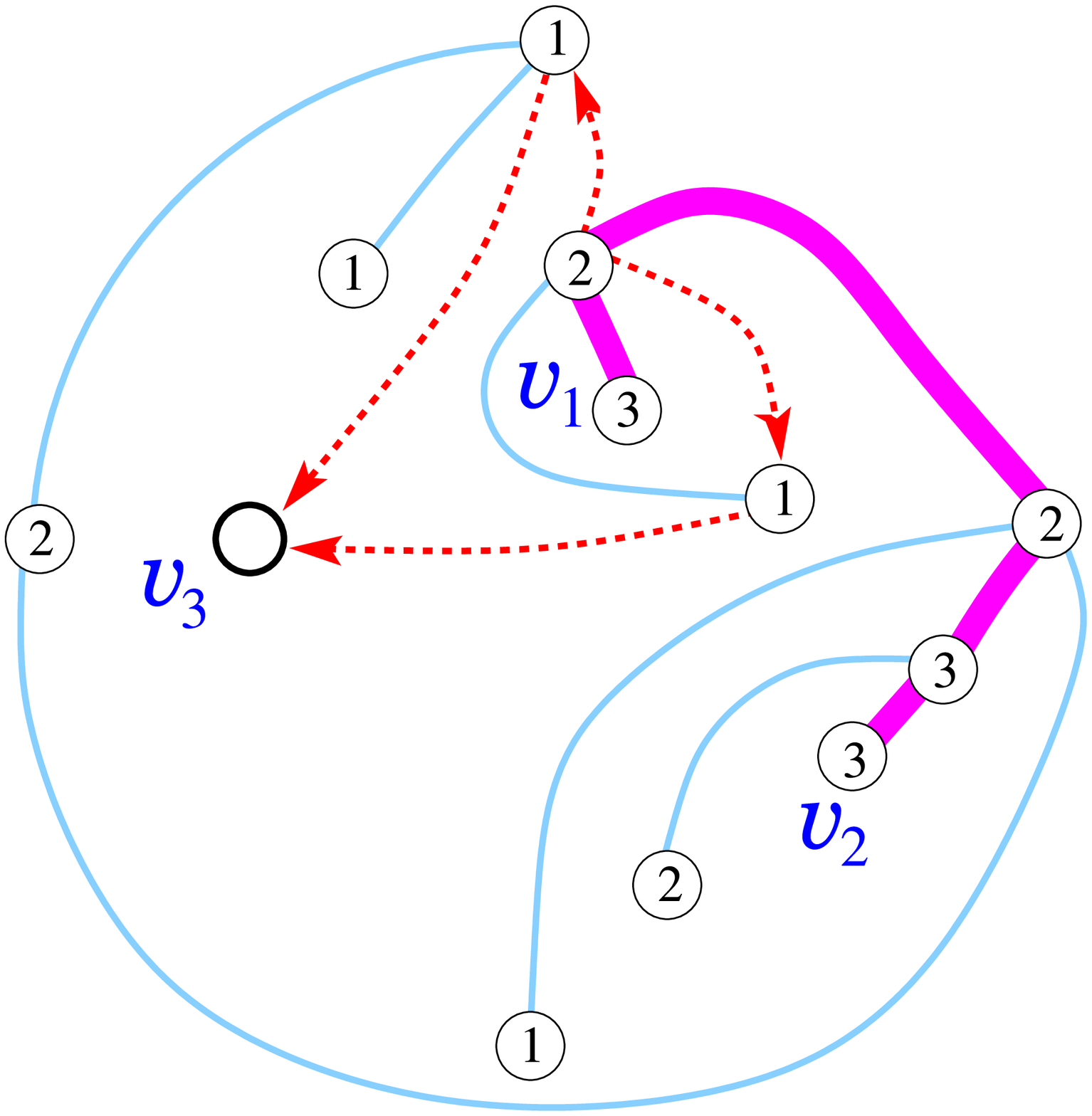}{6.cm}
\figlabel\septree
In the case of a triply-pointed quadrangulation, we can take
$v_3$ as the origin vertex and we end up with a well-labeled tree
with two marked vertices $v_1$ and $v_2$ carrying labels 
$\ell(v_1)=d_{13}$ and $\ell(v_2)=d_{23}$. Let us now explain how
the quantity $l_{123}$ can be read off the tree. Within the tree,
there is a unique branch connecting $v_1$ to $v_2$ (see Fig.~\septree). 
Any loop separating $v_1$ from $v_2$ in the quadrangulation must intersect 
this branch at some vertex $v$. Decomposing the loop into a first part
from $v_3$ to $v$ and a second part from $v$ back to $v_3$, both 
parts have length larger that the distance $\ell(v)$ from $v$ to $v_3$,
and we find that the length of the loop is larger than $2\ell(v)$,
and hence larger than $2u$, where $u$ is the minimal label encountered
along the branch from $v_1$ to $v_2$. This holds in particular for
minimal separating loops, and we therefore have $l_{123}\geq 2u$.
Conversely, a separating loop of length $2u$ is obtained by considering
a vertex with minimal label $u$ on the branch, picking two corners on opposite
sides of the branch and considering the chain of successors of these
two corners which are both paths to $v_3$ of length $u$ 
(see Fig.~\septree). This implies $l_{123}\leq 2u$ and 
therefore $l_{123}=2u$. More generally, any minimal separating loop  
crosses the branch at a vertex with minimal label $u$, and hence it is
made of two geodesic paths of the same length $u$ joining the origin to 
that vertex which they reach from both sides of the branch. 
\fig{The well-labeled tree coding a triply-pointed quadrangulation
with prescribed values of $d_{13}$, $d_{23}$ and $l_{123}$. It has
two marked vertices $v_1$ and $v_2$ with respective labels 
$s=d_{13}-l_{123}/2$ and $t=d_{23}-l_{123}/2$. The minimal label 
on the branch between $v_1$ and $v_2$ is $0$ and the global minimal
label is $1-u=1-l_{123}/2$.}{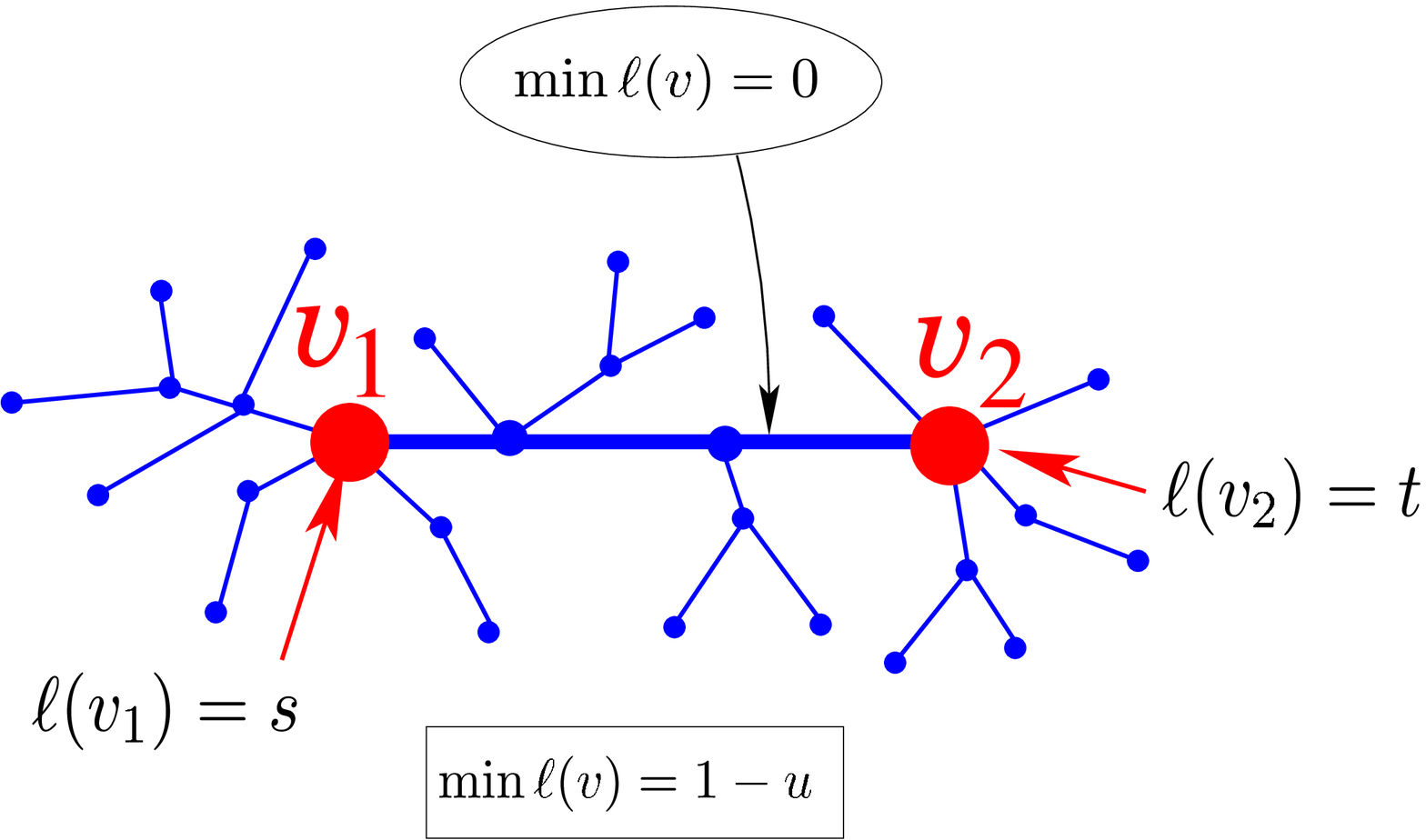}{7.cm}
\figlabel\condtree
For consistency with the alternative approach described below, 
we decide to shift all labels on the well-labeled tree
by $-u$ so that the minimal label on the branch from $v_1$ to $v_2$ 
becomes $0$. The minimal label in the whole tree is now $1-u$, while
$v_1$ and $v_2$ receive respective non-negative labels $s\equiv d_{13}-u$ and
$t\equiv d_{23}-u$ (see Fig.~\condtree\ for an illustration).
To conclude, triply-pointed quadrangulations with prescribed values
of $d_{13}$, $d_{23}$ and $l_{123}$ are in one-to-one correspondence
with well-labeled trees having two marked vertices labeled 
$s=d_{13}-l_{123}/2$ and $t=d_{23}-l_{123}/2$, such that the minimal label
on the branch joining these two vertices is $0$ and the global 
minimal label in the tree is $1-u=1-l_{123}/2$.

\bigskip
\noindent{\it Approach via the Miermont bijection}
\fig{The quadrangulation of Fig.~\separloop\ with three marked vertices 
$v_1$, $v_2$, $v_3$, and its coding (a) by
a well-labeled map (blue thick lines) using the Miermont bijection
with particular delays $\tau_1=\tau_2=-1$ and $\tau_3=-2$. 
The quadrangulation is recovered
from the well-labeled map by connecting each corner to
its successor (dashed red arrows in (b)).}{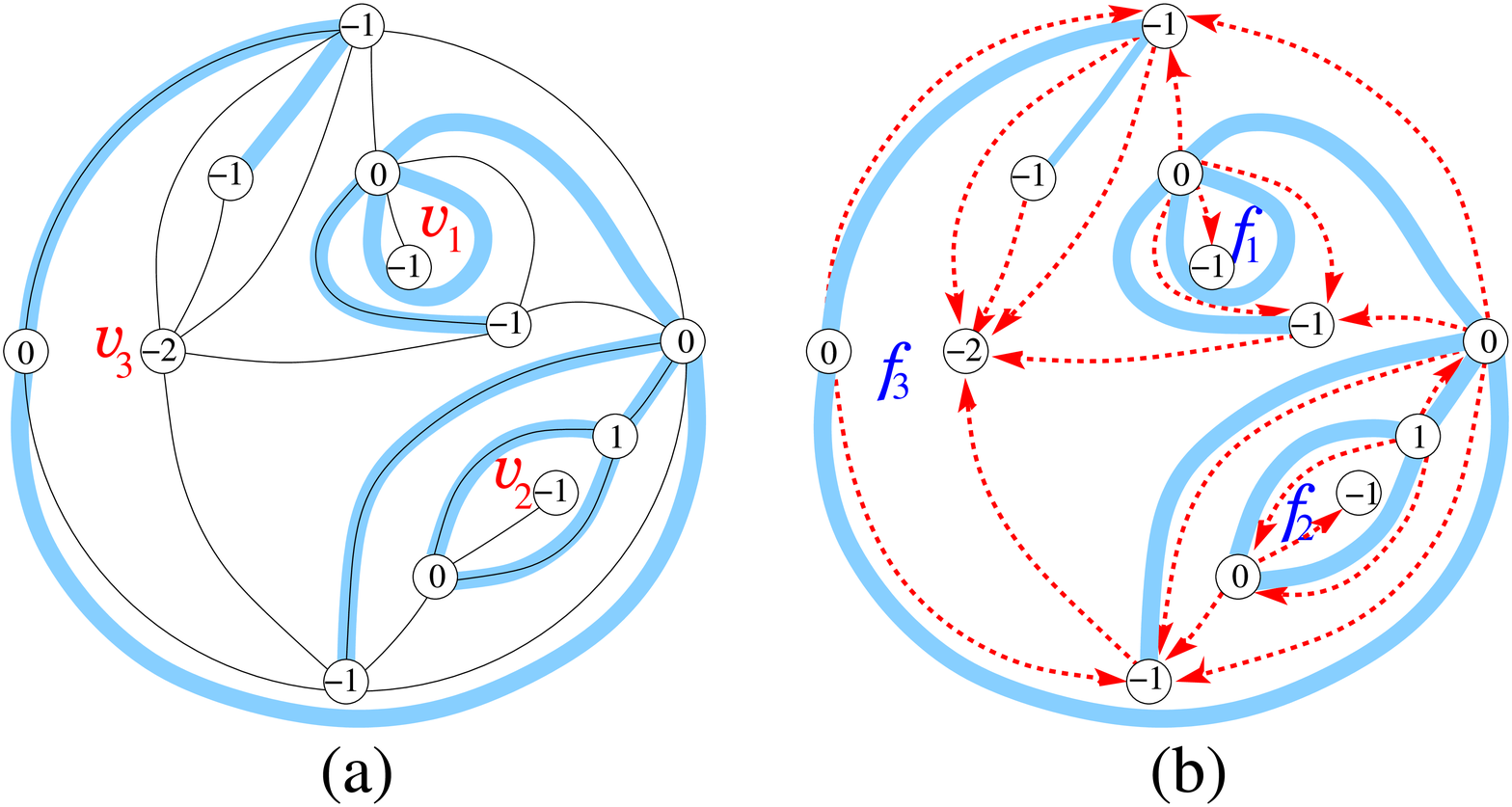}{11.cm}
\figlabel\miermontbij

An alternative approach is based on a bijection by Miermont 
\Mier\ generalizing the Schaeffer bijection to multiply-pointed
planar quadrangulations. More precisely, the Miermont bijection acts
on a quadrangulation equipped with, say $p$ marked vertices 
$v_1, v_2,\ldots v_p$, called {\it sources} and $p$ 
integers $\tau_1,\tau_2,\ldots \tau_p$
called {\it delays}, satisfying the conditions:
\eqn\contdelay{\eqalign{& \vert \tau_i-\tau_j\vert < d_{ij},\quad  
1 \le i\neq j \leq p \ ,\cr & \tau_i-\tau_j+d_{ij} \ \hbox{is even,}\quad
1\leq i,j\leq p\ ,\cr}}
where $d_{ij}$ is the graph distance between $v_i$ and $v_j$.
It results into a planar map with $p$ faces that is well-labeled,
i.e.\ its vertices carry integer labels $\ell$ satisfying 
\eqn\welllabmap{\vert \ell(v)-\ell(v') \vert \leq 1 \ \ \hbox{if
$v$ and $v'$ are adjacent in the map.}}
Again, this map can be drawn directly on the quadrangulation by applying 
local rules which associate with each face of the quadrangulation 
an edge of the map (see Fig.~\miermontbij). 
The map spans all vertices of the quadrangulation
except the $p$ sources and the label of a vertex $v$ is
given by
\eqn\lv{\ell(v)=\min_{j=1,\ldots,p} d(v,v_j)+\tau_j}
where $d(v,v_j)$ is the graph distance from $v$ to the source 
$v_j$ in the quadrangulation.
Each face of the well-labeled map encloses exactly one source
of the quadrangulation and we call the faces $f_1, f_2,\ldots , f_p$
accordingly. We furthermore have the property that, for any vertex $v$
incident to $f_i$, the minimum in \lv\ is attained for $j=i$, i.e.\ 
$d(v,v_i)=\ell(v)-\tau_i$. In particular, the minimal label
among vertices incident to $f_i$ is $\tau_i+1$, corresponding
to nearest neighbors of $v_i$.

Conversely, to recover the quadrangulation from the well-labeled map,
we add inside each face $f_i$ an extra vertex with label $\tau_i$ 
where
\eqn\conlab{\tau_i=\min_{v\ {\rm incident\ to}\ f_i} \ell(v)-1\ ,}
and each corner with label $\ell$ inside $f_i$ is connected by an arch
to its successor, which is the first corner with label $\ell-1$ 
encountered counterclockwise inside the face (corresponding for 
the external face to the clockwise orientation around the map).
The arches form the edges of the quadrangulation and the added vertices
are the sources (see Fig.~\miermontbij-(b)). 

Let us now see how to use the Miermont bijection to address the specific
question of three marked vertices with prescribed values of $d_{13}$,
$d_{23}$ and $l_{123}$. As in Ref.~\THREEPOINT, the idea is to 
supplement the Miermont bijection (here with $p=3$ sources) by a particular
choice of delays related to $d_{13}$, $d_{23}$ and $l_{123}$. This particular
choice will restrict the topology of the resulting well-labeled maps
with $3$ faces, and induce extra conditions on labels. More precisely, 
from the inequality \condu, we may use the following parametrization:
\eqn\param{\eqalign{
d_{13}&=s+u\ , \cr
d_{23}&=t+u\ ,  \cr
l_{123}&=2 u\ , \cr}}
with $s$, $t$, $u$ non-negative integers, and moreover $u\neq 0$.
Our particular choice of delays is:
\eqn\choicedelay{\eqalign{
\tau_1 &= -s = l_{123}/2-d_{13}\ , \cr
\tau_2 &= -t = l_{123}/2-d_{23}\ , \cr
\tau_3 &= -u = -l_{123}/2\ . \cr
}}
Note that this particular choice fulfils the general condition 
\contdelay\ except when we have the equality $l_{123}=2 \min(d_{13},d_{23})$,
i.e.\ when $s$ or $t$ vanishes. This particular case must be treated 
separately, as will be explained below. 
\fig{
The well-labeled map with three faces coding a triply-pointed 
quadrangulation with prescribed values of $d_{13}$, $d_{23}$ and $l_{123}$, 
in the case $l_{123}<\min(d_{13},d_{23})$. The faces $f_1$ and $f_2$
are not adjacent, and their frontiers with the face $f_3$ form two 
cycles $c_1$ and $c_2$, connected by a bridge $b$,
whose edges are adjacent to $f_3$ only. The minimal label for vertices 
incident to $f_1$ (respectively $f_2$ and $f_3$) is $1-s=1-d_{13}+l_{123}/2$ 
(respectively $1-t=1-d_{23}+l_{123}/2$ and $1-u=1-l_{123}/2$). 
The minimal label on the cycle $c_1$ is $0$, as is that on the cycle 
$c_2$ and that on the bridge $b$.}{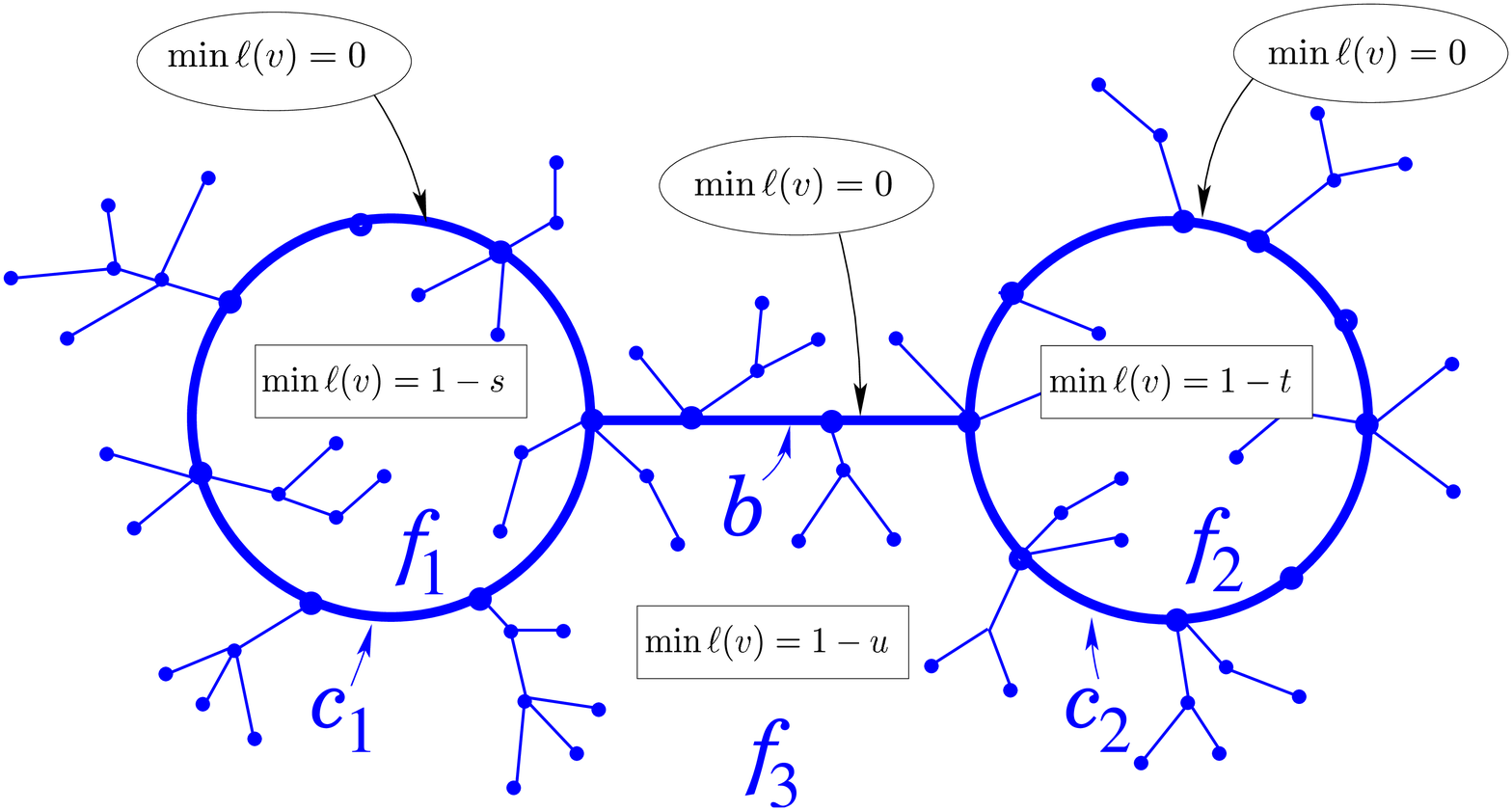}{11.cm}
\figlabel\specialdelays
Assuming $s$ and $t$ strictly positive, a close look at the properties
resulting from the choice of delays \choicedelay\ in the Miermont
bijection shows that the resulting well-labeled map is necessarily 
of the type displayed in Fig.~\specialdelays. In particular, we find that
any minimal separating loop must remain inside the face $f_3$, and hence
{\it the faces $f_1$ and $f_2$ cannot be adjacent}, i.e. cannot 
be incident to a common edge. The 
map can be viewed as made of a skeleton map (thick lines and big dots 
in Fig.~\specialdelays) to which trees are attached. The skeleton
is necessarily made of two {\it cycles} $c_1$ and $c_2$, which form respectively
the frontier between $f_1$ and $f_3$, and between $f_2$ and $f_3$, 
together with a {\it bridge} $b$ connecting $c_1$ to $c_2$, and whose 
edges are only incident to $f_3$. 
Moreover, the labels must satisfy the following constraints (see 
Fig.~\specialdelays):
\eqn\specialcond{\eqalign{&\min_{v\ {\rm incident}\atop {\rm to}\ f_1}
\ell(v) = 1-s
\ , \quad \min_{v\ {\rm incident}\atop {\rm to}\ f_2}
\ell(v) = 1-t \ ,\quad 
\min_{v\ {\rm incident}\atop {\rm to}\ f_3}
\ell(v) = 1-u \ ,\cr
& \min_{v\ {\rm on}\ c_1} \ell(v) = 0 \ ,\quad
\min_{v\ {\rm on}\ c_2} \ell(v) = 0 \ ,\quad 
\min_{v\ {\rm on}\ b} \ell(v) = 0 \ .\cr}}
The first three constraints are general consequences 
of the Miermont bijection and rephrase the general condition \conlab,
while the last three constraints result
from our particular choice of delays, and can be obtained by arguments
similar to those presented in Ref.~\THREEPOINT. More precisely, the
constraint on $c_1$ (respectively $c_2$) ensures that the distance
between $v_1$ and $v_3$ (respectively $v_2$ and $v_3$) is $s+u$ 
(respectively $t+u$), while the constraint on $b$ ensures that 
the length of a minimal separating loop is $2u$. 
Note that the bridge $b$ can be reduced to a single vertex, necessarily
with label $0$. 
\fig{
(a) The well-labeled map with two faces and a marked vertex 
coding a triply-pointed 
quadrangulation with prescribed values of $d_{13}$, $d_{23}$ and $l_{123}$, 
in the case $l_{123}=d_{13}<d_{23}$. The marked vertex $v_1$ is incident
to the face $f_3$ and is connected to the frontier $c_2$ 
between $f_2$ and $f_3$ by a bridge $b$ (whose edges are adjacent 
to $f_3$ only). The label of $v_1$ is $0$ and the minimal label for vertices 
incident to $f_2$ (respectively $f_3$) is $1-t=1-d_{23}+l_{123}/2$ 
(respectively $1-u=1-l_{123}/2$). The minimal label on the cycle $c_2$ is $0$, 
as is that on the bridge $b$.
(b) The well-labeled tree with two marked vertices coding a triply-pointed 
quadrangulation with prescribed values of $d_{13}$, $d_{23}$ and $l_{123}$, 
in the case $l_{123}=d_{13}=d_{23}$. The marked vertices $v_1$ and
$v_2$ are connected by a branch $b$ and have label $0$. 
The global minimal label is $1-u=1-l_{123}/2$, while the minimal label 
on the branch  $b$ is $0$.
}{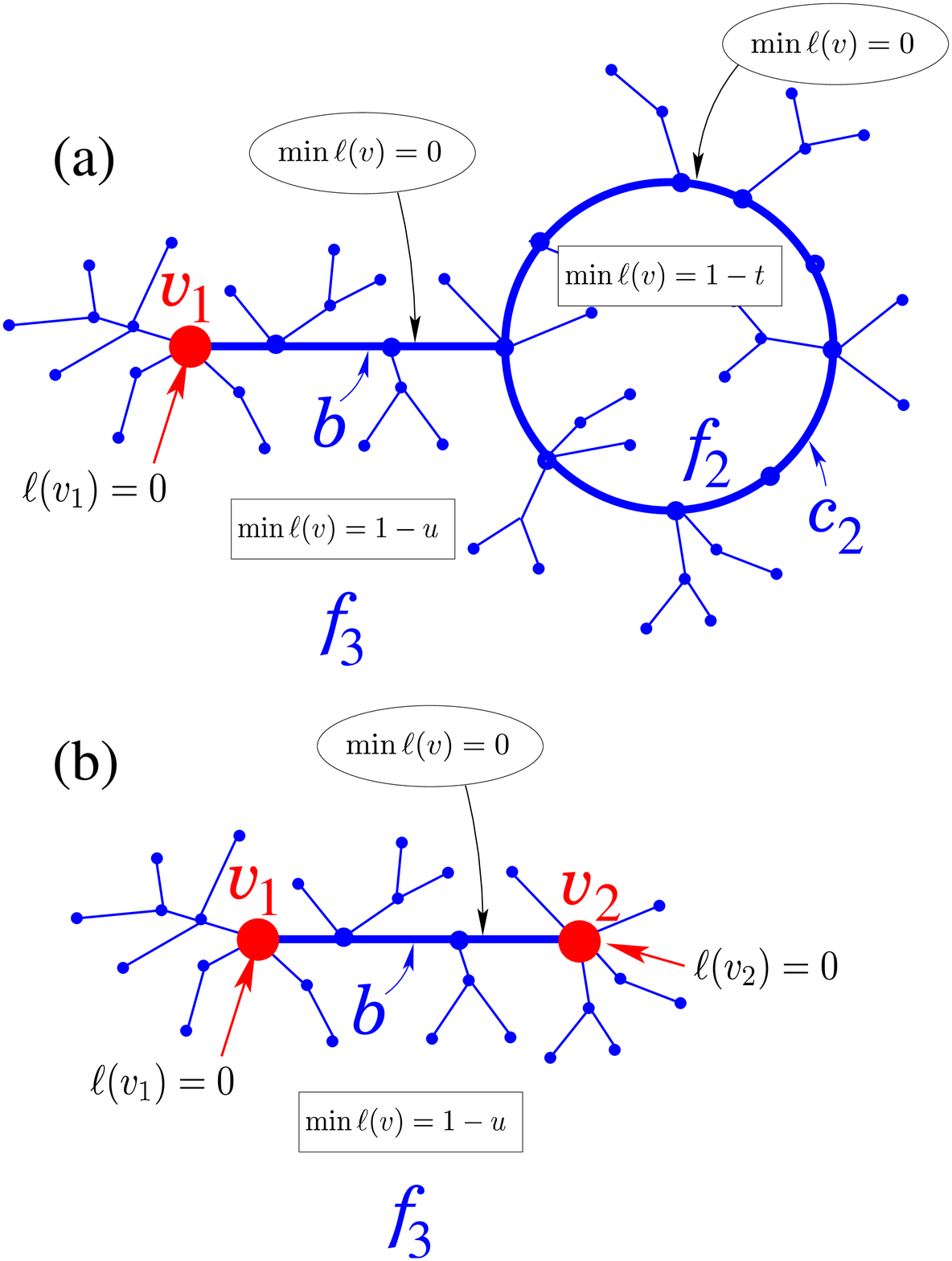}{8.cm}
\figlabel\spectwo
When $s=0$ and $t>0$, we apply the Miermont bijection with $p=2$ sources
only, namely $v_2$ and $v_3$, and delays $\tau_2=-t$, $\tau_3=-u$. 
We obtain a well-labeled map with two faces of the type 
illustrated in Fig.~\spectwo-(a). In particular, $v_1$ is necessarily
incident to $f_3$ and has label $0$, and is connected to 
the frontier between $f_2$ and $f_3$ by a bridge having non-negative
labels only. This can be seen as a degenerate version of the generic 
case displayed in Fig.~\specialdelays, where the face $f_1$ is shrunk
into a single vertex. We have a symmetric picture when $s>0$ and $t=0$.
Finally, if $s=t=0$, we apply the Miermont bijection with $p=1$ source
only (equivalent to the Schaeffer bijection), namely $v_3$, and delay 
$\tau_3=-u$. We then obtain a well-labeled tree on which the vertices 
$v_1$ and $v_2$ have label $0$ and the branch connecting them has 
non-negative labels. Again this is a degenerate case of the generic
situation in which both $f_1$ and $f_2$ degenerate to single vertices.

To conclude, triply-pointed quadrangulations with prescribed values
of $d_{12}$, $d_{13}$ and $l_{123}$ are in one-to-one correspondence
with well-labeled maps of the generic type displayed in Fig.~\specialdelays,
or of its degenerate versions displayed in Fig.~\spectwo.

\fig{A schematic picture of the known generating functions 
$R_\ell$, $X_{s,t}$, ${\tilde X}_{\ell;s,t}$ and $Y_{s,t,u}$ 
(see the text).}{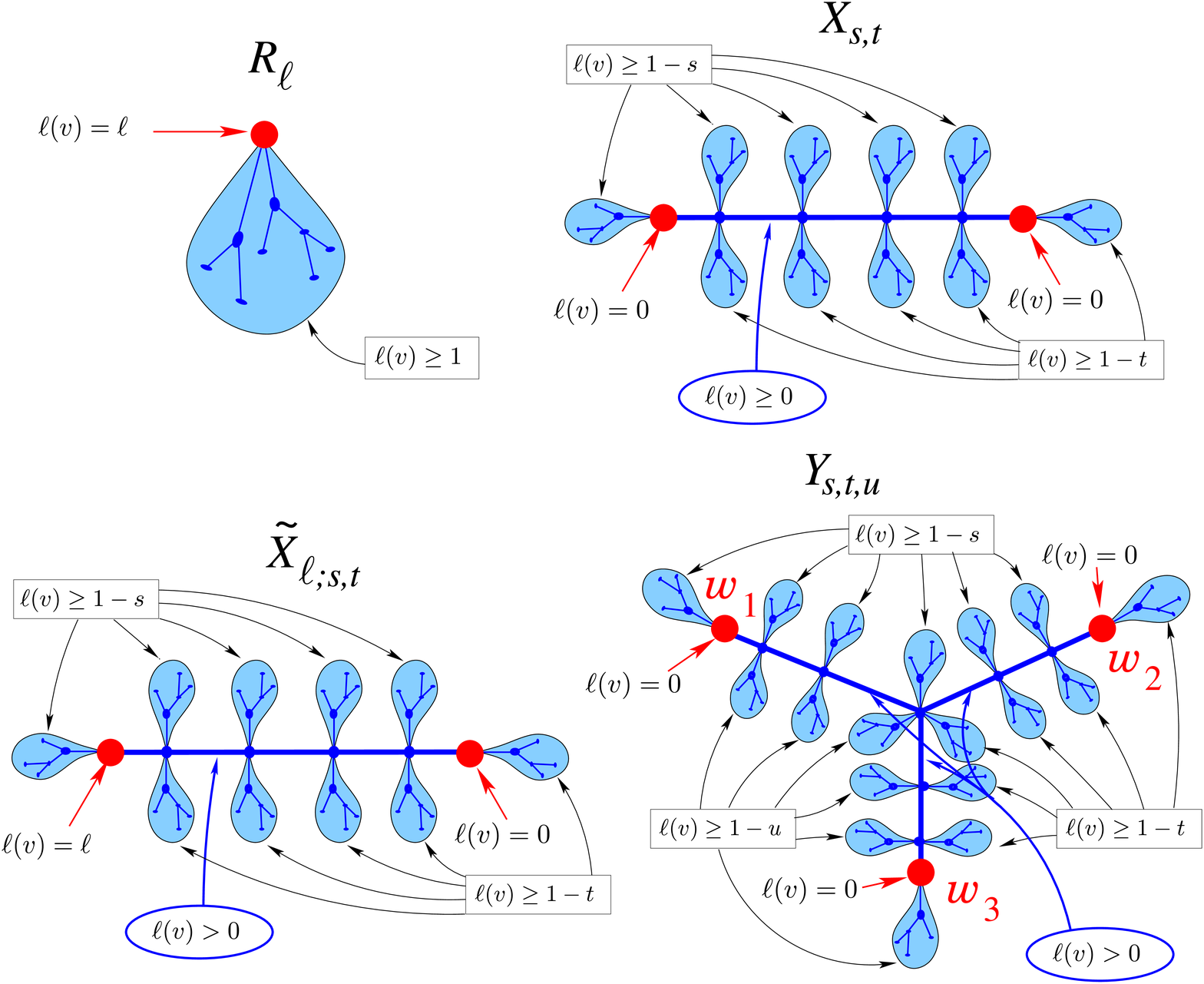}{13.cm}
\figlabel\variousgen
\break
\subsec{Generating functions}

\noindent{\it Known generating functions}

We can now readily relate the generating functions of the various 
well-labeled maps above to those introduced in Ref.~\THREEPOINT.
As usual, we attach a weight $g$ to each edge of a well-labeled map, 
which amounts to a weight $g$ per face of the quadrangulation.
The first generating function is that of well-labeled
trees planted at a corner with label $\ell>0$ and whose labels are
all larger than or equal to $1$ (see Fig.~\variousgen). It reads \GEOD:
\eqn\Rdei{R_\ell= R {\q{\ell}\, \q{\ell+3} \over
\q{\ell+1}\, \q{\ell+2}}} 
where
\eqn\defq{\q{\ell}\equiv {1-x^\ell\over 1-x}}
and where
\eqn\Rxexplicit{\eqalign{R &= {1-\sqrt{1-12g}\over 6 g}\ ,\cr
x &={1-24g-\sqrt{1-12g}+\sqrt{6}\sqrt{72g^2+6g+\sqrt{1-12g}-1}\over
2(6g+\sqrt{1-12g}-1)}\ .\cr}}
Note that $R_{\ell}=1+{\cal O}(g)$ for all $\ell \geq 1$, 
with a conventional weight $1$ for the tree reduced to a single vertex.
The generating function of well-labeled trees planted at a corner with 
label $\ell\geq 0$ and whose labels are all larger than or equal to $1-s$, for
some $s>0$, is then simply given by $R_{\ell+s}$, as obtained by a simple 
shift of all labels by $s$.

The second generating function is that of well-labeled trees
with two distinct marked vertices having label $0$, connected by a branch
with non-negative labels only, and such that the trees attached to
one side of the branch have labels larger than or equal to $1-s$ and those
attached to the other side have labels larger than or equal to $1-t$, 
with $s>0$ and $t>0$ (see Fig.~\variousgen). By convention, the trees 
attached to the marked vertices are assumed to be on opposite sides, 
so that the result is
symmetric in $s$ and $t$. This generating function reads \THREEPOINT:
\eqn\xst{\eqalign{X_{s,t}&=
\sum_{m\geq 0} \quad \sum_{
{\cal M}= (0=\ell_0, \ell_1, \ldots,\ell_{m}=0) \atop
{\rm s.t.}\ \ell_i \geq 0,\ \vert \ell_{i+1}-\ell_i \vert \leq 1,\ i=0,\ldots,
m-1}
\quad
\prod_{k=0}^{m-1} g\, R_{\ell_k+s}\, R_{\ell_k+t}\cr  &
={\q{3}\, \q{s+1}\, \q{t+1}\, \q{s+t+3} \over \q{1} 
\q{s+3}\, \q{t+3}\, \q{s+t+1}}\ .\cr}}
Note that $X_{s,t}=1+{\cal O}(g)$, with a conventional weight $1$
for the tree reduced to a single vertex, which is added for convenience
to the family of trees enumerated by $X_{s,t}$.

We may instead consider well-labeled trees with two marked vertices, 
one with label $\ell > 0$, the other with label $0$, with {\it strictly
positive} labels on the branch inbetween and such that the trees
attached to one side have labels larger than or equal to $1-s$ and those
attached to the other side have labels larger than or equal to $1-t$,
with $s>0$ and $t>0$ (see Fig.~\variousgen). 
The tree attached to extremity with label $\ell$ 
is assumed to have labels larger than or equal to $1-s$ and that
attached to the extremity with label $0$ is assumed to 
have labels larger than or equal to $1-t$. The resulting generating function
reads \THREEPOINT: 
\eqn\tildexlst{\eqalign{\tilde{X}_{\ell;s,t}& = 
\sum_{m\geq \ell} \quad \sum_{
{\cal M}= (\ell=\ell_0, \ell_1, \ldots,\ell_{m}=0) \atop
{\rm s.t.}\ \ell_i > 0,\ \vert \ell_{i+1}-\ell_i \vert \leq 1,\ i=0,\ldots,
m-1}
\quad g\, R_{\ell+s} R_{t}\ 
\prod_{k=1}^{m-1} g\, R_{\ell_k+s}\, R_{\ell_k+t}\cr  &
= {x^{\ell} \q{s+1} \q{s+2} \q{t} \q{t+3} \q{2\ell+s+t+3} \over
\q{s+t+3} \q{\ell+s+1} \q{\ell+s+2} \q{\ell+t} \q{\ell+t+3}}\ .\cr}}
This last formula extends to $\ell=0$ where it yields $X_{0;s,t}=1$,
corresponding again to a conventional weight $1$ for the tree reduced to 
a single vertex.

The final generating function counts well-labeled trees with
three marked vertices, say $w_1$, $w_2$, $w_3$, and with the 
following constraints (see Fig.~\variousgen).
On the tree, the marked vertices are connected by three branches
joining at a central vertex. 
We impose that the branches leading respectively to $w_1$, $w_2$
and $w_3$ appear clockwise around this central vertex. We also
impose that all labels on these
branches be strictly positive, except for $w_1$, $w_2$
and $w_3$, which have label $0$. We further impose that trees attached
to the branch from $w_1$ to $w_2$ on the side opposite to $w_3$
have labels larger than or equal to $1-s$. Similarly, we impose that 
trees attached to the branch from $w_2$ to $w_3$ (respectively from
$w_3$ to $w_1$) on the side opposite to 
$w_1$ (respectively $w_2$) have labels larger than or equal to $1-t$ 
(respectively $1-u$). By convention, the labels on the tree attached
to $w_1$ (respectively $w_2$ and $w_3$) are assumed to be larger
than or equal to $1-s$ (respectively $1-t$ and $1-u$). The corresponding
generating function reads \THREEPOINT:
\eqn\ystu{\eqalign{Y_{s,t,u} & = \sum_{\ell=0}^{\infty} \tilde{X}_{\ell;s,t}
\tilde{X}_{\ell;t,u} \tilde{X}_{\ell;u,s} \cr
& = {\q{s+3} \q{t+3} \q{u+3} \q{s+t+u+3} \over \q{3}
\q{s+t+3} \q{t+u+3} \q{u+s+3}}\ .\cr}}
Again, we have $Y_{s,t,u}=1+{\cal O}(g)$, with a conventional weight $1$
for the tree reduced to a single vertex, which is added for convenience
to the family of trees enumerated by $Y_{s,t,u}$.

\bigskip
\noindent{\it Application to minimal separating loops via the Schaeffer
bijection}
\fig{The cutting of a well-labeled tree of the type of Fig.~\condtree\ 
(with a relaxed constraint on the global minimal label) at the
first and last label $0$ encountered along the branch from $v_1$ to $v_2$.
This results into three pieces, enumerated by ${\tilde X}_{s;u,u}$, 
$X_{u,u}$ and ${\tilde X}_{t;u,u}$ respectively.}{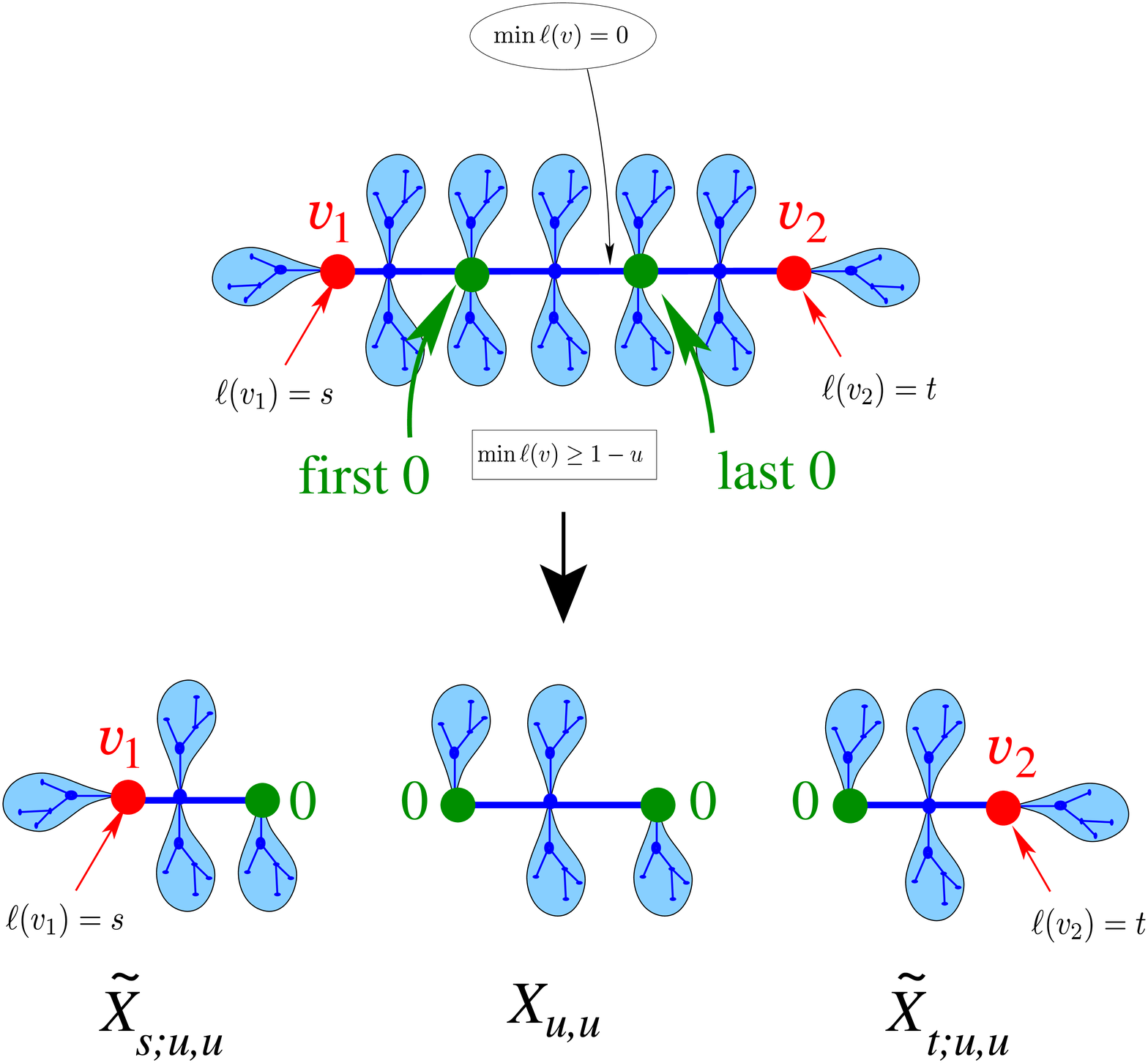}{10.cm}
\figlabel\cutscha
In this approach, we have to enumerate trees 
of the type displayed in Fig.~\condtree. It is convenient to 
first relax the condition on the global minimum, demanding only 
that it be larger than or equal to $1-u$. We can then decompose the
tree by cutting it at the first and last occurrence of the
label $0$ on the branch from $v_1$ to $v_2$, resulting in three
trees counted respectively by ${\tilde X}_{s;u,u}$, $X_{u,u}$ and
${\tilde X}_{t;u,u}$ (see Fig.~\cutscha\ for an illustration). 
The corresponding generating function therefore reads:
\eqn\Hstu{\eqalign{H_{\rm loop}(s,t,u)& ={\tilde X}_{s;u,u}\, X_{u,u}\, 
{\tilde X}_{t;u,u} \cr & =x^{s+t}{\q{3}\q{u}^2\q{u+1}^4\q{u+2}^2
\q{2s+2u+3}\q{2t+2u+3} 
\over 
\q{1}\q{2u+1}\q{2u+3} 
\prod_{k=0}^3 \q{s+u+k} \q{t+u+k}}\ .\cr}}
To restore the condition that the global minimal label be 
exactly $1-u$, we simply have to consider $\Delta_u H_{\rm loop}(s,t,u)$
where $\Delta_u$ is the finite difference operator:
\eqn\finitediff{\Delta_u\, f(u)\equiv f(u)-f(u-1)\ .}
To conclude, the generating function for triply-pointed
quadrangulations with prescribed values of $d_{13}$, $d_{23}$ and $l_{123}$
is given by
\eqn\resultone{\eqalign{& G_{\rm loop}(d_{13},d_{23};l_{123})=
\Delta_u H_{\rm loop}(s,t,u)\cr &{\rm with} \ s=d_{13}-l_{123}/2\ , \ \ 
t=d_{23}-l_{123}/2\ , \ \ u=l_{123}/2 \ .\cr}}

\bigskip
\noindent{\it Application to minimal separating loops via the Miermont
bijection}
\fig{The cutting of a well-labeled map of the type of 
Fig.~\specialdelays\ (with relaxed constraints on labels inside
each face) at the
first and last label $0$ encountered along the cycles $c_1$, $c_2$ and
the branch $b$ (see the text). This results into five pieces, enumerated 
by $X_{s,u}$, $Y_{s,u,u}$, $X_{u,u}$, $Y_{t,u,u}$ and $X_{t,u}$ 
respectively.}{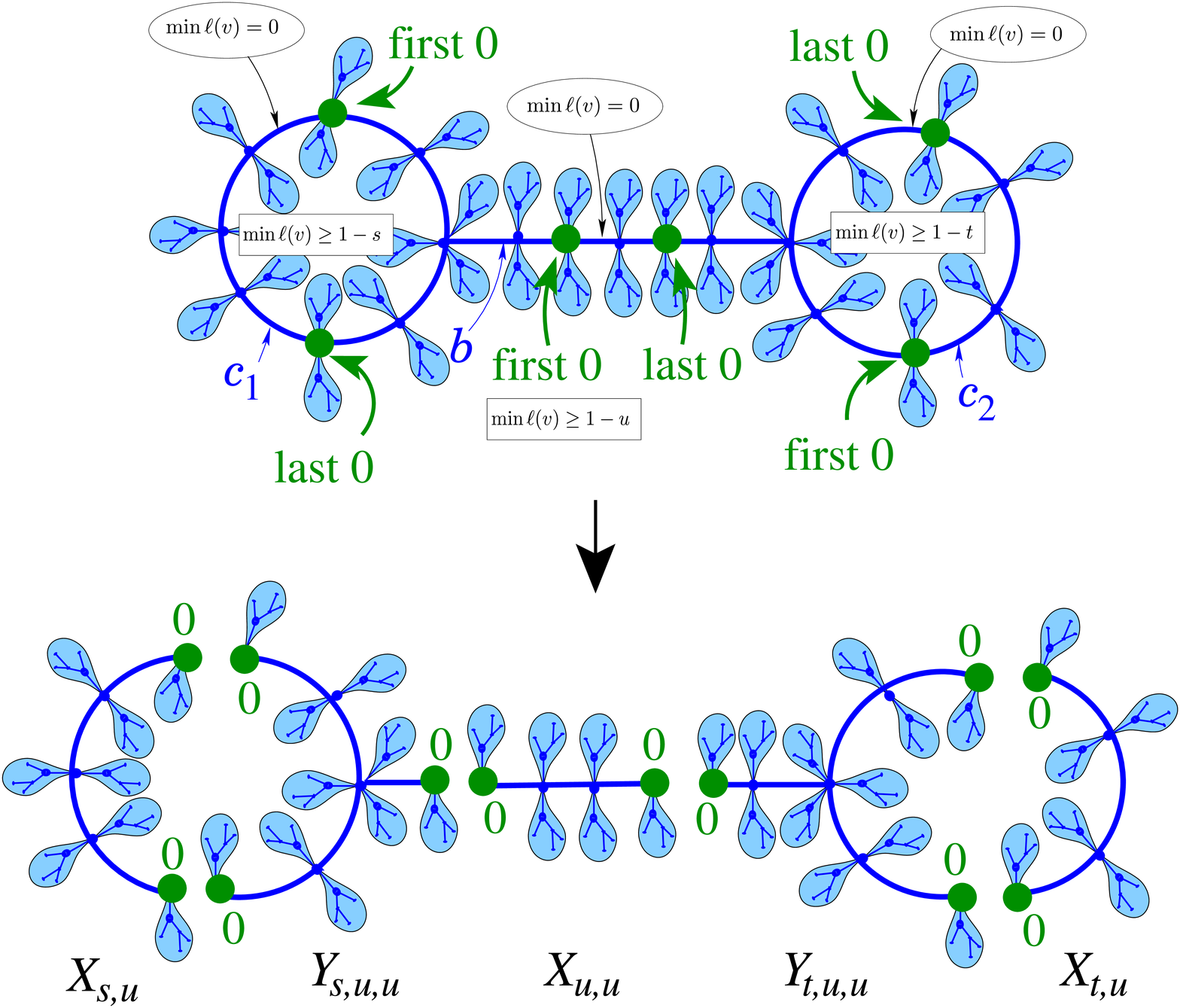}{13.cm}
\figlabel\cutmier
In this approach, we simply have to enumerate maps 
of the type displayed in Figs.~\specialdelays\ and \spectwo. Again, we
relax the conditions on the minimal label within each face, namely
we demand only that it be larger than or equal to $1-s$, $1-t$ or
$1-u$ respectively. In the generic case of Fig.~\specialdelays, 
we can now decompose the map by cutting it 
at the first and last occurrence of the label $0$ on the cycle
$c_1$, starting from the endpoint of the bridge $b$, at
the first and last occurrence of the label $0$ on the cycle
$c_2$, starting from the other endpoint of the bridge $b$, and
finally at the first and last occurrence of the label $0$ on the 
bridge $b$ itself (see Fig.~\cutmier\ for an illustration). 
This results in general into five trees counted
respectively by $X_{s,u}$, $Y_{s,u,u}$, $X_{u,u}$, $Y_{t,u,u}$ 
and $X_{t,u}$. The corresponding generating function therefore reads
\eqn\floopstu{\eqalign{F_{\rm loop}(s,t,u)&= 
X_{s,u}\, Y_{s,u,u}\, X_{u,u}\, Y_{t,u,u}\, X_{t,u} \cr
& = 
{\q{3}\q{s+1}\q{t+1}\q{u+1}^4\q{s+2u+3}\q{t+2u+3}\over
\q{1}^3\q{s+u+1}\q{s+u+3}\q{t+u+1}\q{t+u+3}\q{2u+1}\q{2u+3}}
\cr}}
Note that this formula incorporates the cases where some of the
cutting points above coincide as we added in $X_{s,t}$ and $Y_{s,t,u}$ the 
weight $1$ of the tree reduced to a single vertex. It also 
naturally incorporates the degenerate cases of Fig.~\spectwo: for
instance, the situation of Fig.~\spectwo-(a) 
is properly taken into account by having the two leftmost trees in
the decomposition of Fig.~\cutmier\ reduced to single vertices, 
while the situation of Fig.~\spectwo-(b) is properly taken
into account by having the two leftmost and the two rightmost trees in
the decomposition of Fig.~\cutmier\ reduced to single vertices.
Again, we can restore the constraint that the minimal label
within each face be equal to $1-s$, $1-t$ or
$1-u$ respectively by considering $\Delta_s \Delta_t \Delta_u 
F_{\rm loop}(s,t,u)$. 
We deduce the alternative formula: 
\eqn\resulttwo{\eqalign{& G_{\rm loop}(d_{13},d_{23};l_{123})=
\Delta_s\Delta_t\Delta_u 
F_{\rm loop}(s,t,u)\cr &{\rm with} \ s=d_{13}-l_{123}/2\ , \ \ 
t=d_{23}-l_{123}/2\ , \ \ u=l_{123}/2 \ .\cr}}

Note that the two expressions \resultone\ and \resulttwo\ are consistent
as we have the identity:
\eqn\remident{\Delta_s {\q{s+1}\q{s+2u+3}\over \q{s+u+1}\q{s+u+3}}
= x^s {\q{1}\q{u}\q{u+2}\q{2s+2u+3}\over
\prod_{k=0}^3\q{s+u+k}}}
which can be checked directly from the definition \defq.

A simpler generating function is that of triply-pointed quadrangulations
with a prescribed value of $l_{123}$ only. The corresponding
generating function $G_{\rm loop}(l_{123})$ is obtained by
summing $G_{\rm loop}(d_{13},d_{23};l_{123})$ over all the allowed
values of $d_{13}$ and $d_{23}$ for a fixed $l_{123}$. This amounts
to a summation over all non-negative values of $s$ and $t$, which is 
easily performed upon using the expression \resulttwo\ by
noting that, with the above expression \floopstu, the
quantities $F_{\rm loop}(-1,t,u)$ and $F_{\rm loop}(s,-1,u)$ 
vanish identically, so that:
\eqn\margi{\eqalign{G_{\rm loop}(l_{123})& = \Delta_u
 F_{\rm loop}(\infty,\infty,u)\cr
&= \Delta_u {\q{3}\q{u+1}^4 \over \q{1}^3 \q{2u+1} \q{2u+3}}\  
{\rm with} \ \ u=l_{123}/2\ .\cr}}

\subsec{Continuum limit}

The scaling limit is obtained by letting $g$ approach its
critical value $1/12$ and considering large values of $d_{13}$,
$d_{23}$ and $l_{123}$ with the following scaling:
\eqn\scalone{\eqalign{& g ={1\over 12}\left(1- \Lambda\, \epsilon\right)\cr
& d_{13} =D_{13} \epsilon^{-1/4}\ ,\ 
d_{23} =D_{23} \epsilon^{-1/4}\ ,\ 
l_{123} =L_{123} \epsilon^{-1/4}\ ,\cr
}}
and $\epsilon \to 0$. The quantity $\Lambda$ may be interpreted as
a ``cosmological constant". In this limit, we have:
\eqn\margicont{\eqalign{& G_{\rm loop}(l_{123})\sim \epsilon^{-1/4}\  2 \,
{\cal G}_{\rm loop}
(L_{123}; \alpha) \cr & {\rm where}\ 
{\cal G}_{\rm loop}(L_{123};\alpha)= {1\over 2}\ \partial_U  \left.
{3\over \alpha^2} {\sinh^4(\alpha U) \over\sinh^2(2 \alpha U)}
\right\vert_{U=L_{123}/2}
= {3\over 4 \alpha} {\sinh(\alpha L_{123}/2)\over \cosh^3(\alpha L_{123}/2)}
\ .\cr}}
Here and throughout the paper, we use the notation:
\eqn\valalp{\alpha=\sqrt{3/2}\Lambda^{1/4}\ .}
Note the factor $1/2$ in the definition of ${\cal G}_{\rm loop}$, which
is introduced to compensate the fact that, at the discrete level,
$l_{123}$ can take only even integer values. More generally, we have
\eqn\scalGFloop{\eqalign{& G_{\rm loop}(d_{12},d_{13},l_{123})\sim 
\epsilon^{1/4} \ 2\, {\cal G}_{\rm loop}(D_{12},D_{13},L_{123};\alpha)\ , \cr
& F_{\rm loop}(s,t,u)\sim \epsilon^{-1/2}\ {\cal F}_{\rm loop}(S,T,U;\alpha)
\ , \cr
& H_{\rm loop}(s,t,u)\sim {\cal H}_{\rm loop}(S,T,U;\alpha) 
\ ,\cr}}
where
\eqn\floopcont{\eqalign{& {\cal F}_{\rm loop}(S,T,U;\alpha) = {3\over \alpha^2}
{\sinh(\alpha S) \sinh(\alpha T) \sinh^4(\alpha U) \sinh(\alpha (S+2U)) 
\sinh(\alpha (T+2U)) \over \big(\sinh(\alpha(S+U)) \sinh(\alpha(T+U)) 
\sinh( 2\alpha U)\big)^2}\ ,\cr & 
{\cal H}_{\rm loop}(S,T,U;\alpha)= 3 {\sinh^8(\alpha U)
\sinh(2 \alpha(S+U)) \sinh(2 \alpha (T+U))\over
\sinh^2(2\alpha U)\sinh^4(\alpha(S+U))\sinh^4(\alpha(T+U))}\ ,\cr}}
and where
\eqn\conttroisloop{\eqalign{&{\cal G}_{\rm loop}(D_{12},D_{23},L_{123};\alpha) 
= {1\over 2}\ \partial_S \partial_T \partial_U {\cal F}_{\rm loop}(S,T,U;\alpha)
={1\over 2}\ \partial_U {\cal H}_{\rm loop}(S,T,U;\alpha)\cr 
&{\rm with} \ S=D_{13}-L_{123}/2\ , \ \ 
T=D_{23}-L_{123}/2\ , \ \ U=L_{123}/2 \ .\cr}}
Again the two expressions above for 
${\cal G}_{\rm loop}(D_{12},D_{23},L_{123};\alpha)$ are
consistent as we have the identity
\eqn\remidentcont{\partial_S\left({1\over \alpha}
{\sinh(\alpha S)\sinh(\alpha(S+2U))\over \sinh^2(\alpha (S+U))}\right)
= {\sinh^2(\alpha U) \sinh(2 \alpha(S+U))\over \sinh^4(\alpha(S+U))}\ ,}
which is the continuous counterpart of \remident.

\fig{Plot of the probability density 
$\rho_{\rm loop}(L_{123})$.}{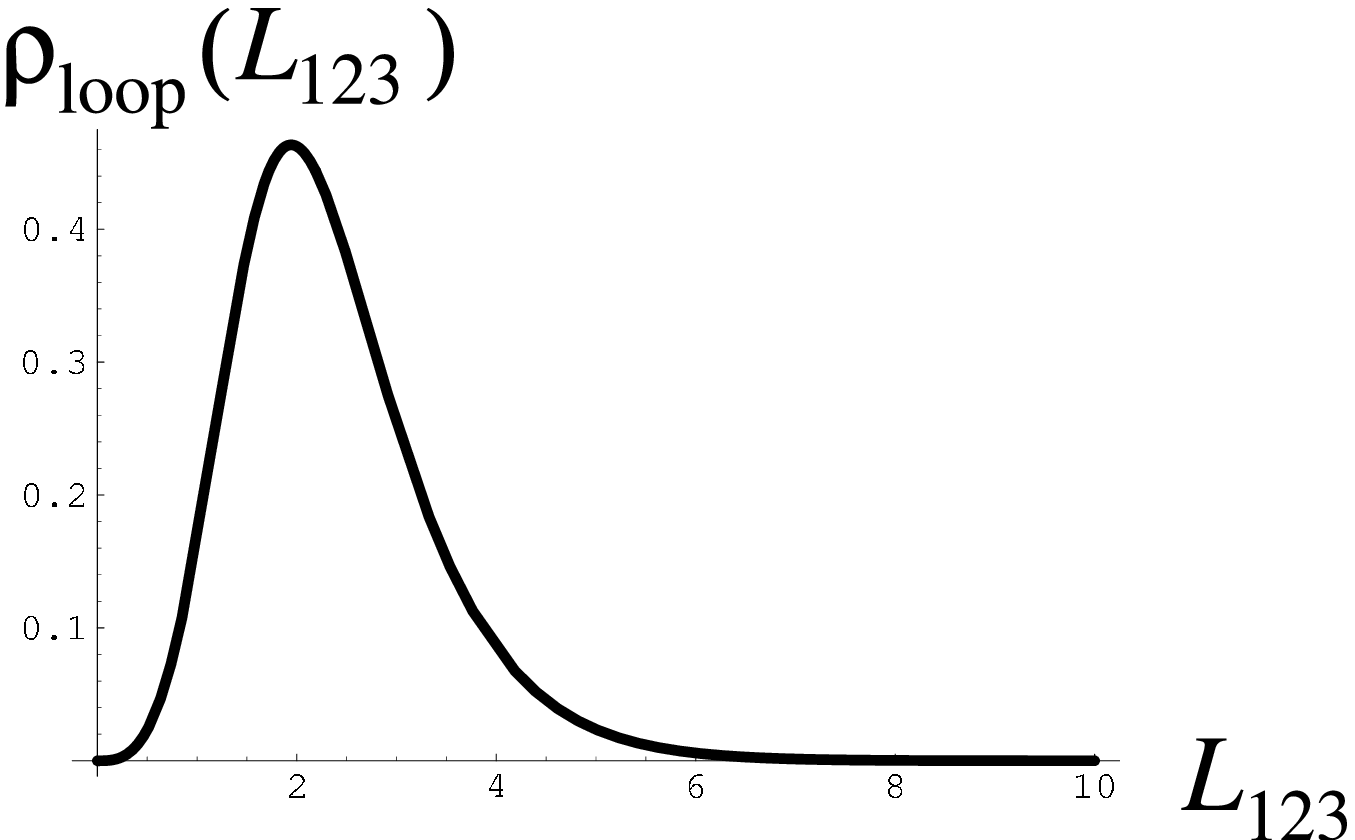}{8.cm}
\figlabel\rhol
The above continuous formulae can be used to capture the statistical 
properties of triply-pointed quadrangulations with {\it fixed size},
i.e.\ with a fixed number $n$ of faces, in the limit $n\to \infty$.
Indeed, fixing $n$ amounts to extracting the $g^n$ term of the various
discrete generating functions at hand. This can be done by a 
contour integral in $g$ which, at large $n$, translates via
a saddle point estimate into an integral over a real variable $\xi$.
More precisely, considering for instance the generating function
$G_{\rm loop}(l_{123})$, we write
\eqn\contintex{G_{\rm loop}(l_{123})\vert_{g^n}=
{1\over 2 {\rm i}\pi} \oint {dg\over g^{n+1}} G_{\rm loop}(l_{123})}
and we perform the change of variables 
\eqn\scalonebis{g={1\over 12}\left(1+{\xi^2\over n}\right)\ , \qquad
l_{123} =L_{123}\, n^{1/4}\ .}
At large $n$, the contour integral becomes at dominant order
an integral over real values of $\xi$ and
we can use the continuous formulae above with $\epsilon=1/n$ and 
$\lambda=-\xi^2$. After a proper normalization by the 
number of triply-pointed quadrangulations with fixed size $n$, 
we obtain the probability density $\rho_{\rm loop}(L_{123})$ for the 
rescaled length $L_{123}$:
\eqn\conttwo{\rho_{\rm loop}(L_{123})= {2\over {\rm i} \sqrt{\pi}} 
\int_{-\infty}^{\infty} 
d\xi\ \xi\, e^{-\xi^2}\, {\cal G}_{\rm loop}(L_{123};\sqrt{-3{\rm i}\xi/2})\ .}
The quantity $\rho_{\rm loop}(L_{123})\,dL_{123}$ is the infinitesimal
probability that the (rescaled) minimal length for loops having origin $v_3$ 
and separating $v_1$ from $v_2$ lies in the range $[L_{123},L_{123}+dL_{123}]$
in the ensemble of triply-pointed quadrangulations with fixed size $n$,
in the limit $n\to \infty$.
This probability density 
is plotted in Fig.~\rhol\ and has the following limiting behaviors:
\eqn\limbe{\eqalign{
&\rho_{\rm loop}(L_{123})\sim {3\over 16} L_{123}^3\  \ {\rm when}\ 
L_{123}\to 0\ ,\cr
&\rho_{\rm loop}(L_{123})\sim {1\over 6^{1/6}} L_{123}^{2/3}\ 
e^{-\left({3\over 4}\right)^{5/3} L_{123}^{4/3}}\  \ {\rm when}\ 
L_{123}\to \infty\ .\cr} }
The associated average value of $L_{123}$ reads
\eqn\avl{\langle L_{123} \rangle ={4\over 3}\, \langle D \rangle =
2.36198\cdots \quad {\rm with}
\quad \langle D \rangle=2 \sqrt{{3\over \pi}}
\Gamma\left({5\over 4}\right)= 1.77148\cdots}
Here and throughout the paper, we decide to express average distances
in units of the average distance $\langle D \rangle$ between two uniformly 
chosen vertices in a large quadrangulation, whose value given above 
was computed in Refs.~[\xref\DELMAS-\xref\JBPHD].
\fig{Plots of the conditional probability density 
$\rho_{\rm loop}(D_{13},D_{23}\vert L_{123})$ for
$L_{123}=2.0$, $L_{123}=1.6$ and $L_{123}=1.0$, from top to 
bottom. For each plot on the left, we display its associated 
contour plot on the right.}{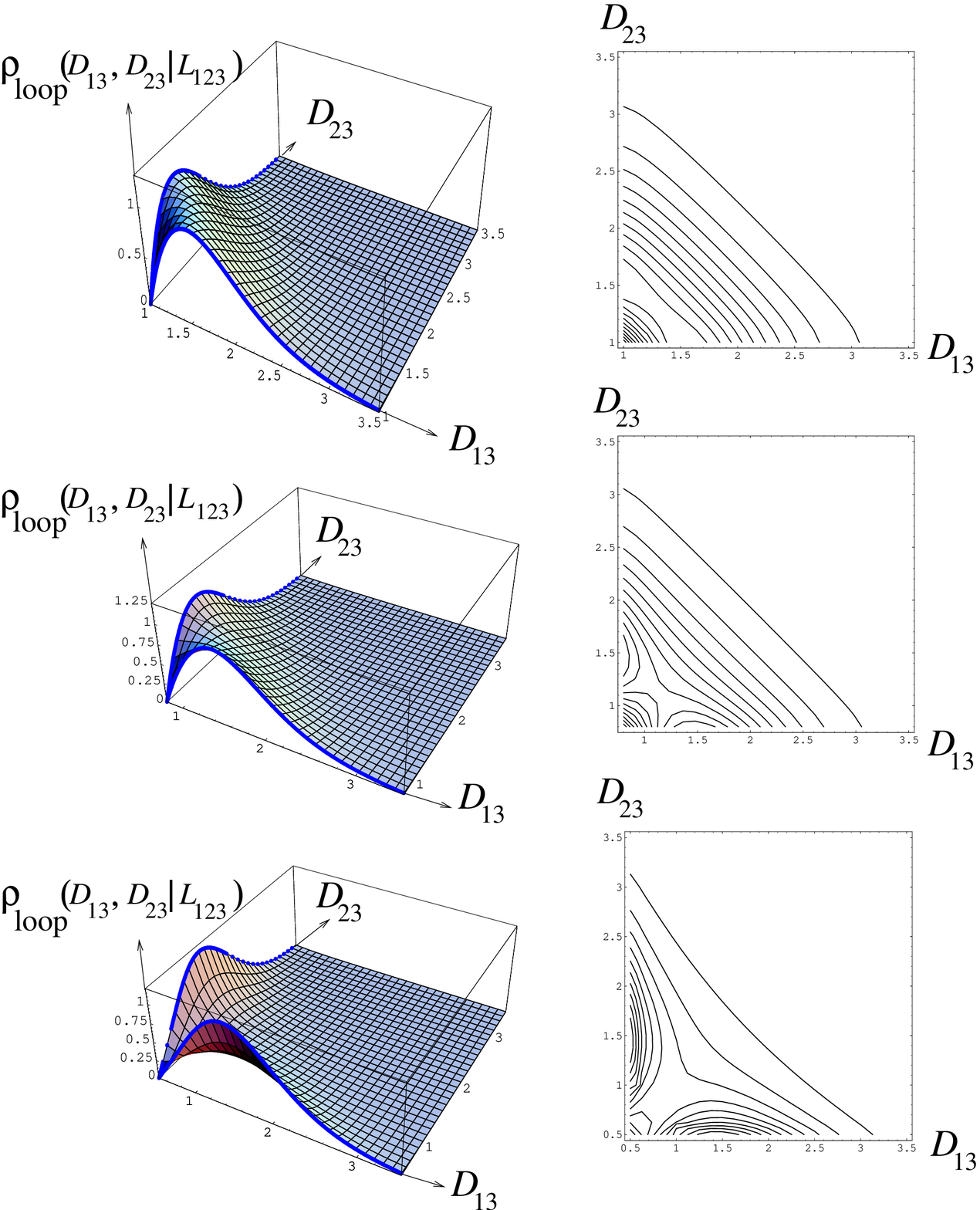}{12.cm}
\figlabel\plots
\fig{Plots of the conditional probability density 
$\rho_{\rm loop}(D_{13},D_{23}\vert L_{123})$ for
$L_{123}=2.0$, $L_{123}=3.0$ and $L_{123}=4.0$, from top to 
bottom.}{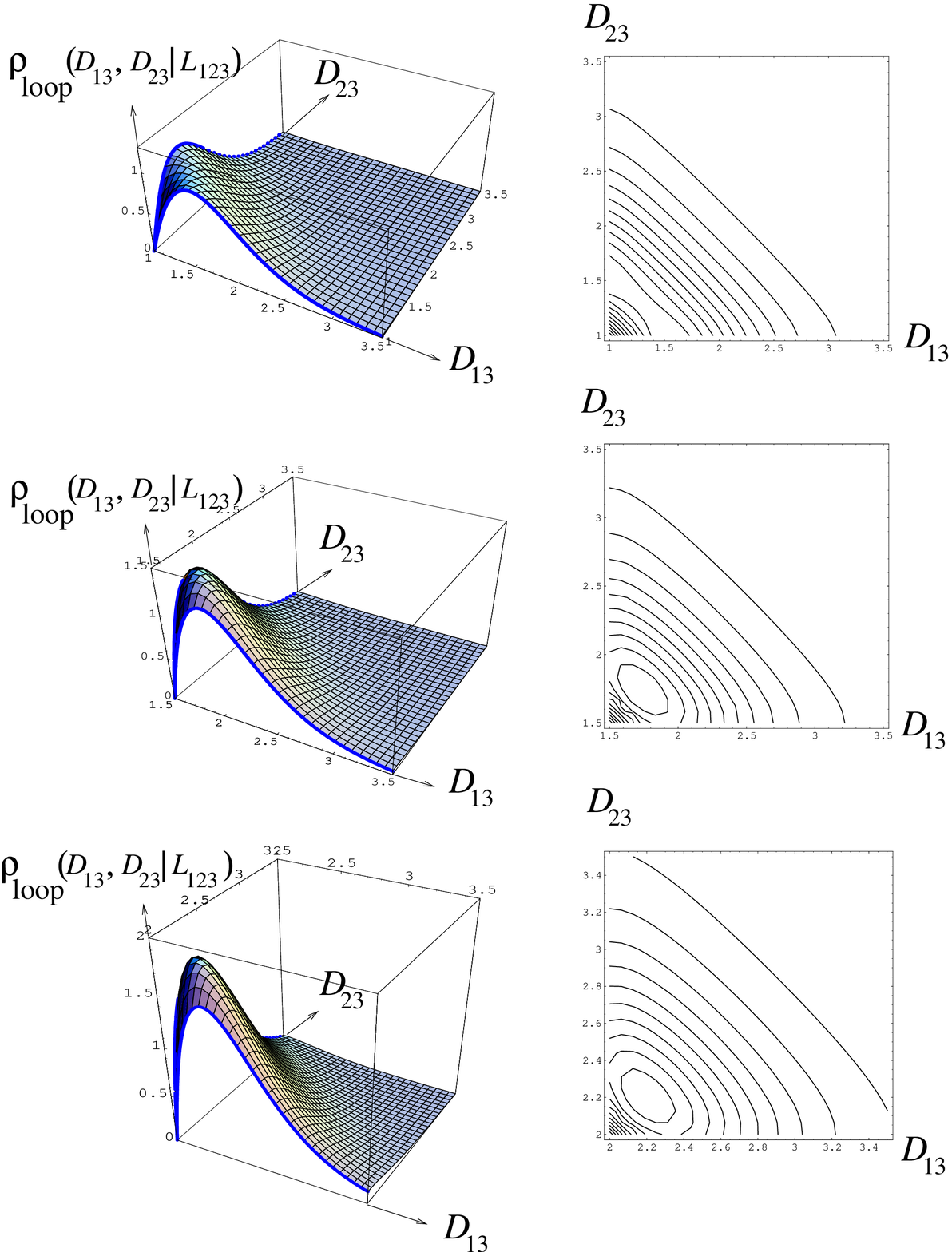}{12.cm}
\figlabel\plotstwo
Similarly, the joint probability density for $D_{13}=d_{13}/n^{1/4}$, 
$D_{23}=d_{23}/n^{1/4}$ and $L_{123}$ reads:
\eqn\contploop{\rho_{\rm loop}(D_{12},D_{13},L_{123})= 
{2\over {\rm i} \sqrt{\pi}} \int_{-\infty}^{\infty} d\xi\ \xi\, e^{-\xi^2}\,
{\cal G}_{\rm loop}(D_{12},D_{13},L_{123};\sqrt{-3{\rm i}\xi/2})\ ,}
while the conditional probability density for $D_{13}$ and $D_{23}$,
given the value of $L_{123}$, simply reads:
\eqn\condpro{\rho_{\rm loop}(D_{12},D_{13}\vert L_{123})={
\rho_{\rm loop}(D_{13},D_{23},L_{123})\over \rho_{\rm loop}(L_{123})}\ .}
This conditional probability density is represented in Fig.~\plots\ for 
decreasing
values of $L_{123}$ (namely $L_{123}=2.0$, $L_{123}=1.6$ and $L_{123}=1.0$),
and in Fig.~\plotstwo\ for increasing values of $L_{123}$ 
(namely $L_{123}=2.0$, $L_{123}=3.0$ and $L_{123}=4.0$). For large
enough $L_{123}$, this joint probability density is maximal for equal
values of $D_{13}$ and $D_{23}$, i.e.\ when the two vertices $v_1$ and
$v_2$ are equally distant from $v_3$. On the contrary, for small enough
$L_{123}$, we observe a symmetry breaking phenomenon with
a probability density being maximal when one of the two vertices 
$v_1$ or $v_2$ lies closer from $v_3$ than the other.
\fig{(a) Plot of the conditional probability density
$\rho_{\rm loop}(D_{13},D_{23}\vert L_{123})$ for a small value of $L_{123}$,
here $L_{123}=0.02$. This density is concentrated in two regions
corresponding to either $D_{13}$ or $D_{23}$ being of order $L_{123}$.
A zoom on the first region is obtained by considering the same plot (b)
with a rescaled abscissa $\omega=2 D_{13}/L_{123}$, or the corresponding
contour plot (c). As apparent
by taking longitudinal and transverse cut views along the thick lines
in (c), the probability density factorizes in this region 
into the product of the density $\psi(\omega)$ (red curve in (d))
and the two-point function $\rho(D_{23})$ (green curve in 
(e)).}{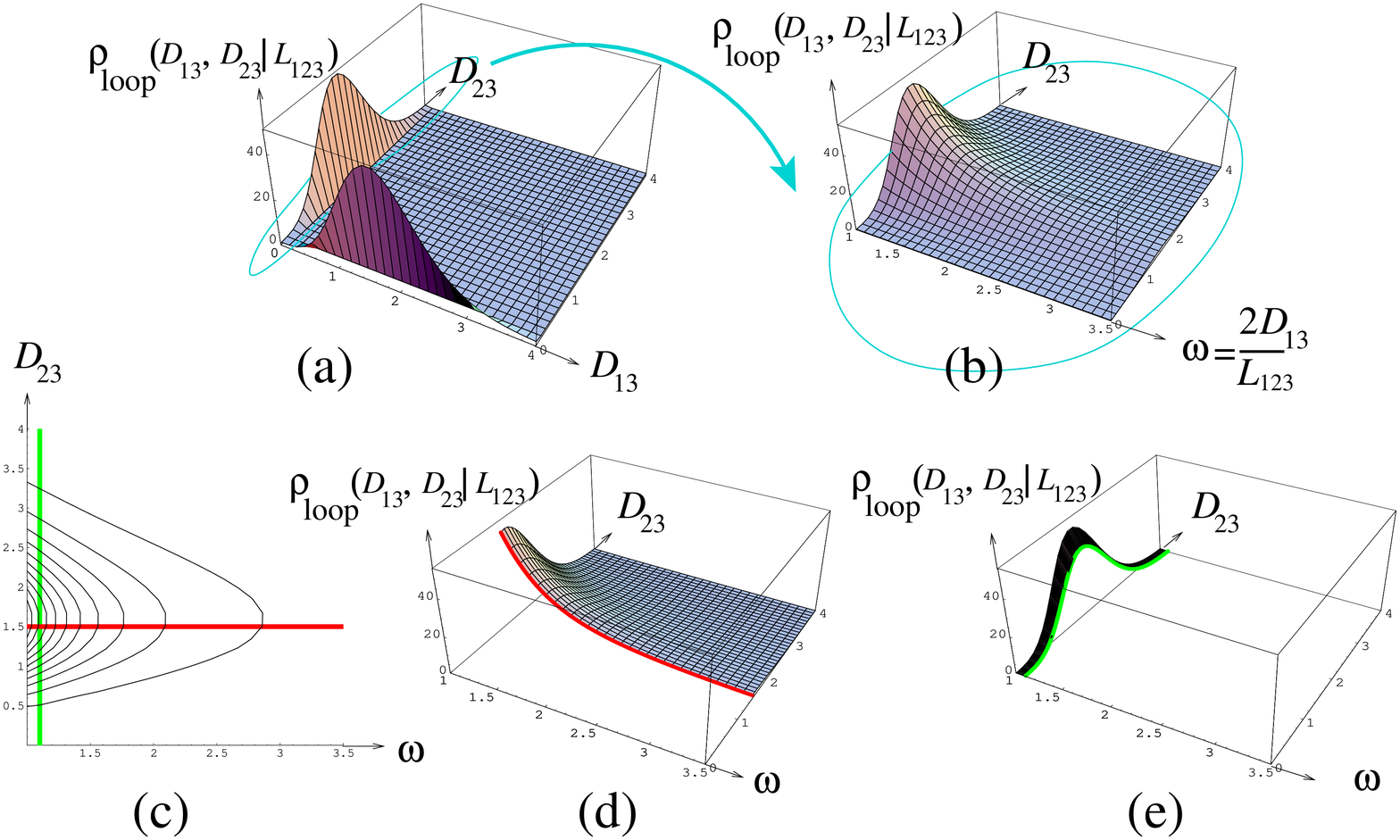}{14.cm}
\figlabel\usmall
This phenomenon increases for smaller $L_{123}$ and, 
when $L_{123}\to 0$, we find that
\eqn\smallubeh{\rho_{\rm loop}(D_{13},D_{23}\vert L_{123})
\sim \rho(D_{13})\times {2\over L_{123}} \psi\left({2 D_{23}\over L_{123}}
\right)
+ \rho(D_{23})\times {2\over L_{123}} \psi\left({2 D_{13}\over 
L_{123}}\right)}
with a scaling function
\eqn\valpsi{\psi(\omega)={3\over 4}\, {2\omega-1\over \omega^4}}
normalized to $1/2$ when $\omega$ varies from $1$ to $\infty$, 
and where $\rho(D)$ is the so called {\it canonical two-point function}, 
which is the probability density for the distance $D$ between two vertices 
picked uniformly at random in a large quadrangulation.
This canonical two-point function is given by a formula
similar to \conttwo: 
\eqn\twopoint{\eqalign{&\rho(D)= 
{2\over {\rm i} \sqrt{\pi}} \int_{-\infty}^{\infty}
d\xi\ \xi\, e^{-\xi^2}\, {\cal G}(D;\sqrt{-3{\rm i}\xi/2})\cr
& {\rm with} \quad {\cal G}(D;\alpha)
= 4 \alpha^3 {\cosh(\alpha D)\over \sinh^3(\alpha D)}\ . \cr}}
The particular form \smallubeh\ expresses that, when $L_{123}$ becomes small, 
one of two vertices $v_1$ or $v_2$, say
$v_1$ necessarily lies in the vicinity of $v_3$, with a distance $D_{13}$
of the order of $L_{123}$ and governed by the density \valpsi\ for 
$\omega=2 D_{13}/L_{123}$, while the other vertex lies at an arbitrary
distance from the two others, with a probability density given simply by the 
two-point function of quadrangulations, as expected. 
This behavior is depicted in Fig.~\usmall, 
for $L_{123}=0.02$. This result corroborates the known property 
of quadrangulations of large size $n$: small loops of length negligible
with respect to $n^{1/4}$ in the quadrangulation necessarily separate it into a
macroscopic domain containing most of the area of the quadrangulation, 
and a small part of negligible size with respect to $n$.
\fig{(a) Plot of the conditional probability density
$\rho_{\rm loop}(D_{13},D_{23}\vert L_{123})$ for a large value of $L_{123}$,
here $L_{123}=20.0$. The same plot (b) and its contour (c)
in the rescaled variables $\mu=(D_{13}-L_{123}/2)(9L_{123}/2)^{1/3}$
and $\nu=(D_{23}-L_{123}/2)(9L_{123}/2)^{1/3}$. In these variables, 
the conditional probability density tends to a limiting 
distribution $\Phi(\mu,\nu)$, as shown in (d).}{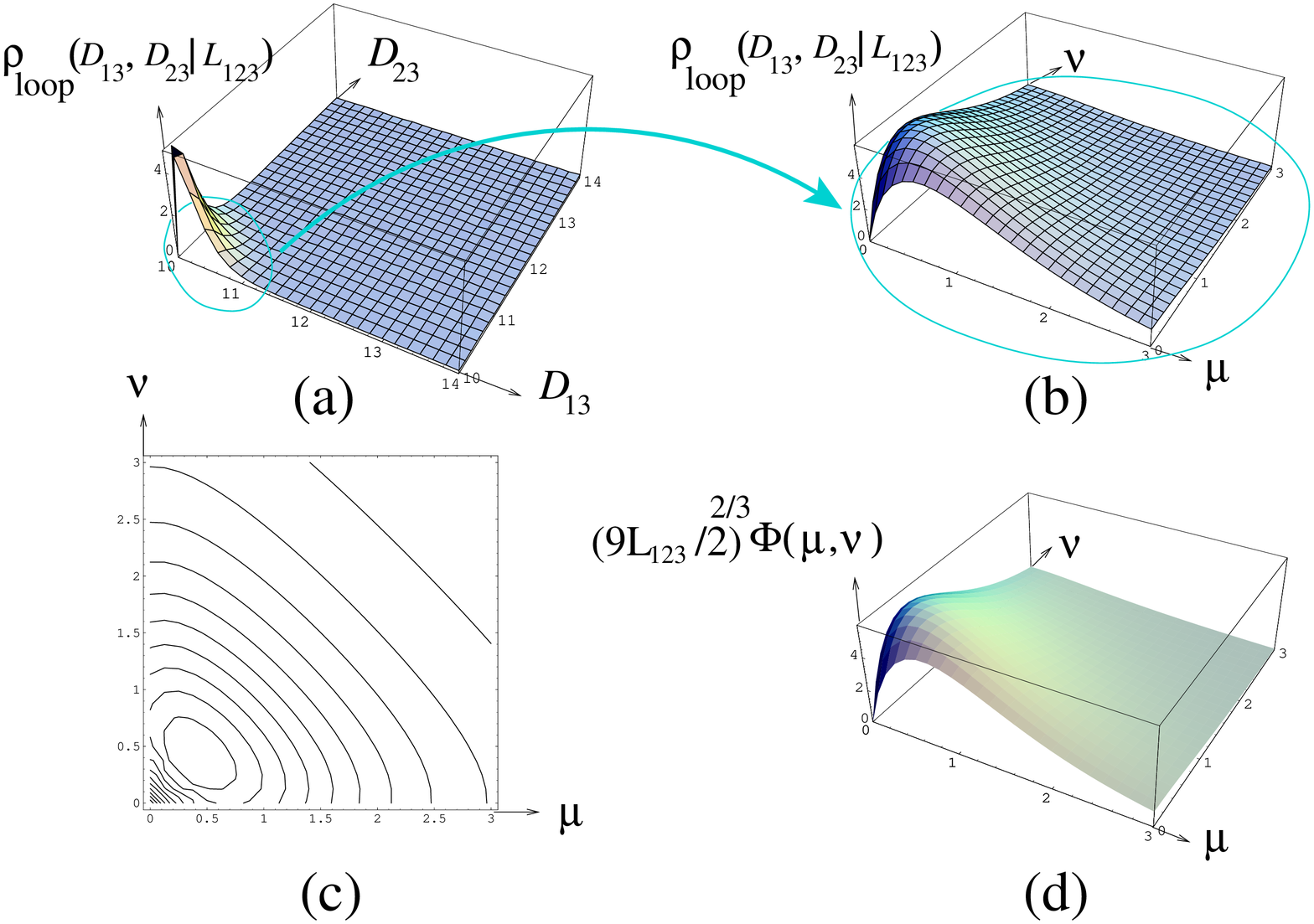}{12.cm}
\figlabel\ularge
In the other limit, i.e.\ when $L_{123}$ becomes large, we find 
the limiting behavior:
\eqn\largeubeh{\eqalign{&\rho_{\rm loop}(D_{13},D_{23}\vert L_{123})
\sim \left({9 L_{123}\over 2}\right)^{2/3}\Phi(\mu,\nu)\cr
&{\rm with}\ \ \mu= \left(D_{13}-{L_{123}\over 2}\right)
\left({9 L_{123}\over 2}\right)^{1/3}
\ ,\ \ \nu=\left(D_{23}-{L_{123}\over 2}\right)
\left({9 L_{123}\over 2}\right)^{1/3}
\ ,\cr}}
with a scaling function
\eqn\valPhi{\Phi(\mu,\nu)= e^{-(\mu+\nu)}\left(2-e^{-\mu}-e^{-\nu}\right)}
properly normalized to $1$ when $\mu$ and $\nu$ vary from $0$ to $\infty$.
At large $L_{123}$, both distances $D_{13}$ and $D_{23}$ are therefore
necessarily of order $L_{123}/2$, with differences 
$D_{13}-L_{123}/2$ and $D_{23}-L_{123}/2$ of order $L_{123}^{-1/3}$, 
governed by the joint probability density \valPhi. This behavior
is depicted in Fig.~\ularge\ for $L_{123}=20.0$.
\bigskip
\bigskip

\newsec{Confluence}
\subsec{Confluence of geodesics}
\fig{A schematic picture of the phenomenon of confluence of 
geodesics. For generic points $v_1$, $v_2$ and $v_3$ and
in the scaling limit of large quadrangulations, the
geodesic from $v_1$ to $v_3$ and that from $v_2$ to $v_3$ 
(represented as thick blue lines) are unique and
have a common part of macroscopic length $\delta$.}{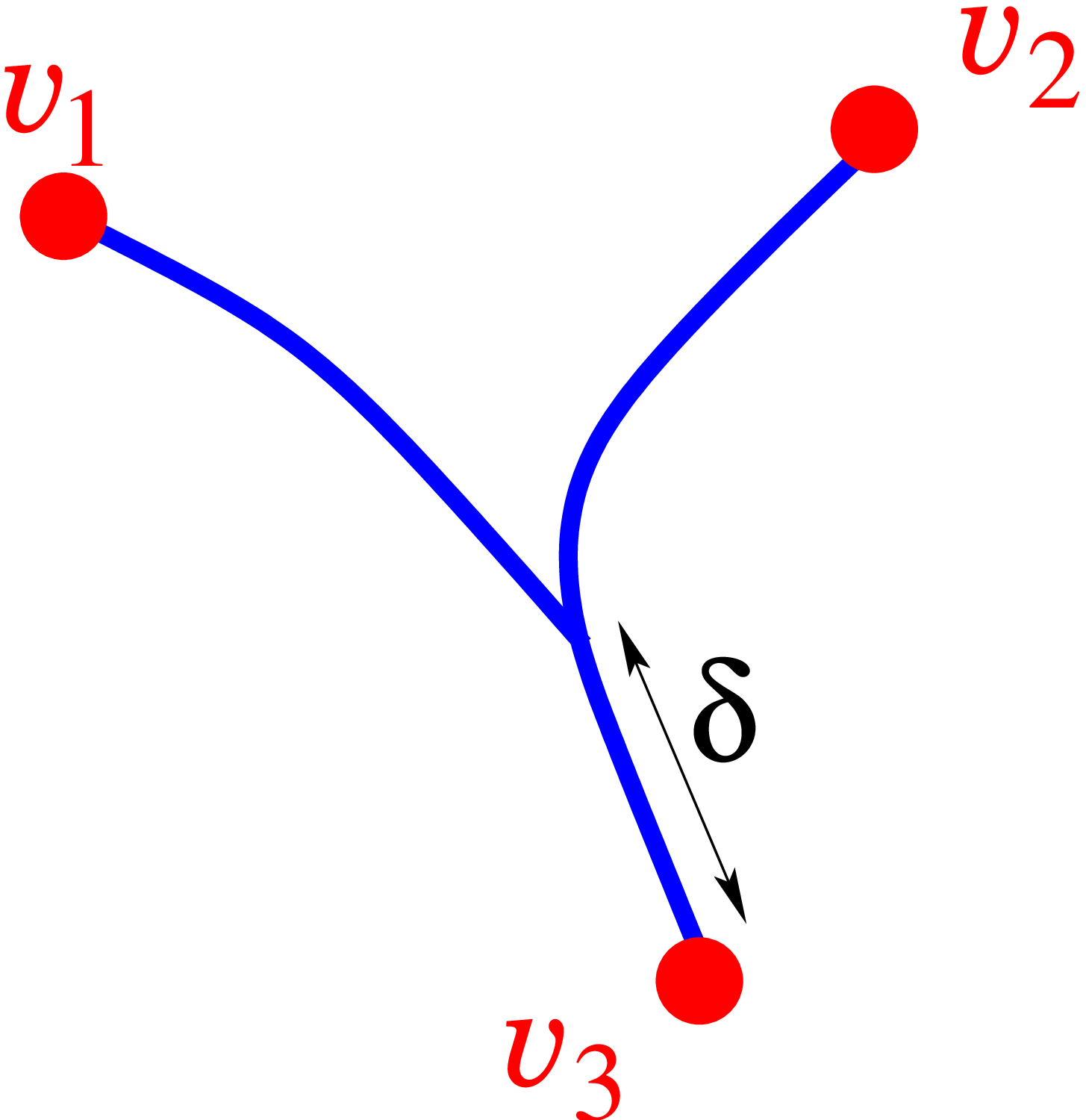}{5.cm}
\figlabel\confgeo
In this section, we explain how we can use the quantities computed
in section 2, or slight generalizations of them, 
to study the phenomenon of {\it confluence of geodesics} in the
scaling limit of large quadrangulations. It was shown by 
Le Gall \LEGALLGEOD\ and Miermont \Mier\ that two typical points in a 
large random quadrangulation are joined by a unique ``macroscopic"
geodesic path. By this, it is meant that, although there is 
a large (extensive in the length) number of geodesic paths between
two points at a discrete level, all these geodesics remain
within a distance negligible with respect to $n^{1/4}$, 
which is the scale at which points can be distinguished in the scaling
limit. Moreover, given three typical vertices $v_1$, $v_2$ and $v_3$, 
the unique macroscopic geodesic from $v_1$ to $v_3$ and the unique
macroscopic geodesic from $v_2$ to $v_3$ merge before reaching $v_3$,
i.e.\ have a macroscopic common part (see Fig.~\confgeo). This is the
phenomenon of confluence of geodesics \LEGALLGEOD\ which raises 
interesting problems,
such as that of the distribution of the length $\delta$ 
of this common part. 

\bigskip
\noindent{\it Approach via the Schaeffer bijection}
\fig{In the well-labeled tree of Fig.~\condtree, we
distinguish (a) the minimal label $1-u'$ on trees 
attached to one side of the branch from $v_1$ to $v_2$ 
and the minimal label $1-u''$ on trees attached to 
the other side of the branch, with $u=\max(u',u'')$. The quantity 
$\vert u'-u''\vert$ measures the length of the common part
of the leftmost geodesics from $v_1$ and $v_2$ to the 
added vertex $v_3$. As apparent in (b), here in
the case  $u'>u''$, these leftmost geodesics are made of two
distinct chains of successors of respective lengths $s-u''$ 
(green long-dashed arrows) and $t-u''$ (red short-dashed arrows),
followed by a common chain of successors of length $u'-u''$
(magenta solid arrows).}{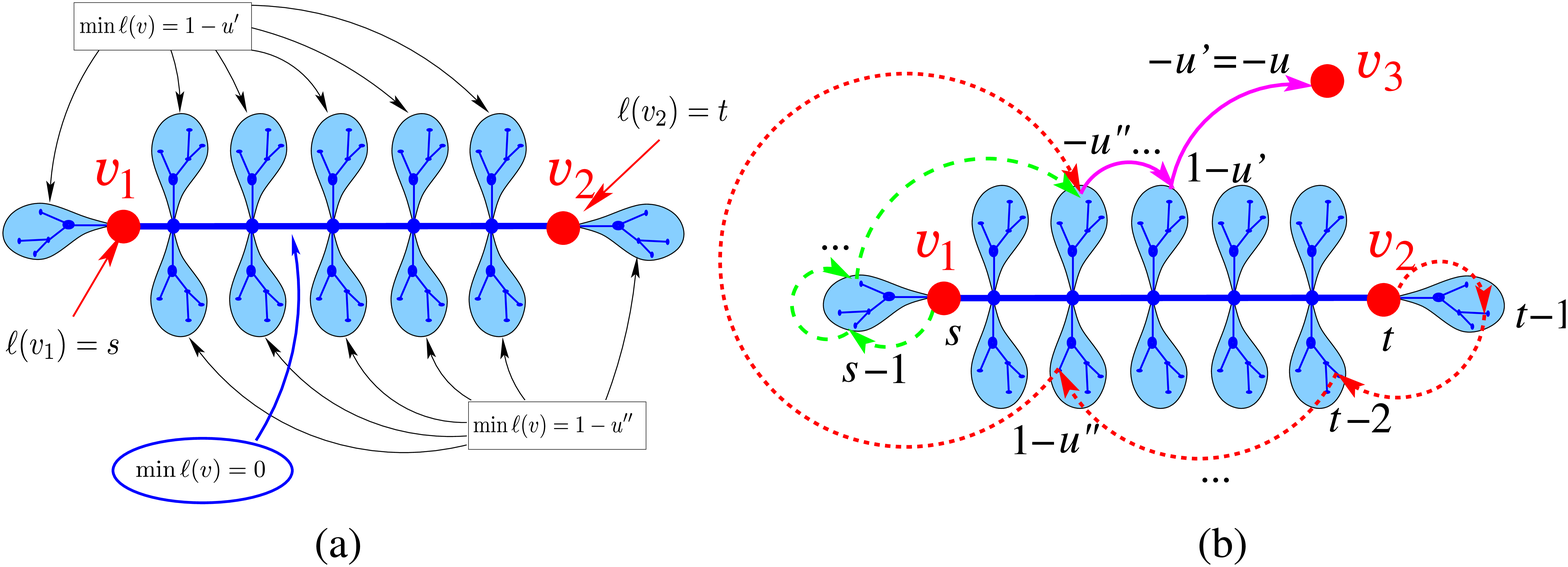}{14.cm}
\figlabel\confscha
At a discrete level, this length can be estimated by a particular
choice of geodesics defined as follows: we start again
with a triply-pointed quadrangulation with marked vertices $v_1$, 
$v_2$ and $v_3$ and consider the associated well-labeled tree obtained 
from the Schaeffer bijection, taking
$v_3$ has the origin. This tree has two marked vertices $v_1$ and $v_2$,
and upon shifting the labels so that the minimal label on the
branch between $v_1$ and $v_2$ is $0$, it is of the type displayed
in Fig.~\condtree\ for some $s$, $t$ and $u$. We can now consider the
{\it leftmost} geodesic from $v_1$ to $v_3$ formed by the chain 
of successors from the corner incident to $v_1$ and lying 
immediately on the right of the branch oriented from $v_1$ to $v_2$.
Similarly, we consider the
leftmost geodesic from $v_2$ to $v_3$ obtained as the chain 
of successors from the corner incident to $v_2$ and lying 
immediately on the left of the branch (oriented again from $v_1$ to $v_2$).
These two geodesics will merge at a point which we characterize
as follows (see Fig.~\confscha\ for an illustration): 
let us call $1-u'$ (respectively $1-u''$) the minimal label
on trees attached to the left (respectively right) side of the
branch oriented from $v_1$ to $v_2$ (with the convention that
the tree attached to $v_1$ lies on the left side of the branch,
and that attached to $v_2$ on the right side), with $u=\max (u',u'')$.
Then the two chosen geodesics have a common part of 
length $\vert u'-u''\vert$. Indeed, assuming without loss
of generality that $u=u' \geq u''$, all the $s+u$ successors of 
the corner chosen at $v_1$ lie on the left of the branch 
until $v_3$ (with label $-u$)
is reached. On the other hand, among the $t+u$ successors of the corner chosen
at $v_2$, the first $t+u''-1$ successors are found on the right of
the branch but the $(t+u'')$-th successor, having label $-u''$, is
on the left of the branch and coincides with the $(s+u'')$-th successor
of the corner chosen at $v_1$. From that point, all remaining successors 
form a common part of length $u'-u''$ (see Fig.\confscha). 
To conclude, there is a correspondence between, on the
one hand, well-labeled trees with fixed values of $s$, $t$, $u'$ and 
$u''$ as defined above and, on the other hand, triply-pointed 
quadrangulations with prescribed values $d_{13}=s+\max(u',u'')$, 
$d_{23}=t+\max(u',u'')$, $l_{123}=2 \max(u',u'')$ and such that 
the leftmost geodesics from $v_1$ to $v_3$ and from $v_2$ to $v_3$ have 
a common part of length $\vert u'-u''\vert$. Note that the 
sign of $u'-u''$ simply accounts for the relative position 
of the geodesics: when $u'> u''$ (respectively $u''>u'$), 
the geodesic from $v_1$ to $v_3$ merges on the right (respectively
on the left) of the geodesic from $v_2$ to $v_3$. 

We now wish to enumerate the above trees. By an immediate generalization
of Eq.~\Hstu, such trees have generating function: 
\eqn\Hstuu{\Delta_{u'}\Delta_{u''} H_{\rm loop}(s,t,u',u'')\ 
{\rm where}\ H_{\rm loop}(s,t,u',u'')
={\tilde X}_{s;u',u''}\, 
X_{u',u''}\, {\tilde X}_{t;u'',u'}\ .}
In the scaling limit, this generating function becomes:
\eqn\Huu{\eqalign{& \partial_{U'}\partial_{U''}
{\cal H}_{\rm loop}(S,T,U',U'';\alpha)  \ {\rm where} \cr
 &{\cal H}_{\rm loop}(S,T,U',U'';\alpha)=\cr
&3{\sinh^4(\alpha U')
\sinh^4(\alpha U'') \sinh(\alpha(2S\!+\!U'\!+\!U''))
\sinh(\alpha (2T\!+\!U'\!+\!U''))\over
\big(\sinh(\alpha (U'\!+\!U''))\sinh(\alpha(S\!+\!U'))\sinh(\alpha(S\!+\!U''))
\sinh(\alpha(T\!+\!U'))\sinh(\alpha(T\!+\!U''))\big)^2}\cr}}
and we expect that any other choice for the geodesics at the
discrete level would lead to the same continuous expression. 
This formula holds in the grand canonical formalism and can 
be transformed via an integral of the type \conttwo\ 
into the canonical normalized joint probability density for 
$D_{13}$, $D_{23}$, $L_{123}$ and the (rescaled)
length $\delta\equiv \vert U'-U''\vert$ for the common part
of the geodesics. 

\bigskip
\noindent{\it Approach via the Miermont bijection}
\fig{In the well-labeled map of Fig.~\specialdelays, we
mark the last occurrence of a label $0$ on each 
(counterclockwise oriented) cycle $c_1$ and $c_2$
and call $1-u'$ the minimal label on trees 
attached to the part of the frontier of the external face made
of: (i) the left side of the branch $b$ (oriented from $c_1$ to $c_2$), 
(ii) the external side of the cycle $c_1$ before reaching the marked label $0$, 
and (iii) the external side of the cycle $c_2$ after passing the marked 
label $0$. We also call $1-u''$ the minimal label on trees attached to 
the complementary part of the frontier, with $u=\max(u',u'')$. The quantity 
$\vert u'-u''\vert$ measures the length of the common part of
two particular geodesics leading from $v_1$ and $v_2$ to 
$v_3$, as apparent in (b), here in the case  $u'>u''$.}{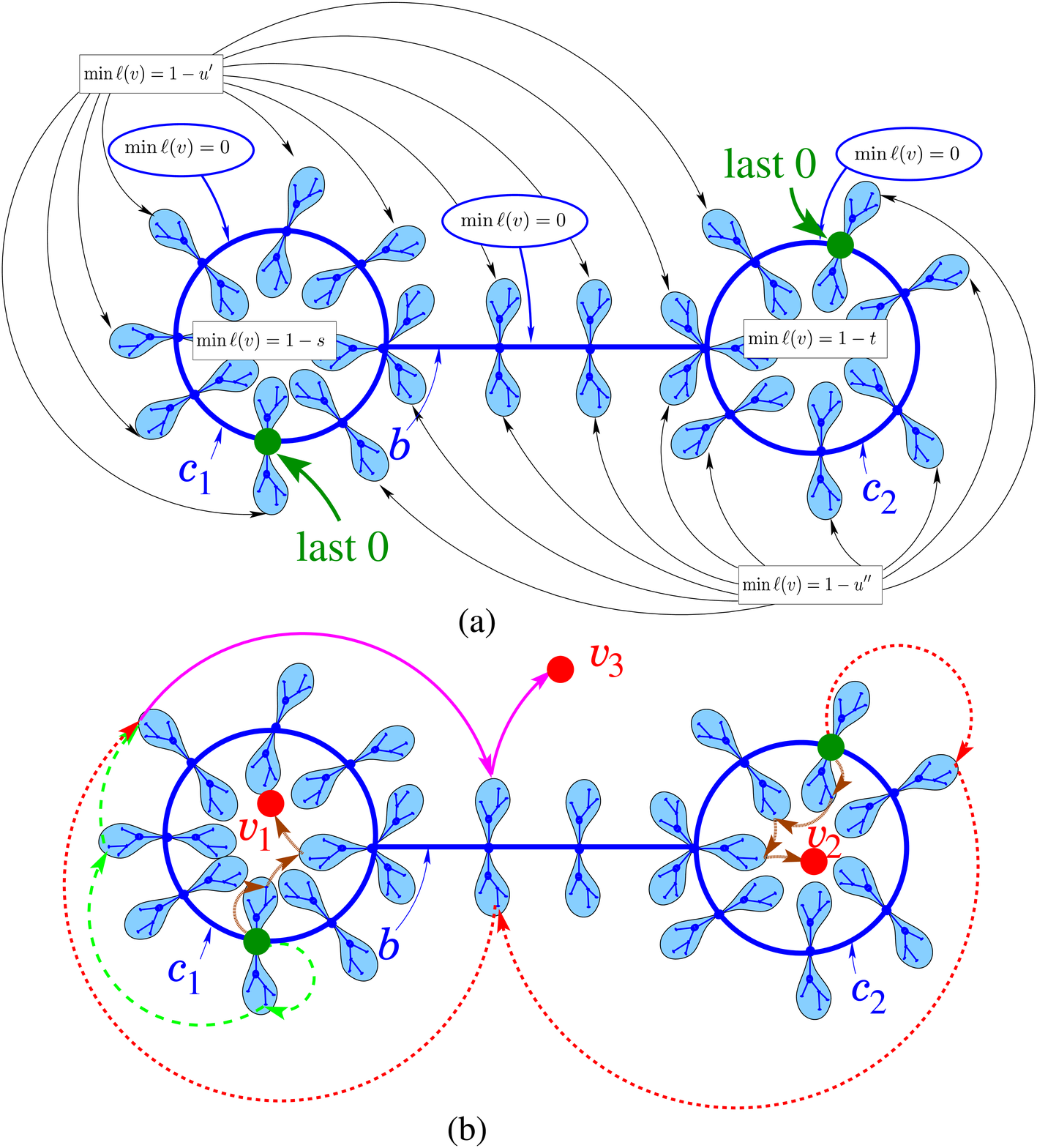}{10.cm}
\figlabel\confmier
As in section 2, a useful alternative expression for the above 
function may be obtained by use of the Miermont bijection for
triply-pointed quadrangulations, leading, for the special
choice \choicedelay\ of delays, to well-labeled maps 
of the type displayed in Fig.~\specialdelays\ (or of its degenerate
versions) for some $s$, $t$, and $u$. A particular geodesic path
from $v_1$ to $v_3$ is obtained by picking say, the last label $0$
on the (counterclockwise oriented) cycle $c_1$, 
looking at the two corners at that vertex lying
immediately on the right of the cycle when we follow the cycle
in both directions, and considering the chains of successors of 
these two corners. The concatenation of these chains forms the
desired geodesic path. A similar geodesic path can be considered
from $v_2$ to $v_3$, passing via the last label $0$ on the 
(counterclockwise oriented) cycle 
$c_2$. Let us now call $1-u'$ the minimal label on trees attached 
to the left side of the branch $b$ (oriented from $v_1$ to $v_2$), 
to the external side of the cycle $c_1$ before the last
occurrence of a label $0$ on this cycle, and to the external side of 
the cycle $c_2$ after the last occurrence of a label $0$ on this
cycle (see Fig.~\confmier\ for an illustration). 
We also call $1-u''$ the minimal label on trees attached 
to the complementary part of the frontier of the face $f_3$,
with $u=\max(u',u'')$. Then by arguments similar to the discussion
above, the two particular geodesics have a common part of length
$\vert u'-u''\vert$. We now have a correspondence between, on the
one hand, well-labeled maps with fixed values of $s$, $t$, $u'$ and 
$u''$ as defined above and, on the other hand, triply-pointed 
quadrangulations with prescribed values $d_{13}=s+\max(u',u'')$, 
$d_{23}=t+\max(u',u'')$, $l_{123}=2 \max(u',u'')$ and such that 
the two particular geodesics considered above 
from $v_1$ to $v_3$ and from $v_2$ to $v_3$ have 
a common part of length $\vert u'-u''\vert$. 
By an immediate generalization of Eq.~\floopstu, such maps
are enumerated by: 
\eqn\floopstuu{\eqalign{&
\Delta_s\Delta_t\Delta_{u'}\Delta_{u''}F_{\rm loop}(s,t,u',u'')\ 
{\rm where}\ \cr &  
F_{\rm loop}(s,t,u',u'')= 
X_{s,u'}\, Y_{s,u'',u'}\, X_{u',u''}\, Y_{t,u',u''}\, X_{t,u''}\cr}}
Note that this generating function is different from that
given by \Hstuu\ as our particular choice of geodesics differs 
in the Schaeffer and in the Miermont bijection approach. 
In the scaling limit however, we expect to recover the same expression
\Huu\ due to the unicity of geodesics at a macroscopic level. Indeed, 
the expression \floopstuu\ translates into:
\eqn\floopuu{\eqalign{&\partial_S\partial_T\partial_{U'}\partial_{U''}
{\cal F}_{\rm loop}(S,T,U',U'';\alpha)\ {\rm where} \cr
&{\cal F}_{\rm loop}(S,T,U',U'';\alpha)=\cr
&{3\over \alpha^2}
{\sinh(\alpha S)\sinh(\alpha T)\sinh^2(\alpha U')\sinh^2(\alpha U'')
\sinh(\alpha(S\!+\!U'\!+\!U''))\sinh(\alpha (T\!+\!U'\!+\!U'')) 
\over \sinh(\alpha(S\!+\!U'))\sinh(\alpha(S\!+\!U''))\sinh(\alpha(T\!+\!U')) 
\sinh(\alpha(T\!+\!U''))
\sinh^2(\alpha (U'\!+\!U''))}\cr}} 
which precisely matches the continuous expression \Huu, namely:
\eqn\conttroisuu{
\partial_S \partial_T \partial_{U'} \partial_{U''} 
{\cal F}_{\rm loop}(S,T,U',U'';\alpha) = \partial_{U'}\partial_{U''} 
{\cal H}_{\rm loop}(S,T,U',U'';\alpha)}
as a consequence of the identity
\eqn\remidentuu{\partial_S\left({1\over \alpha}
{\sinh(\alpha S)\sinh(\alpha(S\!+\!U'\!+\!U''))\over \sinh(\alpha(S\!+\!U'))
\sinh(\alpha (S\!+\!U''))}\right)
= {\sinh(\alpha U')\sinh(\alpha U'') \sinh(\alpha(2S\!+\!U'\!+\!U''))
\over \sinh^2(\alpha(S\!+\!U'))\sinh^2(\alpha(S\!+\!U''))}}

\break
\noindent{\it Marginal law for $\delta$}

It is now a simple exercise to obtain, in this scaling limit, 
the marginal law for $\delta$. We simply have to integrate over all 
positive values of $S$, $T$, $U'$ and
$U''$ with the constraint that $\vert U'-U''\vert=\delta$. This is 
done more easily in the grand canonical formalism first and
by use of the expression \floopuu, namely:
\eqn\margiuu{\eqalign{& 
\int_0^\infty\!dS\!\int_0^\infty\!dT\!
\int_0^\infty\!dU'\!\int_0^\infty\!dU''\! \delta(\vert U'-U''\vert -\delta)\ 
\partial_S\partial_T
\partial_{U'}\partial_{U''} 
{\cal F}_{\rm loop} (S,T,U',U'';\alpha)\cr  & \ \  = 
\int_0^\infty\!dU'\!\int_0^\infty\!dU''\! \delta(\vert U'-U''\vert -\delta)\ 
\partial_{U'}\partial_{U''}{\cal F}_{\rm loop}(\infty,\infty,U',U'';\alpha)
 \cr & \ \  
= \int_0^\infty\!dU'\!\int_0^\infty\!dU''\! \delta(\vert U'-U''\vert -\delta)\ 
\partial_{U'}\partial_{U''} \left({3\over \alpha^2}{
\sinh^2(\alpha U') \sinh^2(\alpha U'') \over \sinh^2(\alpha(U'+U''))}
\right) \cr & \ \  
= \int_0^\infty\!dU'\!\int_0^\infty\!dU''\! \delta(\vert U'-U''\vert -\delta)\ 
 \ 18 \, {\sinh^2(\alpha U')\sinh^2(\alpha U'') \over \sinh^4(\alpha (U'+U''))} 
\cr
& \ \ = 36 \int_\delta^\infty dU \ 
{\sinh^2(\alpha U)\sinh^2(\alpha(U-\delta)) \over 
\sinh^4(\alpha (2U-\delta))} 
\cr &\ \ = 
{3\over 2 \alpha} e^{-2\alpha \delta} \cr
}}
\fig{Plot of the probability density $\sigma(\delta)$ for the 
length $\delta$ of the common part of the two geodesics from
$v_1$ and $v_2$ to $v_3$ in the scaling limit of large
triply-pointed quadrangulations.}{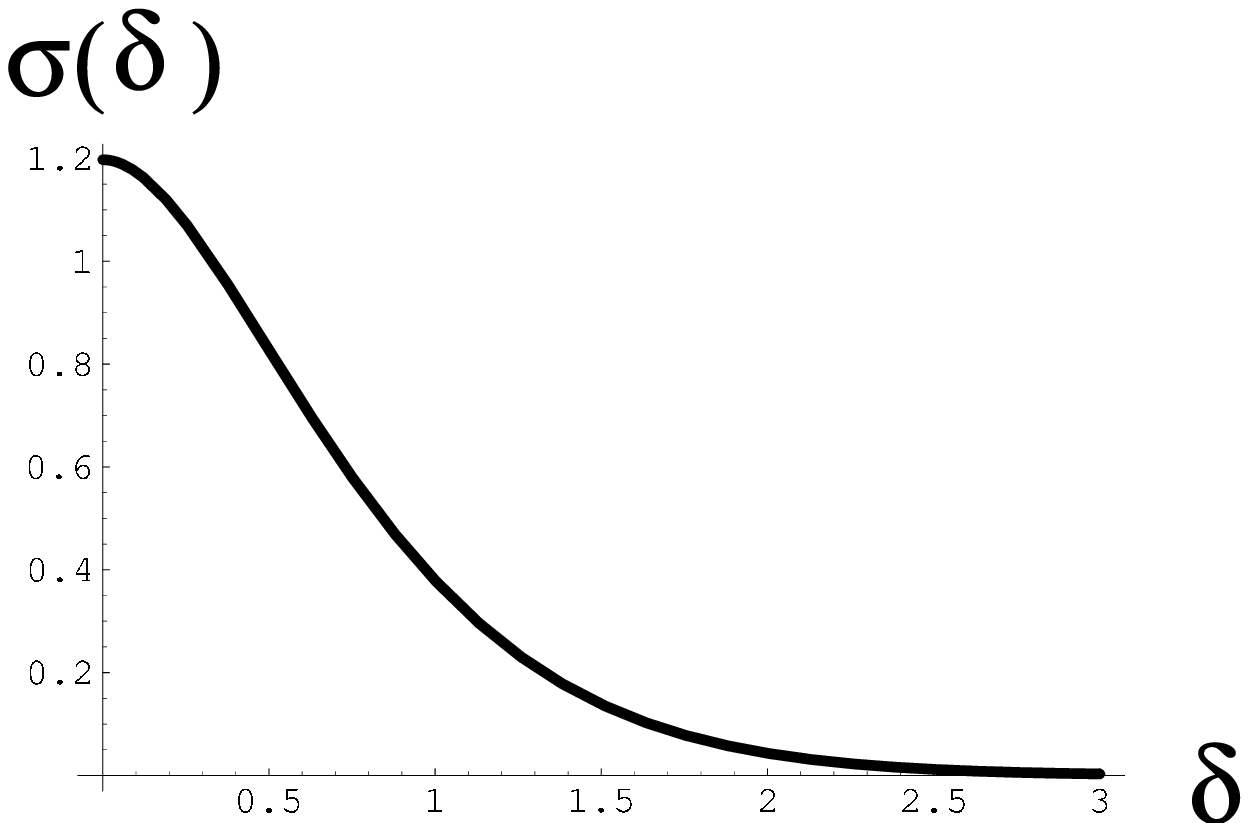}{8.cm}
\figlabel\sigmadelta
As before, we can transform this result into the probability density for 
the (rescaled) variable $\delta$ in the canonical 
ensemble of triply-pointed quadrangulations of large fixed size.
This probability density reads:
\eqn\sigmdelta{\eqalign{\sigma(\delta) & ={2\over {\rm i} \sqrt{\pi}} 
\int_{-\infty}^{\infty} 
d\xi\ \xi\, e^{-\xi^2}\, 
\left.\left({3\over 2 \alpha} e^{-2\alpha \delta}
\right)\right\vert_{\alpha=\sqrt{-3{\rm i}\xi/2}}\cr
&= \sqrt{{3\over \pi}}\left\{
\Gamma\left({3\over 4}\right) {}_0F_{2}\left(\left\{{1\over 4},{1\over 2}
\right\}, -{9 \delta^4\over 64}\right)\right. \cr
& \ \ \ \ \ \ \ \ \ \ \ \  -3 \delta^2 
\Gamma\left({5\over 4}\right) {}_0F_{2}\left(\left\{{3\over 4},{3\over 2}
\right\}, -{9 \delta^4\over 64}\right) \cr
& \left. \ \ \ \ \ \ \ \ \ \ \ \ + \sqrt{3 \pi} \delta^3  
{}_0F_{2}\left(\left\{{5\over 4},{7\over 4}
\right\}, -{9 \delta^4\over 64}\right)\right\} \cr
}}
where
\eqn\zerofdeux{{}_0F_{2}(\{b_1,b_2\},z) \equiv \sum_{k=0}^\infty 
{z^k\over k!} {1\over (b_1)_k (b_2)_k} \ \ {\rm with} \ \ (b)_k\equiv 
\prod_{i=0}^{k-1} (b+i)\ .}
This probability density is plotted in Fig.~\sigmadelta. We have in particular
\eqn\avdelta{\langle \delta \rangle ={1\over 3}
\, \langle D \rangle= 0.590494\cdots}
i.e.\ the common part represents on average one third 
of the length of a geodesic.

\subsec{Confluence of minimal separating loops}
\fig{A schematic picture of the phenomenon of confluence of 
minimal separating loops. For generic points $v_1$, $v_2$ and $v_3$ and
in the scaling limit of large quadrangulations, the
minimal loop originating from $v_3$ and separating $v_1$ from $v_2$ 
(represented as a thick blue line) is unique and
is made of a common part of macroscopic length $\delta_{\rm loop}$
and an open part of macroscopic length $L'_{123}$.}{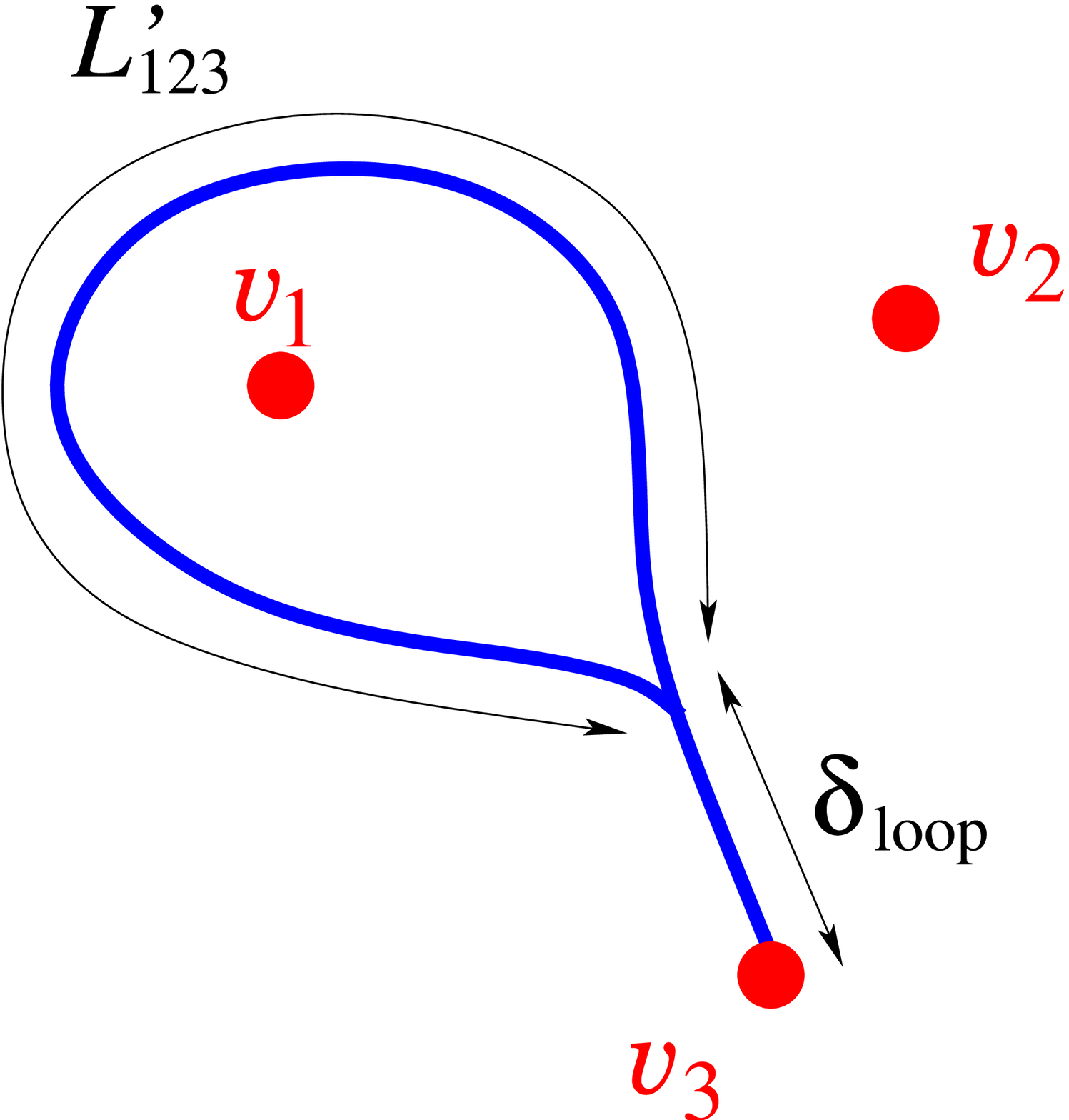}{5.cm}
\figlabel\confsep
The minimal separating loops themselves also exhibit a 
phenomenon of confluence. Indeed, 
a minimal separating loop is made of two geodesics of same length 
emanating from a particular vertex $v$ (with minimal label
on the branch from $v_1$ to $v_2$) and reaching $v_3$. 
In the scaling limit, we expect that the macroscopic
minimal separating loop is unique and moreover, its two constituent
geodesics have a common part of macroscopic length $\delta_{\rm loop}$
(see Fig.~\confsep\ for an illustration).  
Note that, although $v_1$, $v_2$ and $v_3$ are generic points, 
$v$ is a non-typical point as it can be
connected to $v_3$ by two distinct macroscopic geodesics which 
are not confluent at $v$. 
We shall call the 
complementary part the {\it open part} of the loop, with length
$L'_{123}=L_{123}-2\, \delta_{\rm loop}$.
\fig{In the well-labeled tree of Fig.~\condtree, we
mark the first label $0$ on the branch from $v_1$ to $v_2$ and
call $1-u'$ the minimal label on trees attached to  
the part of the branch lying from $v_1$ to the marked label $0$. 
We also call $1-u''$ the minimal label on trees attached to the
complementary part of the branch, with $u=\max(u',u'')$. The quantity 
$\vert u'-u''\vert$ measures the length of the common part
of a particular minimal loop originating from
$v_3$ and separating $v_1$ from $v_2$, as apparent in (b), 
here in the case  $u'>u''$. The length of the open part of
the minimal separating loop is $2\min(u',u'')=2u''$.}{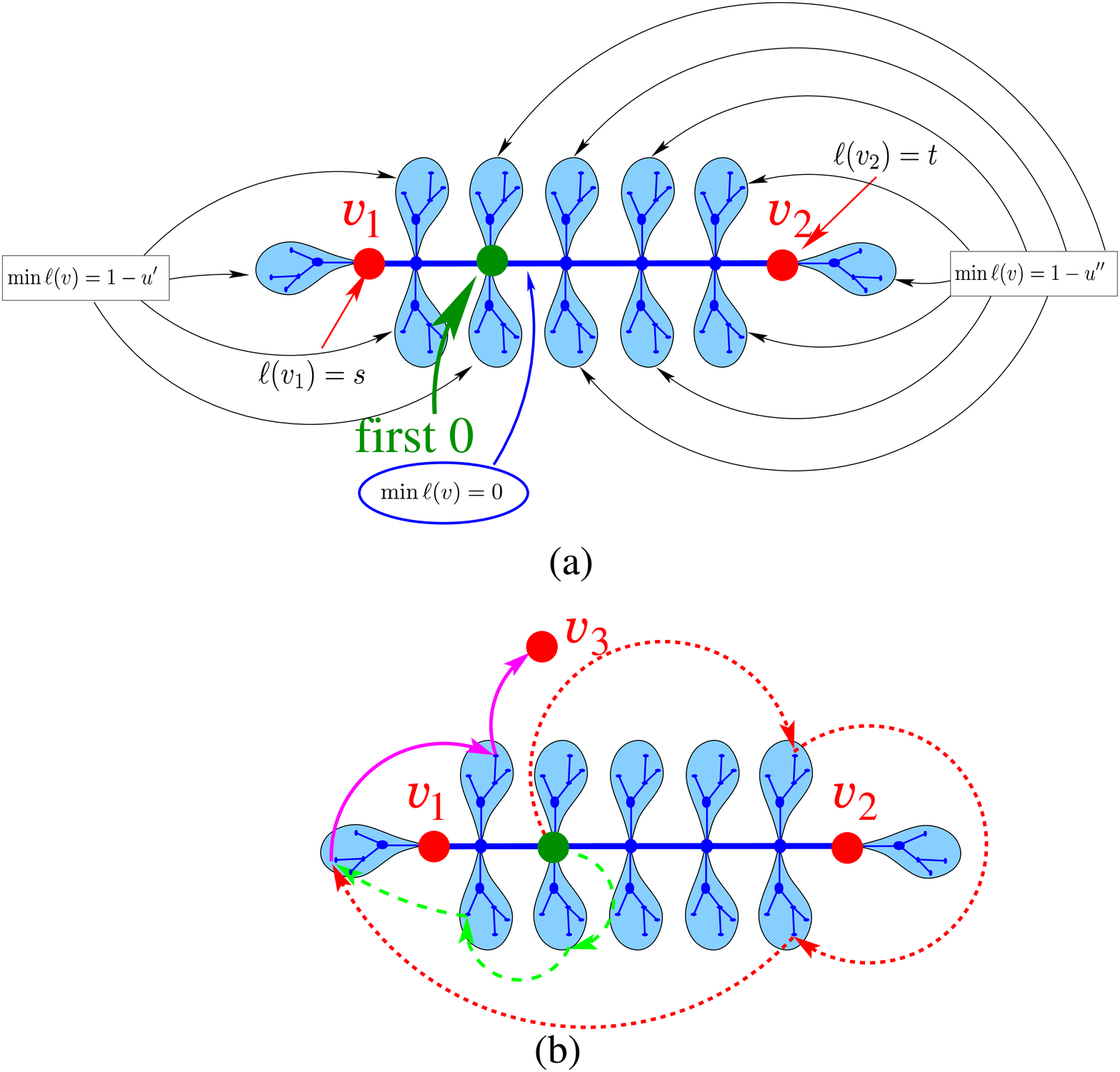}{10.cm}
\figlabel\confuscha
The statistics for $\delta_{\rm loop}$ and $L'_{123}$ can be computed along
the same lines as in section 3.1. In the Schaeffer approach,
on the branch from $v_1$ to $v_2$ in the well-labeled tree, we now 
consider the vertex $v$  with minimal label closest to $v_1$. 
Calling $1-u'$ the minimal label for trees attached the 
the part of the branch from $v_1$ to $v$, and $1-u''$ the
minimal label for trees attached to the complementary part, 
the quantity $\vert u'-u''\vert$
measures the length of the desired common part for a particular
minimal loop formed by two chains of successors starting from $v$
(see Fig.~\confuscha\ for an illustration). As for the length
of the open part of the loop, it is simply measured by $2\min(u',u'')$.
The generating function for the objects above is immediately given
by 
\eqn\Hbarstuu{\Delta_{u'}\Delta_{u''} {\bar H}_{\rm loop}(s,t,u',u'')\ 
{\rm where}\ {\bar H}_{\rm loop}(s,t,u',u'')
={\tilde X}_{s;u',u'}\, 
X_{u'',u''}\, {\tilde X}_{t;u'',u''}\ .}
In the scaling limit, this generating function becomes:
\eqn\Hbaruu{\eqalign{& \partial_{U'}\partial_{U''}
{\bar{ \cal H}}_{\rm loop}(S,T,U',U'';\alpha)  \ {\rm where} \cr
&{\bar {\cal H}}_{\rm loop}(S,T,U',U'';\alpha)=
3{\sinh^4(\alpha U')
\sinh^4(\alpha U'') \sinh(\alpha(2(S\!+\!U'))
\sinh(\alpha (2(T\!+\!U''))\over
\sinh(2 \alpha U') \sinh(2 \alpha U'') \sinh^4(\alpha(S\!+\!U'))
\sinh^4(\alpha(T\!+\!U''))}\cr}}
which yields the joint law for $L'_{123}=2\min(U',U'')$, 
$D_{13}=S+\max(U',U'')$, $D_{23}=T+\max(U',U'')$ and 
$\delta_{\rm loop}=\vert U'-U''\vert$.
Note that the sign of $U'-U''$ indicates which domain delimited 
by the open part contains the common part of the loop. 
As apparent in Fig.~\confuscha, 
this common part lies in the domain containing $v_1$ when $U'>U''$.
\fig{In the well-labeled map of Fig.~\specialdelays, we
mark the first occurrence of a label $0$ on the branch $b$ oriented
from the cycle $c_1$ to the cycle $c_2$ and call $1-u'$ the minimal label 
on trees attached to the part of the frontier of the external face made
of: (i) the part of the branch $b$ lying between $c_1$ and the
marked label $0$ and (ii) the external side of the cycle $c_1$. 
We call $1-u''$ the minimal label on trees attached to 
the complementary part of the frontier, with $u=\max(u',u'')$. The quantity 
$\vert u'-u''\vert$ measures the length of the common part
of a particular minimal loop  
originating for $v_3$ and separating $v_1$ from $v_2$, as apparent in (b), 
here in the case  $u'>u''$ (corresponding to having the common part
in the domain containing $v_1$). The length of the open part
is $2\min(u',u'')$.}{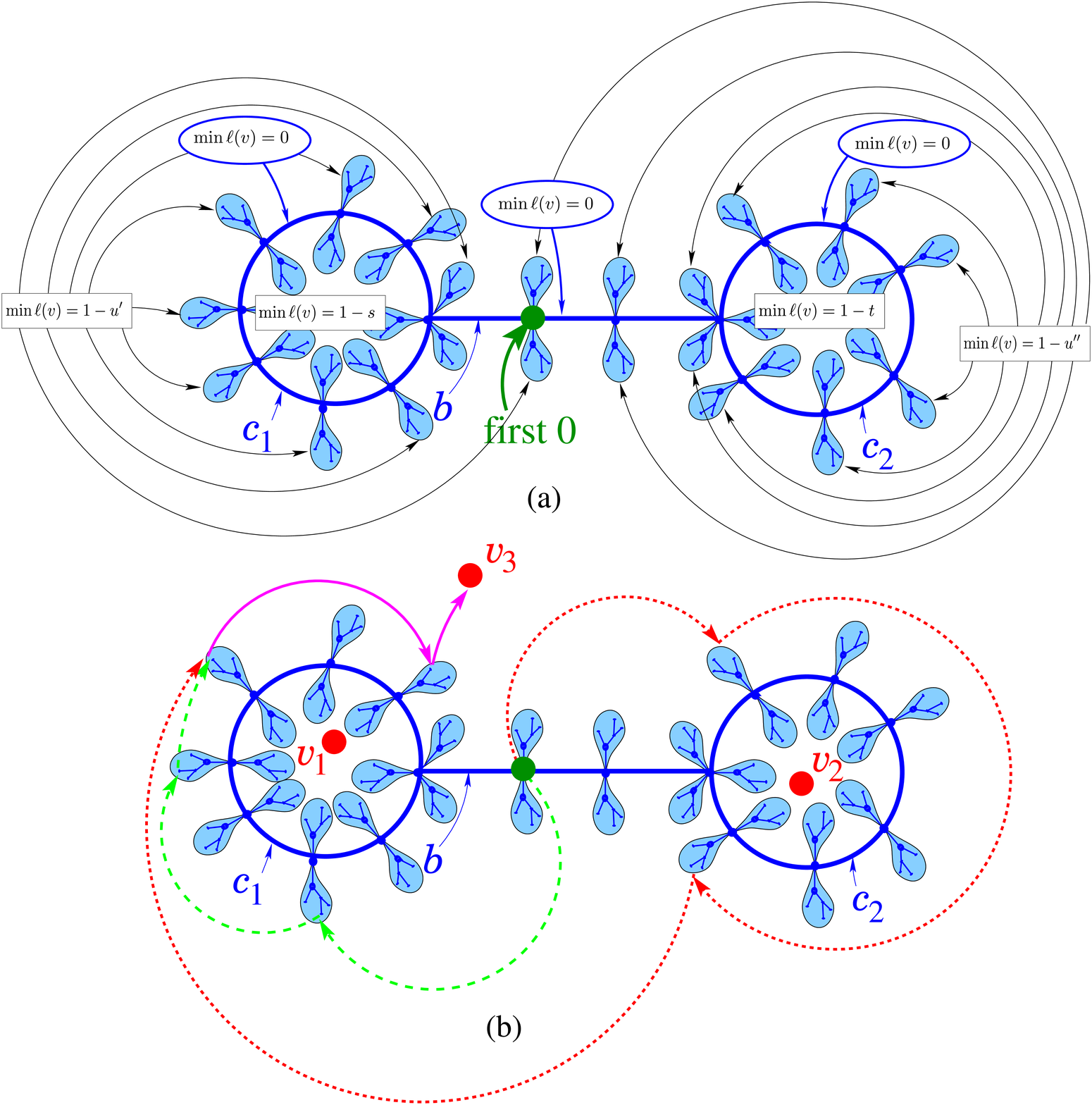}{12.cm}
\figlabel\confumier
An alternative expression is found through the Miermont approach
where we consider well-labeled maps of the type displayed 
in Fig.~\confumier\ using a particular minimal separating loop
passing through the vertex with minimal label on the branch $b$
closest to the cycle $c_1$. 
We find a generating function
\eqn\fbarloopstuu{\eqalign{&
\Delta_s\Delta_t\Delta_{u'}\Delta_{u''}{\bar F}_{\rm loop}(s,t,u',u'')\ 
{\rm where}\ \cr &  
{\bar F}_{\rm loop}(s,t,u',u'')= 
X_{s,u'}\, Y_{s,u',u'}\, X_{u'',u''}\, Y_{t,u'',u''}\, X_{t,u''}\cr}}
whose scaling limit 
\eqn\fbarloopuu{\eqalign{&\partial_S\partial_T\partial_{U'}\partial_{U''}
{\bar {\cal F}}_{\rm loop}(S,T,U',U'';\alpha)\ {\rm where} \cr
&{\bar {\cal F}}_{\rm loop}(S,T,U',U'';\alpha)=\cr
&{3\over \alpha^2}
{\sinh(\alpha S)\sinh(\alpha T)\sinh^2(\alpha U')\sinh^2(\alpha U'')
\sinh(\alpha(S\!+2 U'))\sinh(\alpha (T\!+2 U'')) 
\over \sinh^2(\alpha(S\!+\!U'))\sinh^2(\alpha(T\!+\!U'')) 
\sinh(2\alpha U')\sinh(2\alpha U'')}\cr}} 
matches the expression \Hbaruu\ above. This matching is again
a direct consequence of the identity \remidentcont.

\fig{Plot of the joint probability density $\tau_{\rm loop}(\delta_{\rm loop},
L'_{123})$ for the length $\delta_{\rm loop}$ of the common part and the 
length $L'_{123}$ of
the open part of the minimal loop originating from $v_3$ and separating 
$v_1$ from $v_2$ in the scaling limit of large
triply-pointed quadrangulations.}{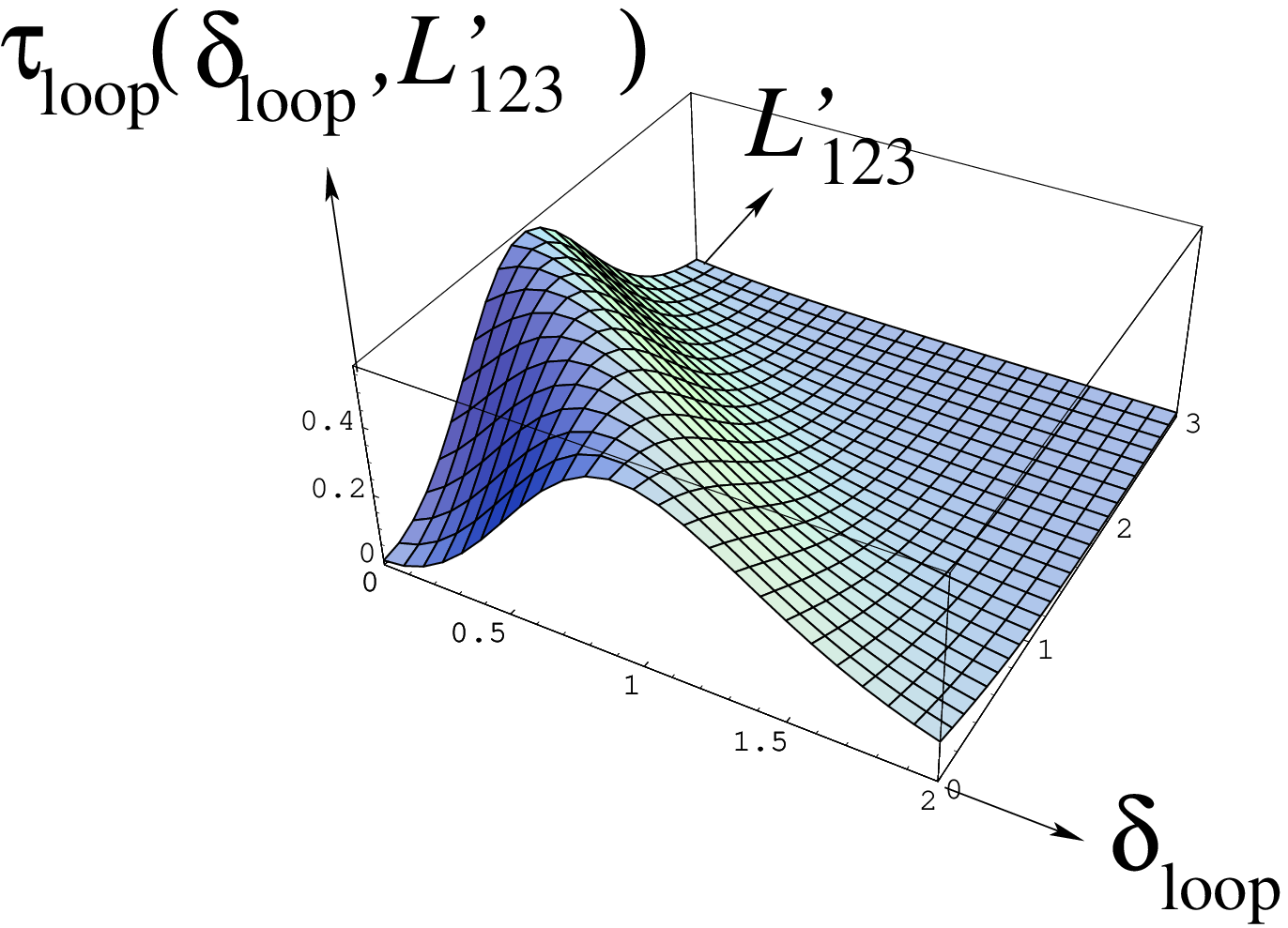}{8.cm}
\figlabel\taudeltaloopLprime
We can integrate over $S$ and $T$ and obtain the marginal law
for $U'$ and $U''$
\eqn\margiuu{{3\over \alpha^2} \partial_{U'}\partial_{U''}
{\sinh^2(\alpha U')\sinh^2(\alpha U'') \over \sinh(2\alpha U')
\sinh(2\alpha U'')}= {3\over 4} {1\over \cosh^2(\alpha U')
\cosh^2(\alpha U'')}\ .}
This can be translated into the marginal joint law
for $\delta_{\rm loop}$ and $L'_{123}$.
For triply-pointed quadrangulations of fixed large size $n$,
we find the joint probability density:
\eqn\lawdeltaL{\tau_{\rm loop}(\delta_{\rm loop}, L'_{123})
= {2\over {\rm i} \sqrt{\pi}} 
\int_{-\infty}^{\infty} 
\!\!\!\!\!d\xi\ \xi\, e^{-\xi^2}\, \left. {3/4\over \cosh^2\left(
\alpha\!\left({L'_{123}\over 2}\!
+\!\delta_{\rm loop}\right)\right)
\cosh^2\left(\alpha\!{L'_{123}\over 2}\right)} \right\vert_{\alpha=\sqrt{-3{\rm i}\xi/2}}\ .}
This probability density is plotted in Fig.~\taudeltaloopLprime.
\fig{Plot of the marginal probability density 
$\tau_{\rm loop}(\delta_{\rm loop})$ for the length of the common part
of the minimal loop originating from $v_3$ and separating 
$v_1$ from $v_2$ in the scaling limit of large
triply-pointed quadrangulations.}{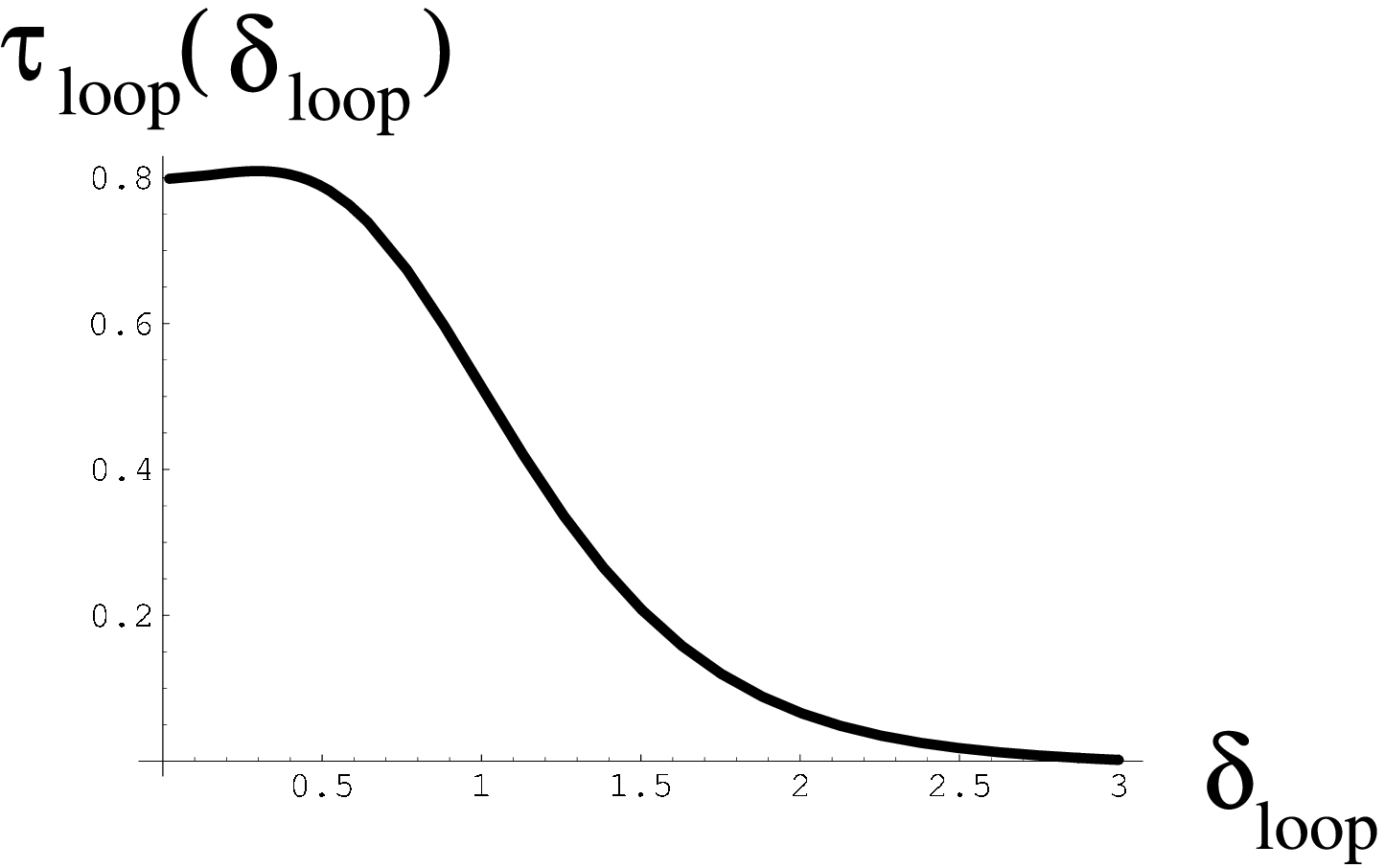}{8.cm}
\figlabel\taudeltaloop
Upon integrating $\tau_{\rm loop}(\delta_{\rm loop},L'_{123})$ over $L'_{123}$,
we get the marginal density distribution for the length $\delta_{\rm loop}$
only, namely:
\eqn\lawdelta{\eqalign{\tau_{\rm loop}(\delta_{\rm loop})
& = {2\over {\rm i} \sqrt{\pi}} 
\int_{-\infty}^{\infty} \!\!\!\!\!d\xi\ \xi\, e^{-\xi^2}
{3\over 2 \alpha} {1\over \sinh^3(\alpha 
\delta_{\rm loop}) \cosh(\alpha \delta_{\rm loop})} \cr 
&\times\! 
\Big\{2\cosh^2(\alpha \delta_{\rm loop})\left(\alpha \delta_{\rm loop}
\!-\!\log(\cosh(\alpha \delta_{\rm loop}))\right)\cr &\left. \ \ -\!
\sinh(\alpha \delta_{\rm loop})\left(\cosh(\alpha \delta_{\rm loop})
\!+\!e^{-\alpha \delta_{\rm loop}}\right)
\Big\}\right\vert_{\alpha=\sqrt{-3{\rm i}\xi/2}}\cr}}
This probability density is plotted in Fig.~\taudeltaloop. We have in
particular 
\eqn\avdeltaloop{\langle \delta_{\rm loop} \rangle =
{2\over 3}\ (2-\log 4)\ \langle D \rangle =0.724779\cdots}
in terms of the average length $\langle D \rangle$ of a geodesic path.
\fig{Plot of the marginal probability density 
${\bar \tau}_{\rm loop}(L'_{123})$ for the length of the open part 
of the minimal loop originating from $v_3$ and separating 
$v_1$ from $v_2$ in the scaling limit of large
triply-pointed quadrangulations.}{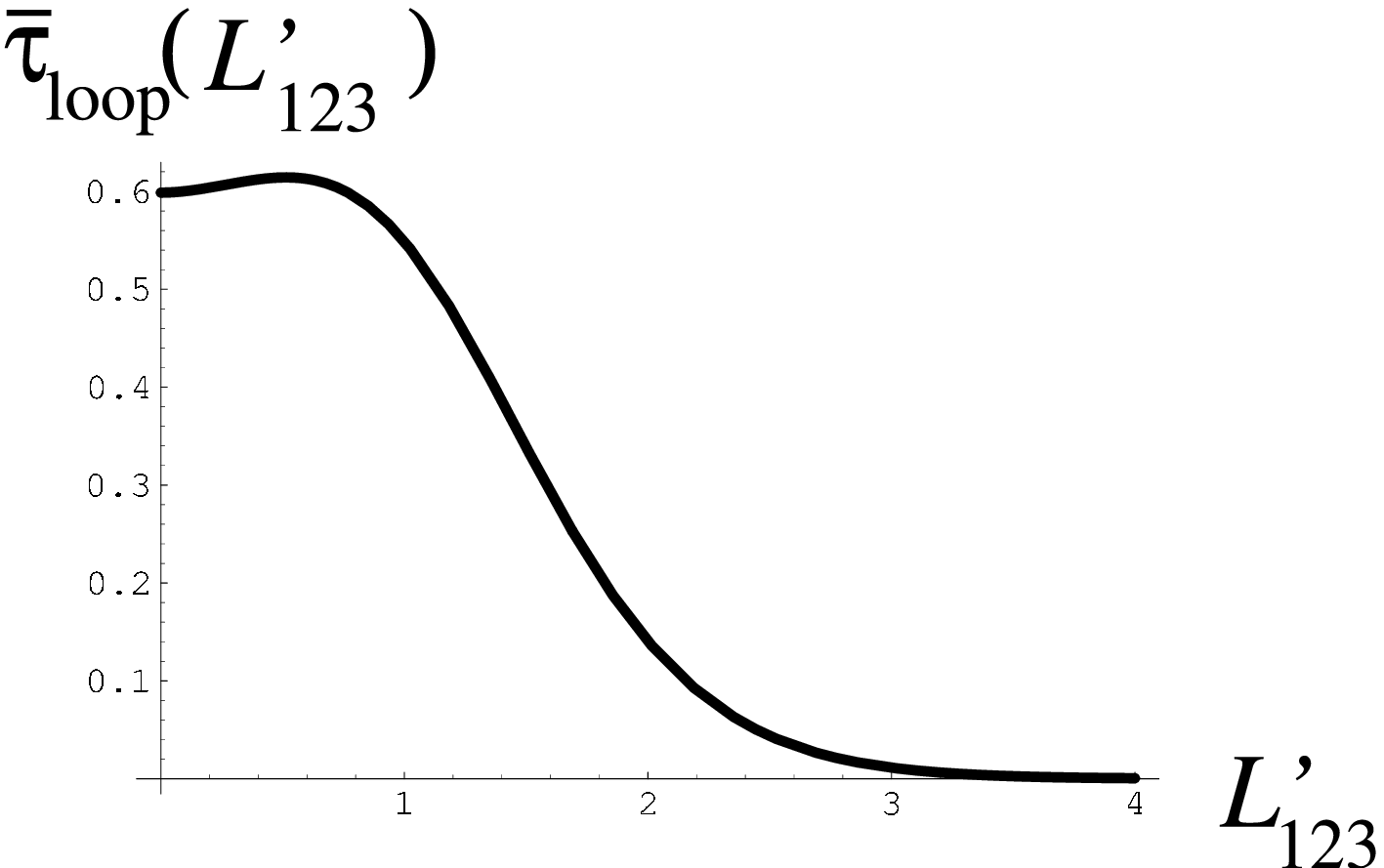}{8.cm}
\figlabel\tauLprime
On the other hand, upon integrating 
$\tau_{\rm loop}(\delta_{\rm loop},L'_{123})$ over $\delta_{\rm loop}$,
we get the marginal density distribution for the length $L'_{123}$
only, namely:
\eqn\lawdelta{{{\bar \tau}_{\rm loop}(L'_{123})
= {2\over {\rm i} \sqrt{\pi}} 
\int_{-\infty}^{\infty} \!\!\!\!\!d\xi\ \xi\, e^{-\xi^2}
\left.{3\over 4 \alpha} {1\over \cosh^3\left(\alpha 
{L'_{123}\over 2}\right) }
e^{-\alpha {L'_{123}\over 2}}\right\vert_{\alpha=\sqrt{-3{\rm i}\xi/2}}\ .}}
This probability density is plotted in Fig.~\tauLprime. We have in
particular 
\eqn\avLprime{\langle L'_{123} \rangle =
{4\over 3}\ (\log 4 - 1)\ \langle D \rangle = 0.912418\cdots \ .}
Note that $2 \langle \delta_{\rm loop} \rangle + \langle L'_{123} \rangle =
{4\over 3} \langle D \rangle $, in agreement with Eq.~\avl.

\subsec{Area enclosed by a minimal separating loop}

Within the above framework, we may easily address the question 
of the partitioning of the area of triply-pointed quadrangulations 
over the two domains separated by a minimal separating loop.
For a given separating loop, we may indeed decide to 
attach a weight per face of the quadrangulation depending on which 
domain it lies in. In the equivalent Miermont picture and for
the particular minimal separating loop considered in 
Fig.~\confumier, this amounts to assigning a weight, say $g_1$ 
(respectively $g_2$) to edges lying in the domain containing $v_1$ 
(respectively $v_2$), which results in the generating function:
\eqn\fbarloopgg{\eqalign{&
\Delta_s\Delta_t\Delta_{u'}\Delta_{u''}{\bar F}_{\rm loop}(s,t,u',u'';g_1,g_2)\ 
{\rm where}\ \cr &  
{\bar F}_{\rm loop}(s,t,u',u'';g_1,g_2)= 
X_{s,u'}(g_1)\, Y_{s,u',u'}(g_1)\, X_{u'',u''}(g_2)\, 
Y_{t,u'',u''}(g_2)\, X_{t,u''}(g_2)\ .\cr}}
Here $X_{s,t}(g_m)$ and $Y_{s,t,u}(g_m)$ denote the generating
functions $X_{s,t}$ and $Y_{s,t,u}$, as given by \xst\ and \ystu, with
$g$ replaced by $g_m$ in \Rxexplicit, for $m=1,2$. 
In the continuum limit, we set 
\eqn\scala{
g_1={1 \over 12}(1-\Lambda_1 \epsilon)
\ \ , \ \
g_2={1 \over 12}(1-\Lambda_2 \epsilon)}
which amounts to having a different cosmological constant
in both domains. The generating function above translates into 
the scaling function
\eqn\fbarloopcont{
\eqalign{& \partial_S\partial_T\partial_{U'}\partial_{U''} 
{\bar {\cal F}}_{\rm loop}(S,T,U',U'';\alpha_1,\alpha_2) 
\ {\rm where}\cr &
{\bar {\cal F}}_{\rm loop}(S,T,U',U'';\alpha_1,\alpha_2)= 
3^3 {\cal Y}(S,U',U';\alpha_1) {\cal Y}(T,U'',U'';\alpha_2)\cr &
{\cal Y}(S,T,U;\alpha)\equiv {1\over 3 \alpha} 
{\sinh(\alpha S) \sinh(\alpha T) \sinh(\alpha U) \sinh(\alpha(S+T+U)) 
\over \sinh(\alpha(S+T)) \sinh(\alpha(T+U)) \sinh(\alpha(U+S))}\cr}}
and $\alpha_m=\sqrt{3/2}\Lambda_m^{1/4}$. Here, 
${\cal Y}$ is the scaling limit of $Y$, while each $X$ tends to $3$
in the scaling limit, irrespectively of its arguments. 

Upon integrating over all possible values of $S$, $T$, $U'$ and $U''$,
we get a function 
\eqn\limigf{{\bar{\cal F}}_{\rm loop}(\infty,\infty,\infty,\infty;\alpha_1,
\alpha_2)={3\over 4 \alpha_1 \alpha_2}\ .}
Returning to the canonical formalism where we fix the sizes of 
the two domains separated by the minimal loop to be respectively 
$n_1$ and $n_2$ (with $n_1+n_2=n$, $n_1 \gg 1$, $n_2 \gg 1$), 
we set $\epsilon=1/n$ and $\Lambda_m=-\xi_m^2$, so that the expression 
\limigf\ tends to $n^{1/2}\ {\rm i} /(2 \sqrt{\xi_1\xi_2})$.
Setting $n_1=\eta n$ and $n_2=(1-\eta) n$, we obtain the probability 
density for $\eta$ as 
\eqn\propeta{\eqalign{\varrho(\eta)& = {2\over {\rm i} \pi^{3/2}}
\int_{-\infty}^{\infty} d\xi_1\ \xi_1\ e^{- \eta \xi_1^2}
\int_{-\infty}^{\infty} d\xi_2\ \xi_2\ e^{- (1-\eta) \xi_2^2}
{1\over 2 \sqrt{\xi_1\xi_2}}\cr &= 
{\sqrt{\pi}\over \Gamma\left({1\over 4}\right)} {1\over \eta^{3/4} 
(1-\eta)^{3/4}}\ .\cr}}
The partitioning $\eta$ of the mass is therefore governed by a simple 
Beta distribution with 
parameters $\{1/4,1/4\}$. In particular the two domains are most
likely of very asymmetric sizes, with a probability density
maximal for $\eta=0$ or $1$. 

More precisely, we can naturally distinguish the two domains 
as exactly one of them contains the common part of the minimal
separating loop. As mentioned above, this information is encoded in
the sign of $U'-U''$. We may integrate \fbarloopcont\ over $S$, $T$, 
$U'$ and $U''$ in the domain $U'> U''$, corresponding
to the case where the common part lies in the domain containing $v_1$. 
This leads to 
\eqn\intuu{\int_0^\infty dU' {3\over 4 \alpha_1\alpha_2}
\left[\partial_{U'} \tanh(\alpha_1 U')\right] \tanh(\alpha_2 U')\ ,}
which, together with the symmetric contribution from the
domain $U''>U'$ (obtained by exchanging $\alpha_1$ and $\alpha_2$),
adds up to \limigf. We can in principle deduce from \intuu\ the
(now asymmetric) law for $\eta$ conditionally on the position of 
the common part. We have not found a compact simple form for this law
but its first few moments can be computed. We find an average 
value 
\eqn\etaasy{\langle \eta \rangle_{U'>U''}=\langle 1-\eta \rangle_{U''>U'}
={1\over 3}(1+\log 4) \sim 79.543 \% }
for the proportion of the total area lying in the same domain as
the common part of the minimal separating loop. 

\newsec{The three-point function revisited}
\fig{A schematic picture of the phenomenon of confluence of 
geodesics for the three geodesics linking three generic points 
$v_1$, $v_2$ and $v_3$ in the scaling limit of large quadrangulations.
These geodesics (represented as thick blue lines) 
have common parts of macroscopic lengths $\delta_1$, $\delta_2$
and $\delta_3$. The remaining open part of the triangle has sides of
macroscopic lengths $D'_{12}$, $D'_{23}$ and $D'_{31}$,
and separates the quadrangulation into two domains. 
The three common parts may lie in the same domain (as represented here)
or in different domains, giving rise to eight possibilities
for the relative position of the three 
geodesics (see Fig.\ 33 below).}{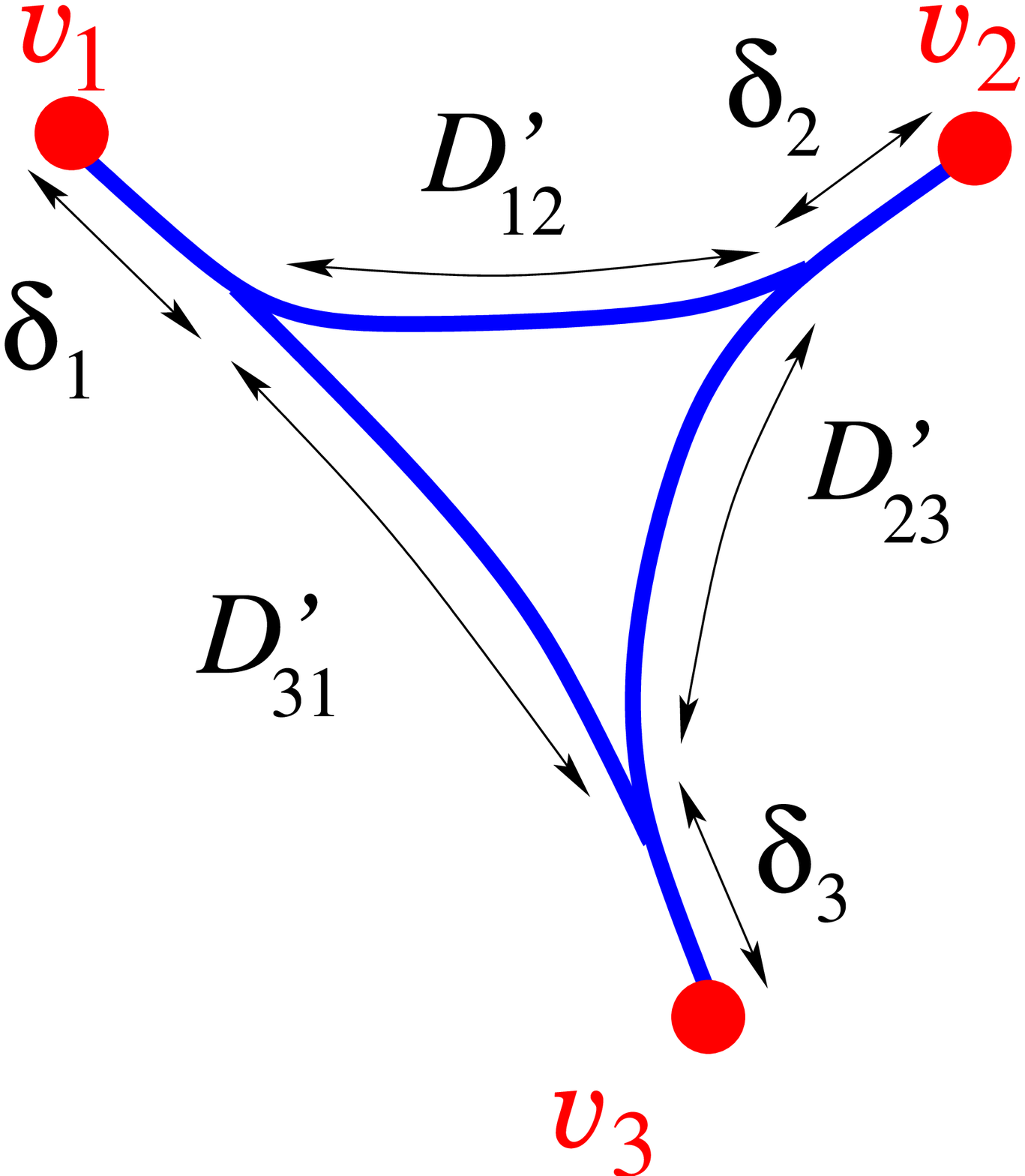}{5.cm}
\figlabel\confall
The three-point function of planar quadrangulations enumerates 
quadrangulations of the sphere with three marked vertices
$v_1$, $v_2$ and $v_3$ {\it at prescribed pairwise distances} $d_{12}$,
$d_{23}$ and $d_{31}$. It was computed in Ref.~\THREEPOINT\ and, in the scaling
limit of quadrangulations of fixed large size $n \to \infty$,  
translates into a universal joint probability $\rho(D_{12},D_{23},D_{31})$ 
for the three rescaled lengths $D_{12}=d_{12}/n^{1/4}$,
$D_{23}=d_{23}/n^{1/4}$ and $D_{31}=d_{31}/n^{1/4}$ of the
three geodesics forming the triangle $(v_1,v_2,v_3)$. As mentioned
in the introduction, a full description of the geometry of this
triangle must incorporate the phenomenon of confluence. We call 
the lengths of the common parts respectively $\delta_1$ 
(for the two geodesics leading to $v_1$), 
$\delta_2$ (for the two geodesics leading to $v_2$) and
$\delta_3$ (for the two geodesics leading to $v_3$). The remaining 
proper parts of the geodesics form 
an open triangle with sides of respective lengths
$D'_{12}=D_{12}-\delta_1-\delta_2$,
$D'_{23}=D_{23}-\delta_2-\delta_3$ and 
$D'_{31}=D_{31}-\delta_3-\delta_1$ (see Fig.~\confall\ for an
illustration). A natural question is that of determining the 
corresponding joint probability density
$\rho(D'_{12},D'_{23},D'_{31},\delta_1,\delta_2,\delta_3)$.
\fig{Structure of the well-labeled maps with three faces coding triply-pointed
quadrangulations with prescribed values of the pairwise distances between
the marked vertices: $d_{12}=s+t$, $d_{23}=t+u$ and $d_{31}=u+s$. These maps 
are easily enumerated by cutting them at the first and last label $0$ 
on each frontier (big green dots).}{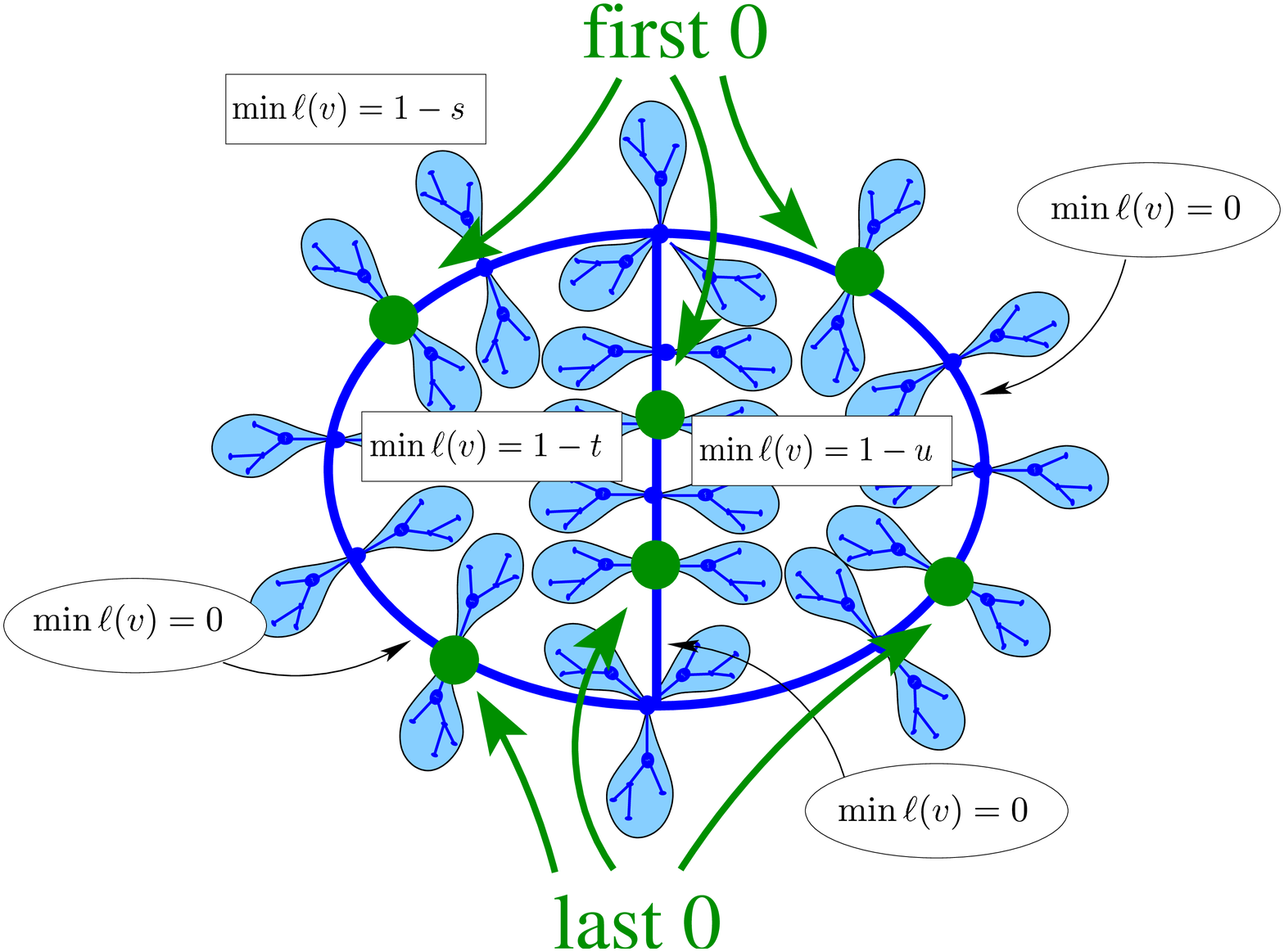}{10.cm}
\figlabel\specdel
As explained in Ref.~\THREEPOINT, triply-pointed quadrangulations
with prescribed values of the pairwise distances $d_{12}$, $d_{23}$
and $d_{31}$ are in one-to-one correspondence with particular 
well-labeled maps with three faces. This is again a consequence
of the Miermont bijection with three sources $v_1$, $v_2$ and $v_3$, 
and with a particular choice of delays, now given by
\eqn\specdelbis{\eqalign{
\tau_1 & = -s \equiv {d_{23}-d_{31}-d_{12} \over 2}
\ , \cr
\tau_2 & = -t \equiv {d_{31}-d_{12}-d_{23} \over 2}
\ , \cr
\tau_3 & = -u \equiv {d_{12}-d_{23}-d_{31} \over 2}
\ , \cr
}}
where we use the parametrization of the pairwise distances
\eqn\parad{\eqalign{
d_{12} & = s+t \ , \cr
d_{23} & = t+u \ , \cr
d_{31} & = u+s \ . \cr}}
As shown in Ref.~\THREEPOINT, the maps obtained for this choice of delays
are now of the type displayed in Fig.~\specdel, or degenerate versions
of this generic form when one of the frontiers between faces or one 
of the faces reduces to a single vertex. By an simple decomposition of
the map in five pieces obtained by cutting the map at the
first and last occurrence of a label $0$ on each frontier, 
we immediately get the generating function for these maps:
\eqn\fstu{\eqalign{&
\Delta_s\Delta_t\Delta_u F(s,t,u)\ 
{\rm where}\ \cr &  
F(s,t,u)= 
X_{s,t}\, X_{t,u}\, X_{u,s}\, (Y_{s,t,u})^2\ .\cr}}
\fig{In the map of Fig.~\specdel, we mark the first label $0$ on each 
frontier. We then call $1-s'$, $1-s''$, $1-t'$, $1-t''$, $1-u'$ and $1-u''$
respectively the minimal label on trees attached to the six frontier sides
delimited by these marked points as shown, 
with $s=\max(s',s'')$, $t=\max(t',t'')$ and
$u=\max(u',u'')$. The quantities $\vert s'-s''\vert$, $\vert t'-t''\vert$
and $\vert u'-u'\vert$ measure the lengths of the common parts
(represented by solid magenta arrows) of three particular geodesics obtained 
from the concatenation of chains of successors of the 
marked labels $0$. The situation represented here corresponds 
to $s''>s'$, $t''>t'$ and $u''>u'$.}{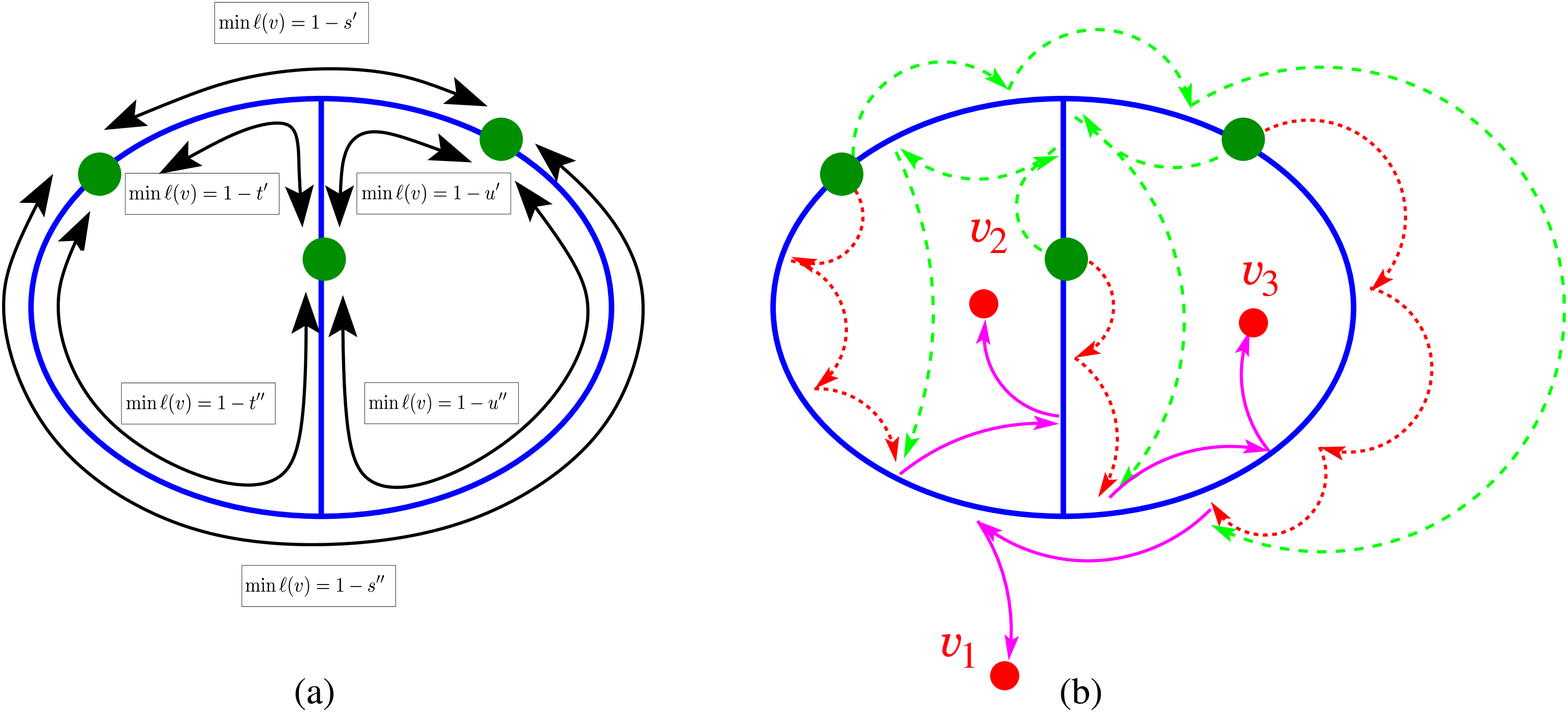}{14.cm}
\figlabel\allconf
If we now consider, say the first label $0$ on each of the three 
pairwise frontiers between faces, the global frontier of each face can
then be divided into two parts lying inbetween the two marked labels $0$ 
on this frontier. We may then distinguish the minimal label on trees attached 
to the first part of the frontier from that on trees attached to 
the second part of the frontier (see Fig.~\allconf-(a) for an illustration).
For instance, the minimal label $1-s$ in the first face corresponds 
to a minimal label $1-s'$ on one part and $1-s''$ on the other part
with $s=\max(s',s'')$. We have similar minima $1-t'$, $1-t''$ 
and $1-u'$, $1-u''$ in the other faces. The quantities 
$\vert s'-s''\vert$, $\vert t'-t''\vert$ and $\vert u'-u''\vert$
measure the lengths of the pairwise common parts of three particular geodesics 
made of chains of successors of corners at the marked labels $0$
(see Fig.~\allconf-(b)).
Similarly, the quantities $\min(s',s'')+\min(t',t'')$, $\min(t',t'')+
\min(u',u'')$ and $\min(u',u'')+\min(s',s'')$ are the lengths of
the proper parts of the same three geodesics. 
With the above refinements, the generating functions
now reads 
\eqn\fstuu{\eqalign{&
\Delta_{s'}\Delta_{s''}\Delta_{t'}\Delta_{t''}\Delta_{u'}\Delta_{u''}
F(s',s'',t',t'',u',u'')\ {\rm where}\ \cr &  
F(s',s'',t',t'',u',u'')= 
X_{s'',t''}\, X_{t'',u''}\, X_{u'',s''}\, Y_{s',t',u'}\, Y_{s'',t'',u''}
\ .\cr}}
and its continuous counterpart reads
\eqn\fssttuu{\eqalign{&\partial_{S'}\partial_{S''}\partial_{T'}
\partial_{T''}\partial_{U'}\partial_{U''}
{\cal F}(S',S'',T',T'',U',U'';\alpha)\ {\rm where} \cr
&{\cal F}(S',S'',T',T'',U',U'';\alpha)= 3^3 {\cal Y}(S',T',U';\alpha)
{\cal Y}(S'',T'',U'';\alpha) \cr &
{\cal Y}(S,T,U;\alpha)=
{1\over 3 \alpha}
{\sinh(\alpha S)\sinh(\alpha T)\sinh(\alpha U)
\sinh(\alpha(S+T+U)) 
\over \sinh(\alpha(S+T))\sinh(\alpha(T+U)\sinh(\alpha(U+S))}
\cr }}
which yields directly the joint law 
$\rho(D'_{12},D'_{23},D'_{31},\delta_1,\delta_2,\delta_3)$
for $D'_{12}=\min(S',S'')+\min(T',T'')$, 
$D'_{23}=\min(T',T'')+\min(U',U'')$, $D_{31}=\min(U',U'')+\min(S',S'')$,
$\delta_1=\vert S'-S''\vert$, $\delta_2=\vert T'-T''\vert$ and
$\delta_3=\vert U'-U''\vert$.
Note that the sign of $S'-S''$ (respectively $T'-T''$, $U'-U''$)
indicates in which of the domains delimited 
by the open part of the triangle the common part leading to $v_1$
(respectively $v_2$, $v_3$) lies. 
Let us now discuss in more details a number of marginal laws
inherited from $\rho(D'_{12},D'_{23},D'_{31},\delta_1,\delta_2,\delta_3)$.
\fig{Plots of the conditional probability density 
$\theta(\delta_1,\delta_2\vert D_{12})$ for the lengths of the two common parts
of a geodesic of fixed length $D_{12}$, here for (a) $D_{12}=1.0$, (b) $2.0$ 
and (c) $5.0$.}{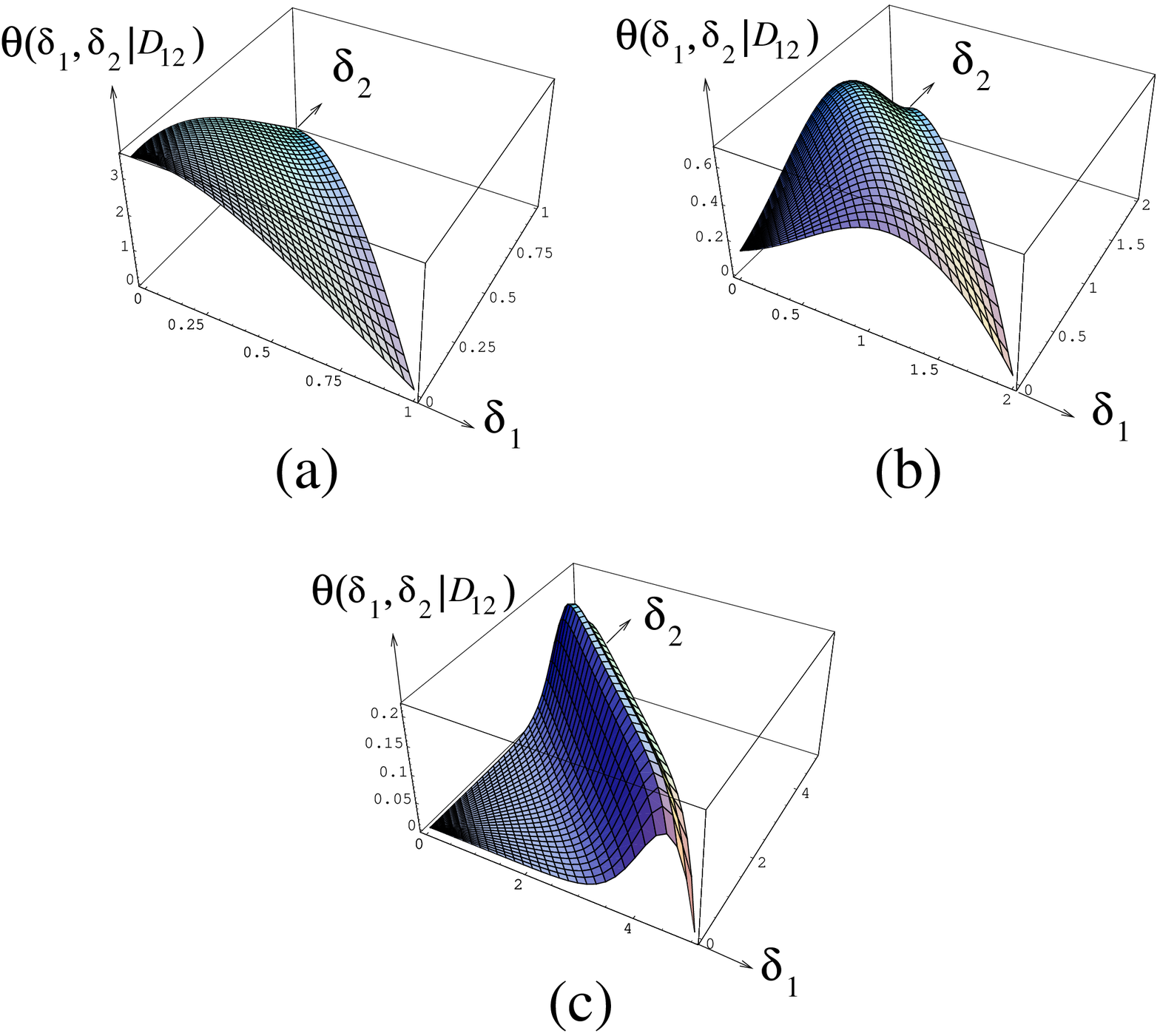}{13.cm}
\figlabel\plotstrois
A first marginal law is that for the lengths $\delta_1$, $D'_{12}$
and $\delta_2$ of the three parts of the geodesic between $v_1$
and $v_2$. It is obtained by first integrating \fssttuu\ over 
$U'$ and $U''$, which yields $\partial_{S'}\partial_{S''}\partial_{T'}
\partial_{T''}{\cal F}(S',S'',T',T'',\infty,\infty)$, 
then integrating over $S'$, $S''$, $T'$ and $T''$ with fixed values
$\min(S',S'')=\sigma$, $\max(S',S'')=\sigma+\delta_1$, $\min(S',S'')=\tau$, 
$\max(S',S'')=\tau+\delta_2$, and finally integrating over $\sigma$ and $\tau$
with the condition $\sigma+\tau=D'_{12}$.
We obtain the grand canonical function
\eqn\threeparts{\eqalign{
{3\over 2} \alpha &
\left\{ {1\over \sinh^3(\alpha D'_{12}) \sinh^3(\alpha(D'_{12}+\delta_1+
\delta_2))}+
{1\over \sinh^3(\alpha (D'_{12}+\delta_1)) \sinh^3(\alpha(D'_{12}+\delta_2))}
\right\} \cr
\times & \Big\{ 2 \alpha D'_{12} \Big( 2 \cosh (\alpha \delta_1) 
\cosh(\alpha \delta_2) + \cosh(\alpha (2D'_{12}+\delta_1+\delta_2))\Big) 
\cr
& + 
2\sinh(\alpha(\delta_1\!+\!\delta_2))\!
-\!2\sinh(\alpha (2D'_{12}\!+\!\delta_1\!+\!\delta_2))
\!-\!\cosh(\alpha(\delta_1\!-\!\delta_2))\!\sinh(2 \alpha D'_{12})
\Big\} \cr }}
from which we can get the canonical joint probability density
$\theta(\delta_1,\delta_2,D'_{12})$ as before. It is interesting to 
consider this probability density conditionally on the value of the total
length $D_{12}$ of the geodesic between $v_1$ and $v_2$, namely:
\eqn\condvarsi{\theta(\delta_1,\delta_2 \vert D_{12})= 
{\theta(\delta_1,\delta_2, D_{12}-\delta_1-\delta_2) \over
\rho(D_{12})}}
where $\rho(D)$ is the canonical two-point function \twopoint.
This conditional probability density is plotted in Fig.~\plotstrois\ for
$D_{12}=1.0$, $2.0$ and $5.0$. 
\fig{The conditional probability density $\theta(\delta_1,\delta_2\vert D_{12})$
for a large value of $D_{12}$ (here $D_{12}=10.0$) becomes uniform in
the ``transverse" direction (corresponding to fixing the value of
$\delta_1+\delta_2$) and characterized by the scaling function 
$\chi(\lambda)$ in the "longitudinal" direction (corresponding to 
varying the value of $\delta_1+\delta_2$), with a scaling
variable $\lambda=(9D_{12})^{1/3} D'_{12}= (9D_{12})^{1/3}(D_{12}
-\delta_1-\delta_2)$.}{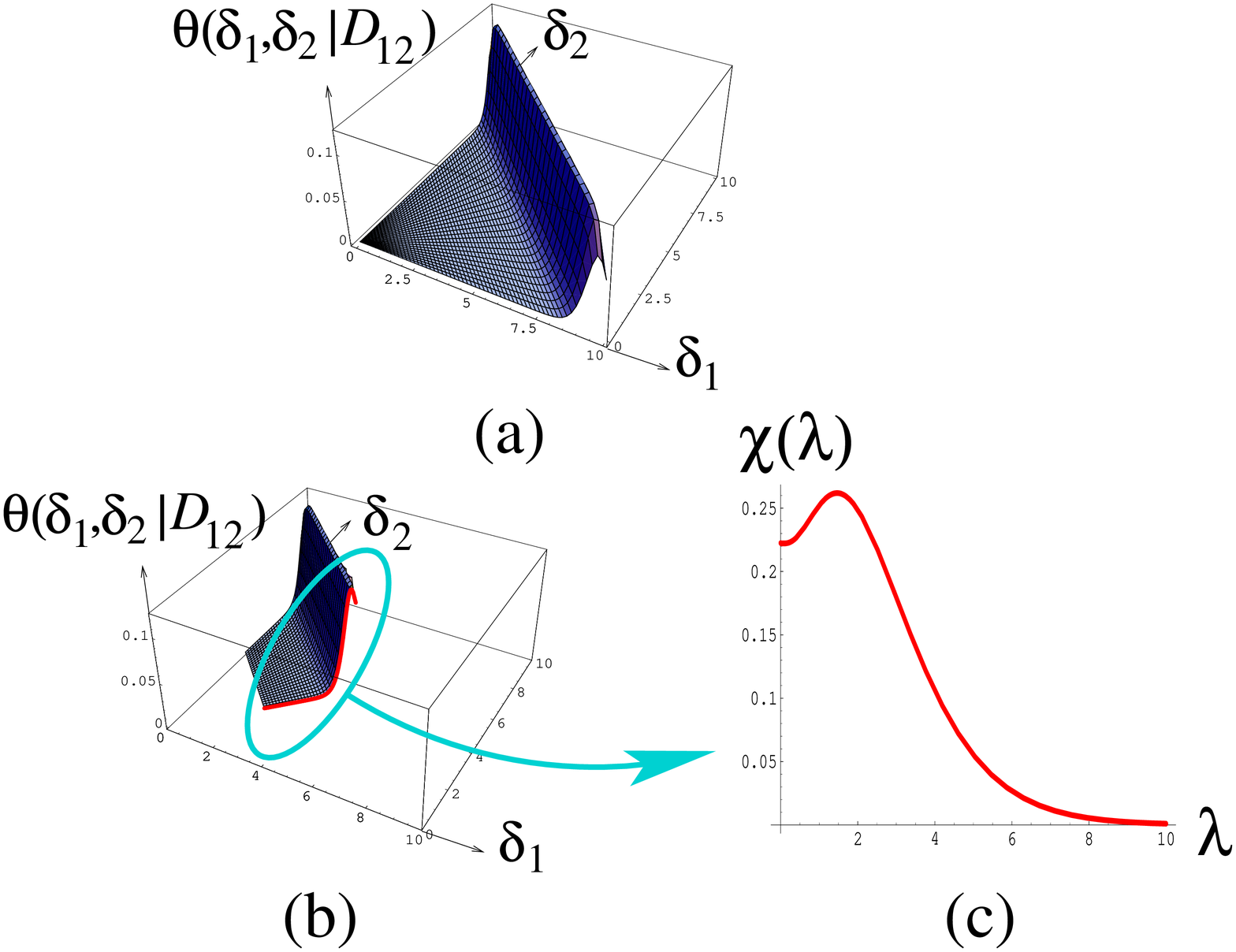}{13.cm}
\figlabel\thetadlarge
At large $D_{12}$, it takes the simple form
\eqn\largeDvarsi{\eqalign{& \theta(\delta_1,\delta_2 \vert D_{12})
\sim {1\over D_{12}} \times (9D_{12})^{1/3} \chi\big(
(9D_{12})^{1/3} (D_{12}-\delta_1-\delta_2)\big)\cr &
{\rm where} \ \ \chi(\lambda)={1\over 3}\left({1\over \sinh^3 (\lambda/2)}+
{8\over e^{3 \lambda/2}}\right)\left(\lambda\cosh(\lambda/2)-2 \sinh(\lambda/2)
\right)\cr}}
In this limit, the geodesic consists mainly of two common parts 
linked by a small open part whose length is of order $D_{12}^{-1/3}$,
with a distribution given by the scaling function $\chi(\lambda)$.
The position of this open part is moreover uniform along the geodesic.
This property is illustrated in Fig.~\thetadlarge.
\fig{Plot of the probability density $\theta(D'_{12})$ for the 
length $D'_{12}$ of the proper part of the geodesic between 
$v_1$ and $v_2$ in the scaling limit of large
triply-pointed quadrangulations.}{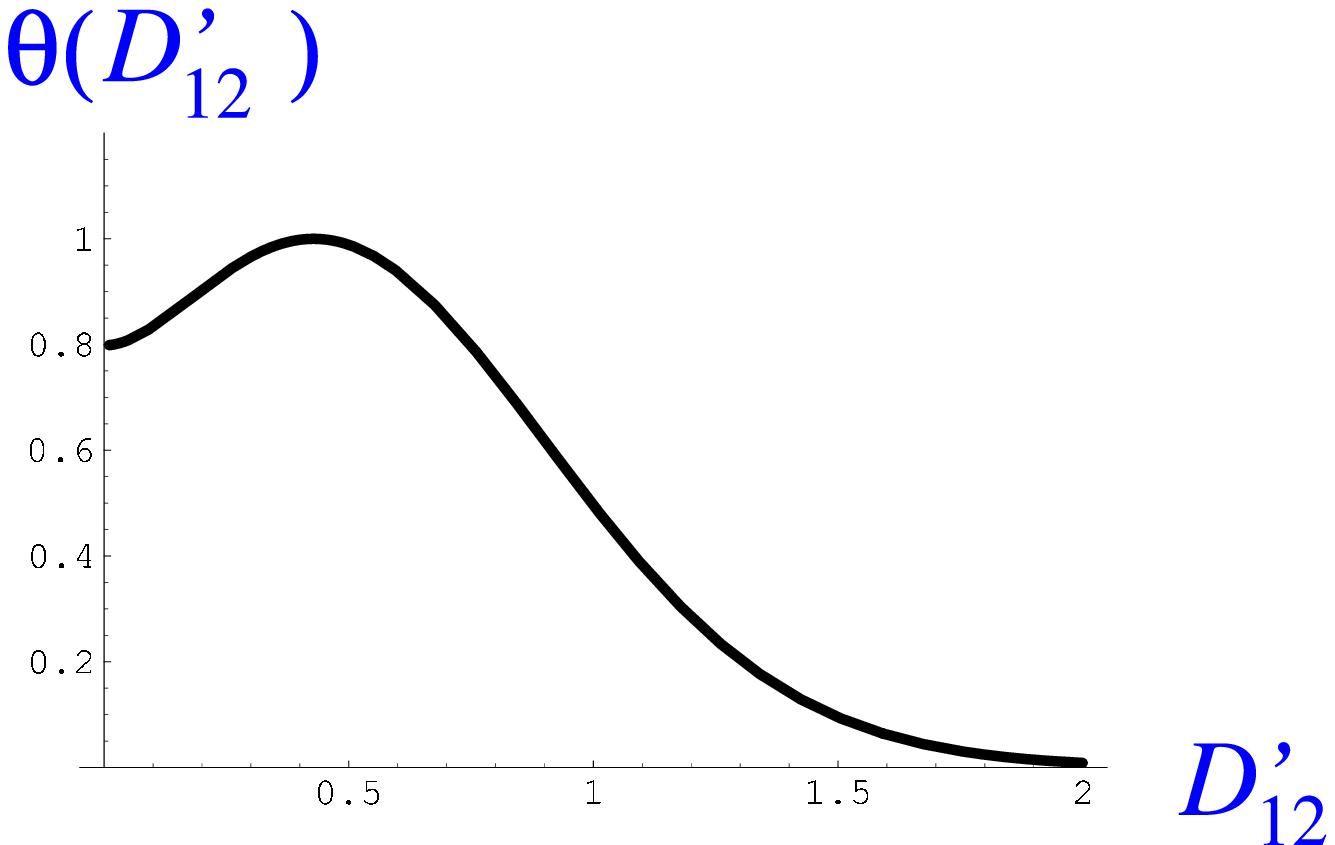}{8.cm}
\figlabel\thetadprime
Upon integrating \threeparts\ over $\delta_1$ and $\delta_2$, we can get
get the marginal law for $D'_{12}$ only. In the grand canonical formalism, 
it reads:
\eqn\lawdprime{\eqalign{{3\over 16 \alpha \sinh^4(\alpha D'_{12})} & 
\Big\{ 2 \alpha D'_{12}\left( 8+13 e^{-2 \alpha D'_{12}} 
-4  e^{-4 \alpha D'_{12}}+ e^{-6 \alpha D'_{12}}\right) \cr
& \ 
- 
\left(1-e^{-2 \alpha D'_{12}} \right)
\left(20-3 e^{-2 \alpha D'_{12}}+e^{-4 \alpha D'_{12}} \right) \Big\}
\cr}}
from which we obtain the canonical probability density $\theta(D'_{12})$.
This probability density is plotted in Fig.~\thetadprime. We have in 
particular 
\eqn\avdprime{\langle D'_{12} \rangle 
={1\over 3}\, \langle D \rangle= 0.590494\cdots}
i.e.\ the length of the open part represents on average one third of
the length of a geodesic, in agreement with \avdelta.
Upon integrating \threeparts\ over $D'_{12}$ and $\delta_2$ and turning
to the canonical formalism, we can recover the marginal law $\sigma(\delta_1)$
of Section 3.1. Similarly, upon integrating \threeparts\ over $D'_{12}$, 
$\delta_1$ and $\delta_2$ with a fixed value of $D_{12}=D'_{12}+\delta_1
+\delta_2$, and upon turning to the canonical formalism, we recover
the two-point function $\rho(D_{12})$, as it should.
\fig{A schematic picture of the eight possible arrangements for
the three common parts of the geodesics with respect
to their open part. In the cases (a) and (h), the three common parts
lie on the same side of the open part. In the 
remaining cases, two of the common parts lie on the same side and the
third one on the other side. Note that, due to the orientation
of the sphere, (a) and (h)  (respectively (b) and (e), (c) and (f), (d) 
and (g)) can be distinguished through the cyclic order of the three sources. 
}{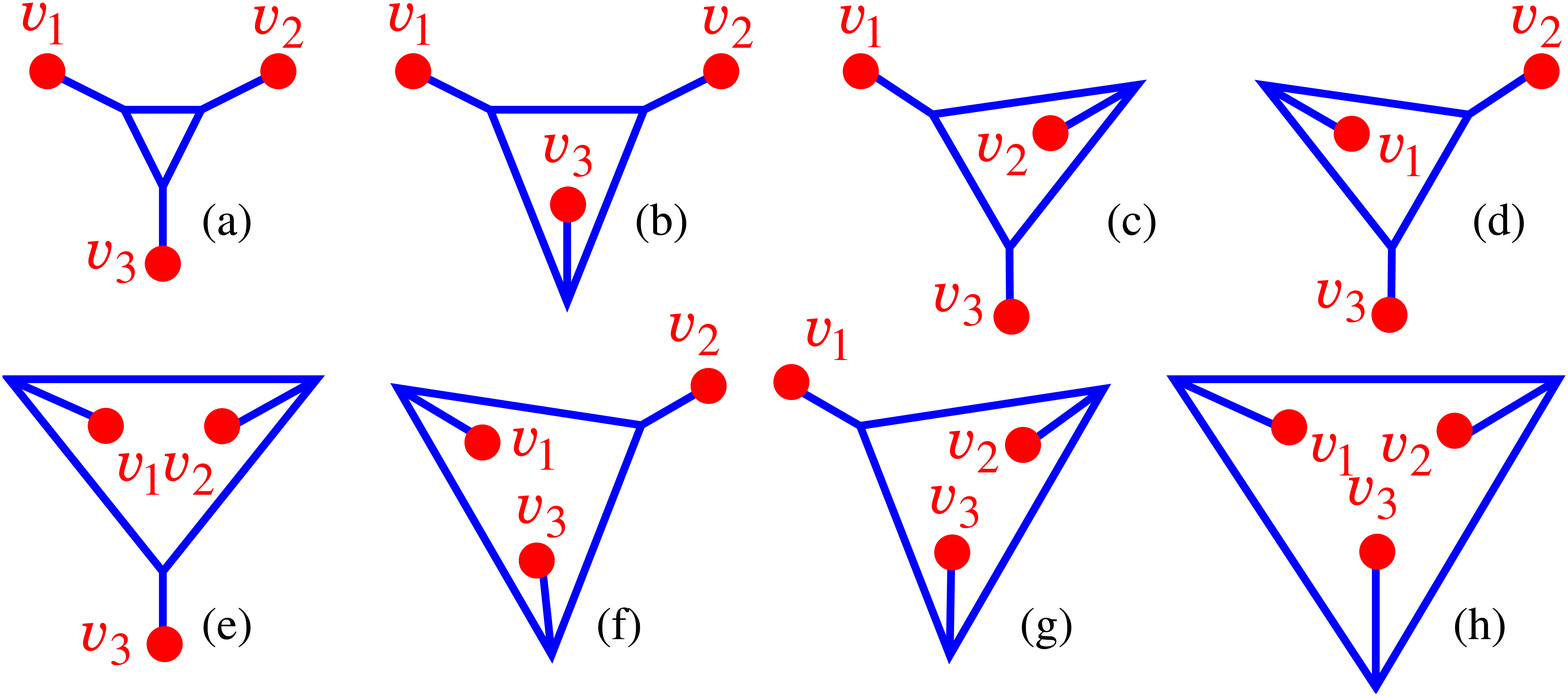}{12.cm}
\figlabel\toposbis
To conclude this section, let us finally discuss the global arrangement of the 
three geodesics on the sphere. As illustrated in Fig.~\toposbis,
there are eight possibilities: in two cases 
((a) and (h) in Fig.~\toposbis), the three common parts lie
in the same domain, while in the remaining six cases 
((b) to (g) in Fig.~\toposbis), two of the common parts
lie in the same domain and the third one in the other domain. 

We may wonder what the probability is of observing
a given arrangement in the canonical ensemble. 
Any of the above arrangements corresponds simply
to a choice of sign for $S'-S''$, $T'-T''$ and $U'-U''$. To obtain, say
arrangement (a), we may integrate \fssttuu\ with the conditions
$S'=\max(S',S'')=S$, $T'=\max(T',T'')=T$, $U'=\max(U',U'')=U$, leading
to the grand canonical function
\eqn\topoa{3^3 \left(\partial_S\partial_T\partial_U 
{\cal Y}(S,T,U;\alpha)\right)\ {\cal Y}(S,T,U;\alpha)\ .} 
To obtain arrangement (b), the third condition must be replaced by
$U''=\max(U',U'')=U$, leading to a grand canonical function 
\eqn\topob{3^3 \left(\partial_S\partial_T 
{\cal Y}(S,T,U;\alpha)\right)\ \left(\partial_U {\cal Y}(S,T,U;\alpha)
\right)\ .}
All the remaining arrangements follow by symmetry and their contributions
add up to the grand canonical three-point function 
\eqn\topoall{3^3 \partial_S\partial_T\partial_U \left(
{\cal Y}(S,T,U;\alpha)\right)^2\ ,}
which is the continuous limit of Eq.~\fstu. 
To obtain the probability of having a given arrangement, we simply have
to integrate its individual contribution over $S$, $T$, $U$, and
divide by the integral of \topoall.  Note that this ratio, obtained
in the grand canonical ensemble yields directly the correct canonical 
probability
since all grand canonical individual contributions integrate to a numerical
constant times the same function $1/\alpha^2$.
A simple calculation shows that each of the arrangements (a) and (h) occurs
with a probability $1/4$, while each of the arrangements (b)-(g) occurs 
with a probability $1/12$. 

As for the partitioning of the area over the two domains, we
find that, if we disregard the particular arrangement at hand, 
the probability density for the proportion $\eta$ of
the total area lying in one of the two domains is again given by 
the symmetric Beta distribution \propeta\ with parameters $\{1/4,1/4\}$. 
On the other hand, if we
consider a particular arrangement, the partitioning of the area 
in no longer symmetric over the two domains. In the case of arrangement
(a) or (h), we find that, on average, $ \sim 94.259\%$ of the total
area lies in the domain containing the three common parts, while,
in the case of the arrangement (b), (c), (d), (e), (f) or (g), 
an average of $\sim 67.224\%$ of the total area lies in the domain
containing the two common parts. 

\newsec{Conclusion and discussion}
In this paper, we derived a number of probability distributions
for the lengths and areas of triangles made of the three geodesics
connecting three uniformly drawn random points, as well as of
minimal separating loops. These laws are expected to be universal
features of the Brownian map and provide quantitative results
characterizing the phenomenon of confluence. This phenomenon 
is remarkable as it places the Brownian map half way between
smooth surfaces and trees. In smooth surfaces, geodesics cannot merge 
and the three sides of a triangle only meet at their endpoints 
so that there are no common parts. In contrast, in trees, 
the three sides of a triangle meet at a central common vertex so
that there is no open part. As for a minimal separating loop
on a tree, it corresponds generically to a back-and-forth travel 
to the above central common vertex and hence has no open part.
On a smooth surface, depending on the shape 
of the surface, a minimal separating loop is either a back-and-forth travel 
along a geodesic, with no open part, or a simple curve
with no common part. Having both open and common parts 
of non-zero length is a peculiarity of the Brownian map.

It is tempting to relate the above results to the so-called 
``baby universe structure" of two-dimensional quantum gravity
well-known in the physics literature [\xref\ADJ,\xref\JainMa]. 
In this picture, a 
baby universe is a region of the surface separated by a small
neck, and a typical surface consists of many such baby universes 
attached to a mother universe and arranged in a tree-like fashion. 
The influence of baby universes on the behavior of the three-point
function was already discussed in Ref.~\Aoki. Qualitatively, the confluence 
phenomenon could simply result from the fact that a
typical point lies in a baby universe and all geodesics leading to it
are forced to pass through the same chain of small necks. The 
length of the common part of geodesics could then be interpreted as a 
measure of the spatial extent of baby universes. More precise statements
would require a rigourous definition of baby universes at a discrete level. 
A first possibility consists in looking only at so-called ``minimum neck baby 
universes" ({\it minbus}) \JainMa. It was shown however that a typical minbu 
remains finite \BaFlScSo, and hence its extent vanishes in the continuum limit. 
One should then look at more general baby universes with larger necks but 
one then faces the problem that there is no canonical decomposition of a 
general map in such baby universes.

Our approach consisted in obtaining discrete results for random 
quadrangulations and taking their scaling limit. So far we 
lack a general formalism  which would allow us to compute the same
results directly in the continuum. 
Despite recent progress [\xref\DSKPZ,\xref\DBKPZ], the so-called 
Liouville field theory
does not yet seem to be able to address such questions. 
Moreover, our results are restricted to the so-called universality class 
of pure gravity. It would be desirable to extend them to other universality
classes of random surfaces coupled to critical matter models \DGZ\ 
(characterized
by their central charge $c$, the pure gravity having $c=0$) such as the 
celebrated Ising model ($c=1/2$) \BOUKA.
Discrete approaches based on bijections with blossom trees 
[\xref\BMS-\xref\HObipar] or labeled trees [\xref\MOB,\xref\FOMAP] 
exist for these problems but those have not been used, so far, to 
extract geometrical information. Some of these models (with a central 
charge $c>1$) are expected to behave like branched polymers, and hence
should have the geometry of trees described above. 

\listrefs
\end